\documentclass[12pt,preprint]{aastex}
\pdfoutput=1

\shorttitle{Star formation feedback and new gas heating mechanisms in merger/starbursts}
\shortauthors{Papadopoulos, Zhang, Xilouris, et al.}
\begin{document}


\title{Molecular gas heating mechanisms, and star formation feedback in merger/starbursts:
 NGC\,6240 and Arp\,193 as case studies}

\author{Padelis \ P.\ Papadopoulos\altaffilmark{1}}
\affil{School of Physics and Astronomy, Cardiff University,\\
 Queen's Buildings, The Parade, Cardiff, CF24 3AA, UK }
\email{Padelis.Papadopoulos@astro.cf.ac.uk}

\author{Zhi-Yu Zhang\altaffilmark{2}}
\affil{Purple Mountain Observatory/Key Lab for Radio Astronomy, 2 West Beijing Road, Nanjing 210008, China}
\email{zyzhang@pmo.ac.cn}

\author{E. M. Xilouris}
\affil{Institute of Astronomy, Astrophysics, Space Applications and Remote Sensing,
I.Metaxa \& Vas.Pavlou str., GR-15236, Athens, Greece}
\email{xilouris@astro.noa.gr}

\author{Axel Weiss}
\affil{Max Planck Institute f\"ur Radioastronomie,  Auf dem H\"ugel 69,  D-53121 Bonn,
 Germany}
\email{aweiss@mpifr-bonn.mpg.de}

\author{Paul van der Werf}
\affil{Leiden Observatory,  Leiden University, P.O.~Box~9513, NL-2300 RA Leiden,\\ The~Netherlands}
\email{pvdwerf@strw.leidenuniv.nl}

\author{F. P. Israel}
\affil{Leiden Observatory,  Leiden University, P.O.~Box~9513, NL-2300 RA Leiden,\\ The~Netherlands}
\email{israel@strw.leidenuniv.nl}

\author{T. R. Greve}
\affil{Department of Physics and Astronomy, University College London, Gower Street, \\
London WC1E 6BT, UK}
\email{tgreve@star.ucl.ac.uk}

\author{Kate G. Isaak}
\affil{Research and Scientific Support Department, European Space Agency,\\ Keplerlaan 1,
2200~AG, Noordwijk, The~Netherlands}
\email{kisaak@rssd.esa.int}

\and

\author{Y. Gao}
\affil{Purple Mountain Observatory, Chinese Academy of Sciences, Nanjing, Jiangsu 210008, China}
\email{pmogao@gmail.com}

\altaffiltext{1}{European Southern Observatory, Headquarters,
Karl-Schwarzschild-Strasse 2, 85748,\\ Garching bei M\"unchen, Germany}
\altaffiltext{2}{The UK Astronomy Technology Centre, Royal Observatory, Edinburgh, 
EH9 3HJ, UK}

\begin{abstract}

 We  used   the  SPIRE/FTS   instrument  aboard  the   Herschel  Space
 Observatory (HSO)  to obtain  the Spectral Line  Energy Distributions
 (SLEDs) of CO from J=4--3  to J=13--12 of Arp\,193 and NGC\,6240, two
 classical merger/starbursts  selected from our  molecular line survey
 of  local  Luminous   Infrared  Galaxies  (LIRGs:  $\rm  L_{IR}$$\geq
 $10$^{11}$\,L$_{\odot}$). The high-J CO  SLEDs are then combined with
 ground-based low-J  CO, $^{13}$CO, HCN,  HCO$^{+}$, CS line  data and
 used  to  probe the  thermal  and  dynamical  states of  their  large
 molecular  gas  reservoirs.   We  find  the  two  CO  SLEDs  strongly
 diverging from J=4--3 onwards, with NGC\,6240 having a much higher CO
 line excitation  than Arp\,193, despite their similar  low-J CO SLEDs
 and  $\rm L_{FIR}/L_{CO,1-0}$,  $\rm L_{HCN}/L_{CO}$  (J=1--0) ratios
 (proxies of  star formation efficiency and dense  gas mass fraction).
 In Arp\,193,  one of the three  most extreme starbursts  in the local
 Universe,  the  molecular  SLEDs   indicate  a  small  amount  ($\sim
 $5\%-15\%) of dense gas (n$\geq$$10^{4}$\,cm$^{-3}$) unlike NGC\,6240
 where most of  the molecular gas ($\sim $60\%-70\%)  is dense (n$\sim
 $($10^4$--$10^5$)\,cm$^{-3}$).   Strong  star-formation feedback  can
 drive  this disparity  in their  dense gas  mass fractions,  and also
 induce extreme  thermal and dynamical  states for the  molecular gas.
 In  NGC\,6240, and  to a  lesser degree  in Arp\,193,  we  find large
 molecular gas masses whose thermal states cannot be maintained by FUV
 photons from Photon Dominated Regions (PDRs).  We argue that this may
 happen often  in metal-rich merger/starbursts,  strongly altering the
 initial conditions  of star formation.   ALMA can now  directly probe
 these  conditions across cosmic  epoch, and  even probe  their deeply
 dust-enshrouded  outcome,  the  stellar  IMF averaged  over  galactic
 evolution.

\end{abstract}


\section{Introduction}

The discovery  of bright CO J=1--0  line emission in the  Orion nebula
(Wilson,  Jefferts, \&  Penzias  1970)  opened up  the  rich field  of
molecular astrophysics with  molecular lines as the  primary probes of
the physical  conditions of  Giant Molecular  Clouds (GMCs),  the most
massive structures in  galaxies and the sites of  star formation.  The
much weaker  rotational transitions from high-dipole  moment molecules
of  CS  and CN,  which  probe  much  denser  gas, were  detected  soon
afterwards (Wilson et al.   1971).  As receiver sensitivities improved
multi-J  transitions of  such  high-dipole molecules  (mostly HCN  and
HCO$^{+}$) were used along with those of CO to probe the full range of
physical conditions of  the molecular gas in  nearby star-forming (SF)
galaxies (e.g.  Solomon et al 1992; Gao \& Solomon 2004; Gracia-Carpio
et  al.   2007; Krips  et  al.   2008;  Greve  et al.   2009).   These
observational   studies,  and   theoretical   investigations  of   the
supersonic turbulent  GMCs either  as individual  entities (Li  et al.
2003; Larson 2005;  Jappsen et al.  2005) or  embedded within galaxies
(Krumholz  \&  McKee 2005),  showed  the  dense molecular  gas  (n$\ga
$10$^{4}$\,cm$^{-3}$)  as the  phase where  stars form.   Its physical
conditions are thus the crucial  input for all star formation theories
and the  resulting stellar Initial  Mass Function (IMF)  (Larson 2005;
Elmegreen et al. 2008).

 However,  the weakness  of high-dipole  moment molecular  lines (e.g.
 HCN J=1--0 is  $\sim $5-100 times fainter than CO  J=1--0) that trace
 high density gas prevented large extragalactic surveys of such lines,
 while strong  atmospheric absorption limits observations  of the more
 luminous CO SLEDs mostly up to  J=3--2 (e.g.  Yao et al.  2003; Leech
 et  al.  2010),  i.e.  the  first CO  transition that  starts tracing
 solely  the   dense  and   warm  SF  gas   ($\rm  n_{crit}$(3-2)$\sim
 $10$^{4}$\,cm$^{-3}$,  $\rm E_{3}/k_B$$\sim  $33\,K).  Such  low-J CO
 line spectroscopy has no diagnostic value regarding the conditions of
 the dense  gas, and little  overlap with the  CO SLEDs of  distant SF
 galaxies where  CO J=3--2, 4--3  and higher-J transitions  are mostly
 detected,  redshifted  into  more  transparent  mm/submm  atmospheric
 windows (e.g.  Solomon \& Vanden Bout 2005; Weiss et al.  2007).

 The importance of  high-J CO lines in probing the  dense gas physical
 conditions was recently underscored by  a small extension of CO SLEDs
 to include J=4--3 and 6--5 for a few local Luminous Infrared Galaxies
 (LIRGs:  $\rm  L_{IR}$$\geq $$10^{11}$\,L$_{\odot}$).  These  revealed
 large dense  {\it and}  warm gas reservoirs  that are  irreducible to
 ensembles    of    Photon-Dominated    Regions   (PDRs)    in    some
 merger/starbursts (Papadopoulos et al.  2012a).  Such conditions were
 also found in Mrk\,231 and Arp\,220  using CO SLEDs from J=1--0 up to
 J=13--12 obtained  with SPIRE/FTS and ground-based  observations (van
 der Werf et al.  2010; Rangwala et al.  2011).  Finally high-J CO and
 heavy rotor molecular lines are necessary for better estimates of the
 $\rm     X_{CO}$=$\rm      M_{tot}(H_2)/L_{CO,1-0}$     factor     in
 merger/starbursts where, unlike isolated spirals, the dense phase can
 contain  much of  their  total molecular  gas  mass (Papadopoulos  et
 al.~2012b).

 The thermal,  dynamical and  chemical state  of the  dense gas  in SF
 galaxies, its mass contribution  to $\rm M_{tot}$(H$_2$), the effects
 of  SF and  AGN feedback,  and complete  CO SLEDs  from J=1--0  up to
 high-J  transitions as  local benchmarks  for high-z  CO observations
 were the key drivers for  our Herschel Comprehensive (U)LIRG Emission
 Survey (HerCULES), an  open time Key program (PI: Paul  van der Werf)
 on the  ESA Herschel Space Observatory  (HSO)\footnote{Herschel is an
   ESA  space   observatory  with  science  instruments   provided  by
   European-led  Principal Investigator  consortia and  with important
   participation from  NASA} (Pilbratt et  al.  2010), augmented  by a
 large ground-based  low-J CO and $^{13}$CO  line survey (Papadopoulos
 et al.   2012a).  Here  we report on  HSO SPIRE/FTS  and ground-based
 observations    of   Arp\,193    and    NGC\,6240,   two    prominent
 merger/starbursts from  HerCULES whose  similar low-J CO  SLEDs, $\rm
 \epsilon_{SF,  co}  $=$\rm  L_{FIR}/L_{co,   1-0}$  (a  proxy  of  SF
 efficiency  SFE=SFR/$\rm  M_{tot}(H_2)$)  and  $\rm  r_{HCN/CO}$=$\rm
 L^{'}_{\rm  HCN,1-0}$/$\rm  L^{'} _{\rm  CO,1-0}$  (a  proxy of  $\rm
 f_{dense}$=M(n$>$10$^{4}$\,cm$^{-3}$)/M$_{\rm  tot}$(H$_2$))  ratios,
 make them good  testbeds for exploring diferences in  their dense gas
 properties.  This  work is structured  as follows: 1) we  present the
 SPIRE/FTS and  ground-based molecular  line data and  their reduction
 (Section  2), 2)  we  construct  the full  CO  SLEDs  from J=1--0  to
 J=13--12 and use  them along with our $^{13}$CO,  HCN, HCO$^{+}$, and
 CS line data to find the average conditions and mass of the molecular
 gas  components  using  radiative  transfer models  (Section  3),  3)
 determine their  thermal states and energy  requirements (Section 4),
 and 4) discuss general implications for the ISM in merger/starbursts,
 and present  our conclusions (Section  5).  We adopt a  flat $\Lambda
 $-dominated  cosmology  with  $\rm  H_0$=71\,km\,s$^{-1}$\,Mpc$^{-1}$
 and~$\Omega_{\rm m}$=0.27.
 
\section{Observations, data reduction, line flux extraction}

 NGC\,6240 and Arp\,193 were observed with the SPIRE/FTS aboard HSO as
 part  of  the  now  completed HerCULES  Key  project.   Arp\,193  was
 observed on November 11, 2010  and NGC\,6240 on February 26-27, 2011,
 using the staring mode with the  SPIRE/FTS aboard the HSO with a high
 spectral       resolution       mode       of       $\rm       \delta
 (\lambda^{-1})$=0.04\,cm$^{-1}$  over  both observing  bands.   These
 were:  a  long  wavelength  band  covering  (14.9--33.0)\,  cm$^{-1}$
 (equivalent    to   $\lambda    $=(671--303)\,$\mu    $m   or    $\nu
 $=(467--989)\,GHz),   and   a    short   wavelength   band   covering
 (32.0-51.5)\,cm$^{-1}$ (equivalent to  $\lambda $=(313--194)\,$\mu $m
 or $\nu $=(959--1544)\,GHz).  For  NGC\,6240 the integration time was
 $\rm  N_{FTS}$=97  repetitions  with on-source  integration  of  $\rm
 T_{int}$=12920\,secs, while for Arp\,193 these were $\rm N_{FTS}$=108
 and $\rm  T_{int}$=14386\,secs.  Dark reference measurements  of $\rm
 N_{FTS}$(NGC\,6240)=113\,scans      (15110\,secs)       and      $\rm
 N_{FTS}$(Arp\,193)=124\,scans (16622\,secs) were used to subtract the
 thermal emission  of the  sky and the  telescope/FTS.  The  data were
 processed and calibrated using  HIPE version 6.0.  Interferometric CO
 J=1--0 (Arp\,193)  and J=2--1  (NGC\,6240) images (Downes  \& Solomon
 1998;  Tacconi  et   al.   1999)  yield  source   sizes  $\rm  \theta
 _{co}$$\sim  $1.5$^{''}$  (Arp\,193)   and  $\sim  $3$^{''}$-4$^{''}$
 (NGC\,6240), much  smaller than even  the smallest SPIRE/FTS  beam of
 $\sim   $17$^{''}$   at   $\sim   $1500\,GHz.   Thus   we   set   the
 aperture-source geometric  coupling factor  $\rm K_c$$\sim  $1 across
 the entire  FTS spectrum\footnote{See  Equation 1 in  Papadopoulos et
   al.   2010 for  the  expression  of $\rm  K_c$}.   Finally the  two
 spectrometer bands  in the  overlap region  $\sim $(32-33)\,cm$^{-1}$
 are well matched and were averaged.

 In  Figure 1  we show  the full  SPIRE/FTS spectra  of  NGC\,6240 and
 Arp\,193.  Besides  the CO  lines, several (6)  water lines  are also
 clearly detected in NGC\,6240, while in both LIRGs the fine structure
 lines of atomic carbon  [CI] ($^{3}$$P _1\rightarrow $$^3$$P_0$ and $
 ^{3}$$P_2\rightarrow $$^3$$P_1$), and of  N\,II are detected.  We use
 the  CO lines  to extract  the redshifts  of  $\rm z_{co}$=0.0245$\pm
 $0.00013 (NGC\,6240) and $\rm z_{co}$=0.0231$\pm $0.00015 (Arp\,193),
 in  good  agreement  with  the  CO-deduced values  from  single  dish
 measurements (Papadopoulos  et al.  2012a).  For these  redshifts the
 luminosity distances are: $\rm D_L$(NGC\,6240)=105.5\,Mpc (with 1$''$
 corresponding  to 487\,pc)  and  $\rm D_L$(Arp\,193)=99.3\,Mpc  (with
 1$''$ corresponding to 460\,pc).

 The velocity-integrated  line fluxes were extracted by:  a) fitting a
 sinc-gaussian  function  to  the  line  profile  (sinc  for  the  FTS
 response, and gaussian for the line profile), b) integrating over the
 entire  line  (=the FTS  response  convolved  to  an underlying  line
 profile)  without  any  assumptions  about its  profile.   Both  give
 similar values, but for Arp\,193 where a non-gaussian line profile is
 present (see Figs 2, 3 in  Papadopoulos et al.  2012a) we adopt those
 obtained via  the second method.   The CO J=1--0, 2--1,  3--2 (ground
 observations),    and    J=4--3    up   to    J=13--12    (SPIRE/FTS)
 velocity-integrated line fluxes and luminosities are in Table 1.

\subsection{HCN, HCO$^{+}$ and CS line observations}

We performed HCN, HCO$^{+}$ J=4--3  and CS J=7--6 line observations of
NGC\,6240  using   the  12-m  Atacama   Pathfinder  EXperiment  (APEX)
telescope\footnote{This publication is based on data acquired with the
  Atacama  Pathfinder  Experiment  (APEX).   APEX is  a  collaboration
  between   the  Max-Planck-Institut   f{\"u}r   Radioastronomie,  the
  European Southern Observatory, and the Onsala Space Observatory.} on
the Chajnantor Plateau in Chile.   Most observations were done in good
($pwv$$<$0.6\,mm)  to median  ($pwv\sim$1\,mm)  weather conditions  in
April and August  2011.  The FLASH receiver was  employed with the LSB
tuned   to  the   CS  $J$=7--6   frequency  ($\nu_{\rm   rest}$$  \sim
$342.883\,GHz),  and   the  USB  covering  the   HCN,  HCO$^+$  J=4--3
transitions.  The  Fast Fourier Transform  Spectrometer (FTS) backends
are employed  in all spectral  observations, with channel  spacings of
$\sim$0.4 MHz, and a bandwidth of  4 GHz for each sideband, yielding a
velocity   coverage  of   $\sim$3500\,km\,s$^{-1}$.    Typical  system
temperatures where $\rm T_{sys}$=(200-240)\,K,  while the beam size at
345\,GHz  is HPBW=$18''$.   Pointing checks  using a  strong continuum
source were made every hour yielding a typical pointing uncertainty of
$2''$-$3''$.  We estimated a main-beam efficiency $\eta_{mb}$=0.7 from
a continuum measurement on  Mars, and a point-source conversion factor
of $\rm  S_{\nu}/T^{*}_a$=40\,Jy/K.  We performed  all observations in
the wobbler switching mode with  a chopping frequency of 1.5\,Hz and a
chop throw of  2$'$ in azimuth (AZ) yielding  flat baselines.  Finally
the APEX-1  SHeFI receiver was used  to observe the  HCN and HCO$^{+}$
J=3--2 lines (see Figure 2) using the same observational setup.

 The  CS  J=2--1, 3--2  spectra  were  obtained  using the  IRAM  30\,m
 telescope at Pico Veleta (Spain)  in December 2011 under good weather
 conditions ($\tau$$<$0.1).  The EMIR receivers E90, E150 observed the
 two   lines   simultaneously,  with   the   fast  fourier   transform
 spectrometer  (FTS) as  the  backend.  To  obtain  flat baselines,  a
 wobbler switching  mode with a frequency  of 2\,Hz and  beam throw of
 120$''$  was  used.  Pointing  checks  were  made  every 30  minutes,
 yielding  pointing  errors  of  $\sim $3$''$  (rms).   The  main-beam
 efficiencies  $\eta_{\rm  mb}$=0.71,  0.63  at 90\,GHz  and  140\,GHz
 respectively.  The beam-sizes (HPBW)  are $\sim $26$''$ (90\,GHz) and
 $\sim  $17$''$ (140\,GHz).   In Figure  3 we  show the  CS  lines for
 NGC\,6240 (Arp\,193  was not detected).  The data  were reduced using
 CLASS, with  each spectrum  inspected by eye  and about 5\%  -10\% of
 them discarded.  For each source, the spectra were co-added, weighted
 by their noise.  The velocity-integrated line fluxes along with those
 from the literature can be found in Table 2.

\subsection{Literature data}

Our  new  HCN  J=3--2,  4--3   fluxes  for  NGC\,6240  from  APEX  are
significantly lower than those reported by Greve et al.  (2009) (whose
HCN  4--3 data  were used  also in  Papadopoulos 2007),  with our  HCN
J=3--2 flux in excellent agreement  with three other such measurements
(Gracia-Carpio et al.  2008, Krips et al.  2008 and Israel 2012).  For
the  HCN J=4--3  line  we  adopt our  new  measurement  since lack  of
adequate baseline  of the  old JCMT spectrum  makes the  flux obtained
highly uncertain.  For  HCN J=3--2 we adopt the mean  of our value and
those reported  in the literature (we  do the same in  all cases where
multiple consistent observations of the same line exist, see Table 2).
Finally  the  CO  J=6--5   velocity-integrated  fluxes  obtained  with
SPIRE/FTS (Table 1) are significantly  larger than those obtained from
the   ground  using   the   James  Clerk   Maxwell  Telescope   (JCMT)
(Papadopoulos et  al.  2012a,  their Table 5).   We attribute  this to
pointing offsets of  the JCMT, now known to have  affected some of the
CO  J=6--5 measurements  despite its  overall good  pointing (pointing
rms$\sim  $2$''$)  and  the  corrections  made  to  account  for  flux
reduction due  to pointing errors\footnote{Equation 4  in Papadopoulos
  et al. 2012a}.

\section{The state of the molecular gas reservoirs}

In Figure 4 we show the CO SLEDs of the two LIRGs, normalized by their
far-IR      luminosities:      $\rm     L^{(n-IR)}      _{J+1,J}$=$\rm
L_{J+1,J}/L_{FIR}$,  and by  the continuum  at the  corresponding line
rest frequency: $\rm L^{(n-\nu)} _{J+1,J}$=$\rm L_{J+1,J}/[\nu_{J+1,J}
  L_{IR}(\nu _{J+1,J})]$.   These SLEDs  remain similar up  to J=3--2
(within factors  of $\sim $1.2-2)  but then diverge  significantly for
higher-J lines by factors of $\sim  $5-10. The lower excitation of the
high-J CO lines of  Arp\,193 comes as a surprise for  a galaxy that is
one of the three ULIRGs that harbor the most extreme starburst regions
in the local Universe (the other  two being Arp\,220 and Mrk\,273; see
Downes \& Solomon 1998, hereafter DS98).

The high-J CO  SLED divergence and the different HCN  ratios (Table 2)
imply different dense gas conditions and/or dense gas mass fractions
for   these   two  mergers.    This   occurs   despite  similar   $\rm
r_{HCN/CO}$=$\rm L^{'} _{HCN}/L^{'}_{CO}$  (J=1--0), and $\rm \epsilon
_{SF}$=$\rm L_{FIR}/L_{CO,1-0}$  ratios ($\rm  r_{HCN/CO}$$\sim $0.073
(NGC\,6240)  and   $\sim  $0.055   (Arp\,193),  while   $\rm  \epsilon
_{SF}$$\sim  $$10^{6}$ for  both galaxies),  quantities often  used as
proxies  of   $\rm  f_{dense}$=M(n$\geq  $10$^{4}$\,cm$^{-3}$)/M$_{\rm
  tot}$(H$_2$) and SF efficiency SFE=SFR/M(H$_2$) in~galaxies.

 \subsection{The dense gas phase: the HCN, HCO$^{+}$ and CS lines}
 
We start the radiative transfer modeling with the HCN, HCO$^{+}$ lines
since (HCN/HCO$^{+}$)-rich  gas is where  the {\it minimal}  high-J CO
SLEDs are  set.  Any additional  high-J CO line luminosity  would come
from  warmer  (FUV)-heated  outer  layers of  GMCs  where  high-dipole
molecules  like HCN  are  mostly dissociated  because  of their  lower
dissociation potential than of CO (Boger \& Sternberg 2005), confining
them  deeper inside  GMCs, beyond  the reach  of strong  FUV radiation
fields  (attenuated  by  the  outer  GMC  layers).   In  such  regions
supersonic  turbulence has  mostly  dissipated, and  only cosmic  rays
(CRs) heat the gas and control its ionization and chemical state (e.g.
Bergin \& Tafalla  2007).  The strong molecular line  cooling of dense
gas  ($\rm  \Lambda  _{line}$$\propto   $$\rm  [n(H_2)]^2$),  and  the
gas-dust thermal coupling with cold dust (warmed only by the feeble IR
fields able to penetrate that deep into GMCs) $''$thermostate$''$ such
regions around a narrow range of states that serve also as the initial
conditions of  star formation in  galaxies (Larson 2005;  Elmegreen et
al.~2008).

  We   used   the   public   Large  Velocity   Gradient   (LVG)   code
  RADEX\footnote{Extensive  runs   with  RADEX   revealed  convergence
    problems (i.e. the solutions  obtained changing significantly even
    after many iterations).  Our solutions  have been checked for this
    problem.}  (van der Tak et al.  2007) to map the $\rm [n, T_{kin},
    K_{vir}]$ parameter  space compatible  with the heavy  rotor lines
  available  for   NGC\,6240  and   Arp\,193  (Table  2),   with  $\rm
  K_{vir}$=$\rm     \left(dV/dR\right)/\left(dV/dR\right)_{vir}$
  parametrizing the dynamical state of the gas.  We first examine only
  self-gravitating states (0.5$<$$\rm  K_{vir}$$<$2) which are typical
  for dense gas regions inside  Galactic GMCs.  Strictly speaking only
  states with  $\rm K_{vir}$=1 can  be called self-gravitating  but we
  consider a  slightly wider range (by  a factor of 2)  to account for
  the  uncertainties   of  cloud   density  profiles,   geometry,  and
  [CO/H$_2$] abundance that are used  in extracting $\rm K_{vir}$ from
  the   RADEX  input   parameter  N(CO)/$\rm   \Delta  V$=n(H$_2$)$\rm
  [CO/H_2]/[K_{vir}(dV/dR)_{vir}]$ (see Appendix A  for details of our
  model).

 In Figure 5 we show the probability density functions (pdfs) for $\rm
 [n, T_{kin}]$ as constrained by the  HCN line ratios, the heavy rotor
 molecule with  the larger  number of  available transitions  for both
 galaxies.   For  NGC\,6240 where  several  HCO$^{+}$  lines are  also
 available, a nearly identical solution space is recovered (see Figure
 6).  The  HCN SLEDs corresponding to  the solution space in  Figure 5
 are shown  in Figure  7, along  with those for  the HCO$^{+}$  and CS
 lines available  for NGC\,6240.  For  this LIRG the  HCN-derived $\rm
 [n,  T_{kin}]$  solution range  is  a  good  fit  for {\it  all}  the
 high-dipole    moment   molecular    lines,    with   densities    of
 $10^{4}$\,cm$^{-3}$$\leq$n$\leq  $$10^{5}$\,cm$^{-3}$  for  most  LVG
 solutions within the 15\,K$\leq $$\rm T_{kin}$$\leq $150\,K interval.
 Nevertheless,  as   expected  for  radiative  transfer   modeling  of
 optically thick lines  like those of HCN  and HCO$^{+}$, considerable
 degeneracies remain. However, with the current HCN and HCO$^{+}$ line
 data containing  the SLED  turnover, additional measurements  of even
 one higher-J line (e.g.  HCN and/or HCO$^{+}$ J=5--4) can much reduce
 them (see Figure 7).  For  NGC\,6240 our results differ somewhat from
 those Greve et al.  2009 as  our new lower HCN J=4--3 line luminosity
 (and a much lower CS J=7--6  upper limit) now exclude any significant
 gas masses with $\rm n$$>$10$^{5}$\,cm$^{-3}$.

 The low HCN  line ratios of Arp\,193 on the  other hand correspond to
 significantly lower average gas densities  than in NGC\,6240 over the
 entire  $\rm [n,  T_{kin}]$ solution  range (see  Figure~8).  Indeed,
 quite unlike  NGC\,6240 where  solutions within the  15\,K$\leq $$\rm
 T_{kin}$$\leq  $150\,K interval  have n$\ga  $10$^{4}$\,cm$^{-3}$, in
 Arp\,193 most have n$<$10$^{4}$\,cm$^{-3}$, i.e.  {\it are compatible
   even with  an absent dense gas  component in one of  the three most
   prominent  starbursts  in  the  local  Universe.}   This  can  then
 naturally yield its  low-excitation high-J CO SLED.   We note however
 that the  degeneracy of  HCN-derived LVG  solutions still  leave such
 HCN$\rightarrow  $CO   SLED  extrapolations   rather uncertain.   For
 example  the HCN-extrapolated  CO  SLEDs for  NGC\,6240 and  Arp\,193
 while indeed divergent  beyond J=4--3 as proposed  by Papadopoulos et
 al.  2010,  do not  correspond well to  those actually  observed with
 SPIRE/FTS.  Only  the {\it minimal}  CO SLEDs of  a SF galaxy  can be
 confidently deduced in this manner (Geach \& Papadopoulos 2012).

\subsubsection{Unbound states for the dense gas}

Requiring  $\rm  K_{vir}$$\sim  $1  for   the  LVG  solutions  of  the
(HCN/HCO$^{+}$)-line emitting gas is appropriate for the dense regions
of ordinary GMCs in spirals.  In  mergers however GMCs, if they remain
as individual  entities at all, may  be far from ordinary  (Solomon et
al.   1997;  DS98). Moreover  strong  gas  outflows (i.e.   bona  fide
unbound gas) have been found in  extreme starbursts even for dense gas
(Aalto et  al.  2012) while  dense {\it  and} unbound gas  states have
been  inferred for  the Galactic  Center and  the nucleus  of NGC\,253
(Bradford et al.  2005, 2003).  We  thus search also for unbound dense
gas states compatible  with the heavy rotor molecular  SLEDs.  The new
$\rm [n,  T_{kin}]$ pdfs  for 0.5$<$$\rm  K_{vir}$$<$20 (Fig.   9) now
encompass higher densities  for any given $\rm  T_{kin}$ (compare with
Fig.  5), and  extend towards colder/denser states  (whether these are
possible is  discussed in 3.2).   Figure 10 shows similar  effects for
the  (HCO$^{+}$)-constrained $\rm  [n,  T_{kin}]$  pdfs of  NGC\,6240,
while  the HCN-determined  pdfs  of the  two  LIRGs now  significantly
overlap (Fig.  11), especially  for $\rm T_{kin}$$\ga $30\,K, (compare
with Fig.~8).

\subsection{Beyond PDRs: the extraordinary states of the dense gas in mergers}

   Before extending our molecular SLED  modeling to include the two CO
   SLEDs it  is worth revisiting the  paradigm of how these  emerge in
   the  ISM  and  whether  it  remains adequate  in  the  extreme  ISM
   conditions in merger/starbursts.  In  the standard picture low-J CO
   line emission  (J=1--0, 2--1  and some  fraction of  J=3--2) arises
   from the  entire molecular gas  mass distribution, while  high-J CO
   lines  mainly  from  the   FUV-illuminated  warm  outer  layers  of
   molecular clouds  near O,  B stars, the  so-called Photon-Dominated
   Regions (PDRs).  Deeper inside GMCs, where the (HCN/HCO$^{+}$)-rich
   gas  resides,  lack of  strong  FUV  radiation (which  allows  such
   molecules to survive), and strong gas-dust thermal coupling with an
   increasingly colder dust  reservoir, leaves the gas  too cold ($\rm
   T_{kin}$$\sim $10\,K  in Galactic  HCN-bright cores) to  radiate in
   high-J  CO lines.   Turbulent heating  does not  fundamentaly alter
   this picture since is significant  also in the outer, lower density
   GMC  layers (where  turbulence is  thought to  be injected),  while
   dissipating in  the inner denser  regions.  In ordinary  GMCs these
   contain  $\sim $(1-2)\%  of  their  mass, and  set  the SF  initial
   conditions and the IMF mass scale.

  This simple  picture of  GMCs$\sim $[PDRs]+[cold  FUV-shielded dense
    cores], with most  of their mass in PDRs,  was recently challenged
  in some merger/starbursts whose ISM states indicate large amounts of
  dense  ($\geq $$10^{4}$\,cm$^{-3}$)  {\it  and}  warm ($\ga  $80\,K)
  molecular gas (Papadopoulos et al.   2012a).  Even in the absence of
  such  observations, a  general  argument  for extraordinary  thermal
  and/or dynamic states for the  dense gas in merger/starbursts can be
  made starting  from their  typically large  HCN/CO J=1--0  ratios of
  $\rm r_{HCN/CO}$$\sim $0.1--0.2 ($\rm r_{HCN/CO}$$\sim $0.01-0.03 in
  galactic  disks).   For  n$\sim $5$\times  $$10^4$\,cm$^{-3}$,  $\rm
  T_{k}$$\sim   $(10--15)\,K,   and   $\rm  K_{vir}$$\sim   $1   (i.e.
  Galactic-type dense  gas conditions) the HCN-bright  component would
  have:  $\rm   X_{HCN}$$\sim  $(20--60)\,$\rm   X_l$  (\footnote{$\rm
    X_l$=M$_{\odot}$(K\,km\,s$^{-1}$\,pc$^2$)$^{-1}$})          (using
  expressions in Papadopoulos  et al.  2012b for  a thermalized line).
  Then  for $\rm  L^{'}  _{HCN, 1-0}$=(0.5--2)$\times  $$10^{9}$\,$\rm
  L_{l}$  (\footnote{$\rm L_{l}$=K\,km\,s$^{-1}$\,pc$^2$})  typical in
  merger/starbursts:      $\rm     M_{dense}$$\sim      $(1-12)$\times
  $10$^{10}$\,M$_{\odot}$, large  enough to dominate and  even surpass
  the typical  $\rm M_{dyn}$  of their  CO-bright regions  (see DS98).
  The dense  gas in such  HCN-bright galaxies  must then be  {\it much
    warmer and/or  in unbound  states} (i.e. $\rm  K_{vir}$$>$1) whose
  smaller $\rm X_{HCN}$ factors would  then yield $\rm M_{dense}$ well
  within  their   $\rm  M_{dyn}$.    Actually  $\rm   M_{dense}$  even
  approaching a large fraction of  $\rm M_{dyn}$ of a merger/starburst
  would be worrying as one would expect $\sim $50\% of $\rm M_{dense}$
  to be newly-formed  stars (for a typical dense-gas  SF efficiency of
  $\sim $50\%), with some room left for lower density gas and an older
  stellar population.   Thus the high  HCN/CO (J=1--0) line  ratios of
  such galaxies  can mean either  unreasonably large amounts  of cold,
  dense, self-gravitating  gas, or {\it very  different thermal and/or
    dynamical  dense gas  states.}   These, as  we  will argue  later,
  cannot be maintained by FUV photons from~PDRs.

 The aforementioned  reasoning assumes that in  both merger/starbursts
 and isolated spirals HCN lines  are collisionally excited.  Only then
 their very different HCN/CO J=1--0 ratios reflect different dense gas
 mass fractions  (modulo any  differences of  their average  dense gas
 conditions).  Large scale  IR pumping of HCN levels  occuring only in
 merger/starbursts  could upset  this by  yielding high  global HCN/CO
 J=1--0 ratios  and well-excited  HCN rotational transitions  in these
 galaxies,   without  large   amounts  of   their  molecular   gas  at
 n$\geq  $$10^4$\,cm$^{-3}$.  While  this  may occur  in some  extreme
 systems  (Aalto et  al.  2012),  it could  not be  generaly important
 without   $''$breaking$''$   the   tight  linear   HCN-$\rm   L_{IR}$
 correlation found  across the entire  (U)LIRG population and  down to
 individual  Galactic GMCs  (Gao \&  Solomon 2004;  Wu et  al.  2005).
 Large scale IR  pumping of HCN lines only  in merger/starbursts would
 instead  produce a  broken,  non-linear,  $\rm L_{HCN}$-$\rm  L_{IR}$
 correlation towards  ULIRGs, with a larger  dispersion, and certainly
 not extending smoothly down to  individual GMCs where such pumping is
 negligible.  The same picture of large dense gas tracer luminosity in
 ULIRGs with respect to isolated  gas-rich spirals, and a tight linear
 $\rm L_{IR}$-L$_{\rm  line}$ correlation is  found also for  CS lines
 (Zhang et  al.  2013), and benchmarked  for SF regions in  the Galaxy
 (Wu et al.~2010).   The simplest explanation is that both  HCN and CS
 are collisionally excitated in most~LIRGs.

\subsection{The CO SLEDs of NGC\,6240 and Arp\,193: an inside-out
 decomposition }
 
Our previous  discussion makes  clear that  a significant  dense, warm
and/or unbound gas component can  be a general feature of HCN-luminous
merger/starbursts such as Arp\,193 and NGC\,6240.  Such a component is
then {\it  bound to significantly  contribute to their high-J  CO line
  luminosities.}  The  practical importance  of this  is that  the LVG
solution space defined by the HCN line ratios (Figures 5, 9) (also the
solution space for the CS and HCO$^{+}$ lines in NGC\,6240) can now be
used  to   model  also  the   high-J  CO  SLEDs  of   these  galaxies.
Incorporating the  (HCN/HCO$^{+}$/CS)-rich dense gas into  the CO SLED
modeling is  an imperative for  merger/starbursts as it can  contain a
large fraction of their $\rm  M_{tot}(H_2)$ (Solomon et al.  1992; Gao
\&  Solomon  2004; Greve  et  al.   2009),  and  its omission  may  be
responsible for  the often  contradictory conclusions  regarding their
molecular gas conditions  found in the literature (e.g.   Greve et al.
2009  versus   Rangwala  et   al.   2011  regarding   Arp\,220).   For
HCN-constrained   CO   SLED   decompositions,   $\rm   M_{dense}$/$\rm
M_{dyn}$$<$1 can be used as  an additional constraint on the parameter
space possible  (see Appendix  B for details on the aforementioned
  fitting procedure and the constraints).

  From Figures  12, 13 is obvious  that for $\rm K_{vir}$$\sim  $1 and
  $\rm  T_{kin}$$<$30\,K, the  high  $\rm X_{HCN}$  values yield  $\rm
  M_{dense}$$\sim  $(2-3)$\times  $$10^{10}$\,M$_{\odot}$  (NGC\,6240)
  and  $\rm M_{dense}$$\sim  $$10^{10}$\,M$_{\odot}$ (Arp\,193).   For
  NGC\,6240  this   surpasses  its  $\rm   M_{dyn}$($\rm  r_{co}$$\leq
  $0.60\,kpc)$\sim  $10$^{10}$\,M$_{\odot}$   within  its   CO  region
  (Tacconi   et  al.    1999).   However   for  Arp\,193   where  $\rm
  M_{dyn}$($\rm            r_{co}$$\leq            $1.3kpc)=1.6$\times
  $10$^{10}$\,M$_{\odot}$   (DS98)   no   such   constraint   can   be
  placed\footnote{In  our  detailed   HCN/CO  SLED  decompositions  we
    actually use $\rm r_{dyn}$=$\rm M_{tot}(H_2)/M_{dyn,vir}$$\la $1.3
    since in  strongly evolving gas-rich galaxies  where molecular gas
    has  yet to  settle in  circular  motions (e.g.   in mergers)  the
    dynamical   mass   computed   under  the   assumption   of   exact
    virialization may  underestimate the true  one. The value  of $\rm
    r_{dyn}$=1.3 rather  than 1 is deduced  from numerical simulations
    of isolated gas-rich  disks (see Daddi et al.   2010, Equation 2).
    Its  value in  mergers would  most likely  be higher  still}.  For
  super-virial dense gas states, the resulting lower $\rm X_{HCN}$ and
  $\rm M_{dense}$ (see  Figures 14, 15) can relax  such dynamical mass
  constraints  considerably by  allowing unbound  states even  for the
  high density gas in these galaxies.

 The  high-J CO SLED  for both galaxies  can be decomposed  to two
  components  (A) and  (B) drawn  from the  HCN-defined LVG  parameter
  space (for  NGC\,6240 also  (HCO$^{+}$/CS)-compatible) and  its $\rm
  T_{kin}$$>$30\,K  sub-region  (Figures  16, 17).   A  lower  density
  component (C) from  outside this parameter space  can then reproduce
  the remaining  low-J CO SLED  up to  J=4--3, 5--4 (beyond  which its
  contribution   subsides).   It   typically  has:   $\rm  n(C)$$\sim
$(400--$10^{3}$)\,cm$^{-3}$, $\rm T_{kin}(C)$$\sim $(50--400)\,K, with
the  warmest   ($\sim  $(60--400)\,K)  and  lowest   densities  ($\sim
$(400-500)\,cm$^{-3}$) in NGC\,6240.  In Arp\,193  most gas is in this
component ($\rm  f_m(C)$$\sim $75\%--95\%), while for  NGC\,6240: $\rm
f_m(C)$$\sim $10\%--30\%.   This disparity of the  dense gas mass
  fraction between the two galaxies  remains for all their CO/HCN SLED
  decompositions (see Table~3).

\subsubsection{The $^{13}$CO lines, the  [CO/$^{13}$CO]  abundance ratio, the IMF and ALMA  }

High r=[CO/$^{13}$CO]  abundance ratios  of r$\geq $150  are necessary
for  reproducing  the weak  $^{13}$CO  lines  of both  galaxies.   For
Arp\,193 r$\sim  $150 while  r$\sim $300-500  is needed  for NGC\,6240
(see Figures 16,  17) as its $^{13}$CO lines are  the weakest found in
(U)LIRGs.   Degeneracies regarding  the range of [CO/$^{13}$CO] do
  remain, especially  between the  virial ($\rm K_{vir}$$\sim  $1) and
  super-virial (0.5$\la  $$\rm K_{vir}$$\la $20)  SLED decompositions.
  However, as  long as HCN-constrained  states are used to  obtain the
  high-J CO  SLEDs, such  high abundance  ratios are  necessary.  This
  becomes more evident from Figure 18  where the $^{13}$CO SLED of the
  most  mass-dominant  component of  the  decomposition  is shown  for
  NGC\,6240 and r=80 (the highest  abundance ratio in the Galaxy).  The
  corresponding $^{13}$CO lines are  $\sim $2.5-5 times brighter than
  observed.

 The CO line emission samples most  of the metal-rich molecular gas in
 galaxies  and thus  the  aforementioned  large [CO/$^{13}$CO]  ratios
 concern  the  bulk  of  the  molecular  gas  in  the  two  metal-rich
 merger/starbursts.   A  lower  limit of  [C/$^{13}$C]$\geq  $150  was
 placed  also for  the inner  500\,pc  of the  starburst M\,82,  using
 several  groups of  C/$^{13}$C isotopologues  (Martin et  al.  2010).
 Selective photodissociation  of $^{13}$CO with respect  to CO, infall
 of  unprocessed   gas  with  large  $\rm   [CO/^{13}CO]$  from  large
 galactocentric radii of the progenitor spirals, or nucleosynthesis by
 a larger number  of massive stars of a top-heavy  stellar IMF can all
 yield the  high CO/$^{13}$CO line  ratios in (U)LIRGs (Casoli  et al.
 1992; Henkel  \& Mauersberger 1993). However  rather strong arguments
 against selective dissociation of $^{13}$CO  as a principal cause can
 be found  in Casoli et al.   (1992).  Moreover this process  can only
 take place near  PDRs, which may contain only small  fractions of the
 molecular  gas  in  merger/starbursts  (see 3.2).   Infall  of  large
 amounts   of   unprocessed   gas    to   the   central   regions   of
 merger/starbursts is difficult given the molecular gas {\it outflows}
 in starburst galaxies like NGC\,6240 (Cicone et al.  2012).

  Selective nucleosynthesis  by massive stars (i.e.   a top-heavy IMF)
  can yield  high [CO/$^{13}$CO] abundance  ratios.  However to  do so
  for the large amounts of molecular  gas sampled by the CO lines such
  a top-heavy IMF would have to be sustained long enough as to recycle
  most  of the  molecular gas  into being  dominantly enriched  by the
  different   [CO/$^{13}$CO]   abundance   ratios.   Alternatively   a
  top-heavy IMF may  have been there from the  beginning of a galaxy's
  evolution.  Either scenario calls for a top-heavy {\it time-averaged
    stellar IMF (t-IMF)}  rather than such an IMF  prevailing only for
  the starburst episode occuring at the observed epoch.

   A CR-driven  mechanism inducing a  top-heavy IMF, and  triggered by
   the high  SFR densities  of merger/starbursts, has  been identified
   (Papadopoulos et al.  2011).  Such an IMF would then strongly alter
   the isotopic ratios of carbon, oxygen, nitrogen, and sulfur (Henkel
   \&  Mauersberger  1993).  Such  atoms  are  then $''$locked$''$  in
   numerous molecules  whose rotational transitions span  a vast range
   of  $\rm  n_{crit}$  and  $\rm  E_{ul}/k_B$  values,  and  thus  of
   molecular gas phases.  A large $\rm [C/^{13}C]$ ratio injected by a
   top-heavy IMF  prevailing during the  main SF episodes in  a merger
   would then $''$contaminate$''$ most of these phases, and thus boost
   {\it all}  the corresponding isotopologue line  ratios probing them
   (e.g.  $\rm CO/^{13}CO$, $\rm HCN/H^{13}CN$).   However whether
     such a top-heavy IMF indeed prevails during most main SF episodes
     of  a merger  and whether  it can  $''$erase$''$ the  likely more
     ordinary isotopologue abundance  ratios in the ISM  of the spiral
     disk progenitors remains an open question.

 How  much  $\rm [C/^{13}C]$  is  boosted  by  selective massive  star
 nucleosynthesis remains  unclear (Henkel \& Mauersberger  1993) as is
 the effect of the delayed  release of $^{13}$CO by longer-lived lower
 mass stars with respect to  $^{12}$CO synthesized by the massive ones
 (Wilson \& Matteucci 1993; Henkel et al. 2010).  These issues must be
 tackled  before isotopologue  ratios  can be  used  to constrain  the
 observed-epoch  IMF and  the  t-IMF in  the  heavily dust  enshrouded
 environments of merger/starbursts.  In  that regard the several other
 isotopologues  available (e.g.   based on  N, O,  and S)  can provide
 powerful constraints on the  galaxy chemical evolution models used to
 interpret their  relative abundances  in terms of  the IMF  and t-IMF
 (Wilson   \&  Matteucci  1993;   Matteucci  2013).    Using  mm/submm
 isotopologue lines  to measure isotope  ratios and then use  those to
 set  constraints on  the  IMF and  t-IMF  of heavily  dust-enshrouded
 star-forming galaxies in  the Universe is however {\it  the next best
   thing other  than using their faint  dust-absorbed starlight.}  The
 huge  leap in  sensitivity and  correlator flexibility  of  ALMA will
 allow  determination  of any  given  isotopic  abundance ratio  using
 several   molecular   lines   of  its   isotopologues.    Double-rare
 isotopologues  (e.g.   $\rm  [^{12}CO^{18}O/^{13}C^{18}O]$) can  much
 reduce  the  optical   depth  uncertainties  of  extracting  relative
 abundances from  such line ratios  (Langer \& Penzias 1990).   On the
 practical side, the similar  frequencies of same-J isotopologue lines
 greatly facilitate  the imaging of  their intensity ratios  with ALMA
 since  such  observations  will  always  have  nearly  identical  u-v
 coverage.

\subsection{The   molecular gas  in Arp\,193: terminal SF feedback at work? }

It is  remarkable that in  Arp\,193 there seems  to be little  mass at
n$\geq $$10^{4}$\,cm$^{-3}$  given that: a)  such high density  gas is
the actual SF  $''$fuel$''$ in galaxies, and b) this  merger hosts one
of  the three  most extreme  starbursts in  the local  Universe.  Star
formation and its feedback (via  radiative pressure and/or SNR shocks)
may have  rapidly dispersed/consumed  much of its  initial (pre-burst)
dense gas  reservoir during a  nearly $''$coherent$''$ SF  event whose
O,B stars are  still present (thus maintaining the  large $\rm L_{IR}$
that makes Arp\,193 a LIRG).    In this non-steady state situation
  the dense gas reservoir within  the compact SF~region of Arp\,193 is
  yet to be resupplied by the ongoing merger, with SF expected to shut
  down as  a result until this  happens.  In this regard  this extreme
  starburst  may  have been  $''$caught$''$  during  a rare  phase  of
  a terminal feedback event.

Strong  SF  feedback  may  have acted  as  both  a  SF-synchronization
$''$trigger$''$ and a fast disperser of the dense gas reservoir inside
the  central  SF   region  of  Arp\,193.   The   solutions  with  $\rm
K_{vir}$$>$1 (see  Table 3, Figure 17)  may then be the  more relevant
ones  since  unbound  average  gas  motions  are  possible  if  strong
mechanical or radiative SF feedback affects  the bulk of the dense gas
reservoir to  the point of unbinding  it.  Such extreme events  may be
the  triggers   of  the  strong   molecular  gas  outflows   found  in
merger/starbursts (Cicone  et al.   2013).  Interestingly  Arp\,193 is
one of the few merger/starbursts where  a ring with an inner radius of
$\sim  $220\,pc rather  than  a  disk best  fits  its  CO J=2--1  line
velocity fields  (DS98), a geometric configuration  expected if strong
SF feedback cleared out the inner parts of its gas~disk.

 The short  lives of  O,B stars  will make  such starbursts  very rare
 since   such    $''$coherent$''$   SF   events   must    have:   $\rm
 T_{*}/T_{SN}$$<$1  where $\rm T_{*}$  is the time  interval during
 which most O,B stars of the  burst were formed, and $\rm T_{SN}$$\sim
 $(1-5)$\times $10$^{6}$\,yrs  the time  for them  to become  SNs (and
 thus cease to be FUV emitters).  The expected rareness of (U)LIRGs in
 such an  evolutionary stage can be  set in prespective using  the ISM
 evolution  model for  LIRGs by  Baan et  al.  2010,  and specifically
 their  Figure 1c  that  shows  the short  timescale  during which  IR
 luminosities remain high  (so the object is selected  as a (U)LIRG)),
 while the dense gas mass is being rapidly depleted.  A random ergodic
 sampling of the ISM evolution curve in their Figure 1c by a large set
 of evolving mergers will then yield  very few galaxies during such a
 stage.  We  note however that  unlike Baan et al.   our $''$delay$''$
 between $\rm  L_{IR}$ and  the dense gas  mass fraction  evolution is
 $\sim$$\rm  T_{*}$  (i.e. not due to any delayed  impact of low
 mass stars on $\rm L_{IR}$).

  Synchronizing  star formation and  the dispersal/consumption  of the
  dense gas  reservoir that fuels  it over timescales  of T$_{*}$$\sim
  $(1-5)\,Myr requires  a triggering mechanism with at  least as short
  of  a crossing  time over  the SF  region of  Arp\,193.   Using $\rm
  r_{co}$$\sim $(740--1300)\,pc for the gas disk revealed by CO J=2--1
  imaging  (DS98),  means  that  the speed  for  a  SF-synchronization
  $''$signal$''$ initiated  from a star-formation burst  in the disk's
  center would have to be: $\Delta $$\rm V_{SF}$$\ga$$\rm r_{co}$/$\rm
  T_{SN}$$\sim  $(240--420)\,km\,s$^{-1}$  (for $\rm  T_{*}$=3\,Myrs).
  Such  high velocities  correspond to  $\sim $$\alpha  $$\times $$\rm
  V_{rot}$ with $\alpha $=1--2  ($\rm V_{rot}$ the rotational velocity
  of the CO-bright gas disk (or  ring) in Arp\,193, see DS98).  Such a
  starburst would then consume/disperse  its molecular gas disk within
  1/(2$\pi  \alpha$)$\rm   T_{rot}$$\sim  $(0.08--0.16)$\rm  T_{rot}$!
  This  is possible  for strongly  evolving  merger/starburst systems,
  especially  if  SF feeback  initiates  strong non-gravitational  gas
  motions  ($\rm K_{vir}$$>$1)  that  act to  both  trigger fast  star
  formation  and  unbind/disperse  the  gas  disk.   Moreover,  recent
  observations indicate  that the SF  region of Arp\,193 is  much more
  compact than indicated by low-J  CO imaging (which trace both SF and
  non-SF  molecular gas).   Indeed  recent e-MERLIN  and VLBI  imaging
  revealed recent  starburst activity indicated by SNe  and SNR taking
  place   within  a   region  with   a  radius   of   r$\sim  $120\,pc
  (Romero-Ca\~nizales  et  al.   2012).   For such  compact  region  a
  $''$coherent$''$ starburst event becomes much easier to induce.

  The extremity of a starburst event that can disperse/consume the
   massive dense  gas reservoir  neccesary to  fuel star  formation in
   LIRGs like  Arp\,193 can be viewed  also in terms of  the molecular
   outflow this would imply.  A  typical pre-burst dense gas reservoir
   of   $\rm  M_{dense}$$\sim   $(1/10--1/5)$\rm  M_{total}(H_2)$$\sim
   $(1-2)$\times $$10^{9}$ (for $\rm M_{tot}(H_2)$$\sim 10^{10}$\,$\rm
   M_{\odot}$), with $\sim$1/2$\rm M_{dense}$ consumed by the SF burst
   itself (i.e.  a Galactic SFE$_{\rm dense}$ of 50$\%$), leaves $\sim
   $(0.5--1)$\times $$10^{9}$\,$\rm M_{\odot}$ of dense gas mass to be
   dispersed within  $\rm T_{*}$$\sim $3\,Myr.  The  implied molecular
   gas outflow  then is: $\sim  $(165-330)\,$\rm M_{\odot}$\,yr$^{-1}$
   (which can be  higher still if the SFE$_{\rm  dense}$ under extreme
   SF  feedback conditions  is less  than its  Galactic value).   Such
   strong molecular outflows from merger/starbursts have been recently
   observed (Cicone  et al.  2012).   High resolution imaging  of HCN,
   H$^{13}$CN, and high-J CO, $^{13}$CO  line emission in Arp\,193 can
   be used  to verify both  the compactness of  its SF region  and the
   presence of  a strong molecular gas  outflow affecting high-density
   gas.

Finally,  Arp\,193  and  NGC\,6240   with  their  similar  $\rm  L^{'}
_{HCN}/L^{'} _{CO}$ J=1--0 ratios but  different actual dense gas mass
fractions caution against using such  ratios as simple proxies of $\rm
f_{dense}$($\rm H_2$)  for galaxy-galaxy  comparisons.  The  same goes
for $\rm  L_{FIR}/L_{HCN}$ often used as  a proxy of the  dense gas SF
efficiency.   Nevertheless  the  expected  scarcity  of  objects  like
Arp\,193,  would  retain  the  {\it statistical}  usefulness  of  such
proxies for large galaxy samples.

\section{The heating sources  of the molecular gas in NGC\,6240 and Arp\,193}

 The CO and HCN SLED decomposition for these two LIRGs can now be used
 to examine in  detail whether FUV photons from PDRs  can maintain the
 thermal  state of  the components  that make  up their  molecular gas
 reservoirs.   The energy  balance equation  for molecular  gas heated
 only by photoelectric heating and cooled by line emission and gas-dust
 coupling~is:

\begin{equation}
\rm \Gamma _{pe} = \Lambda _{CO} +  \Lambda _{g-d} + \Lambda _{H_2} + 
\Lambda _{CII\,158\mu m} + \Lambda _{OI\,63\mu m} 
\end{equation}

\noindent
where  $\rm  \Gamma _{pe}$  denotes  photoelectric  heating, and  $\rm
\Lambda _{CO}$,  $\rm \Lambda _{H_2}$, $\rm  \Lambda _{CII\,158\mu m}$
and $\rm \Lambda  _{OI\,63\mu m}$ the CO, H$_2$, C\,II,  and O\,I line
cooling  (see Appendix  C).   For the  dense  FUV-shielded regions  of
Galactic GMCs  where $\rm T_{kin}$$\sim$10\,K neither  the $\rm H_{2}$
lines ($\rm E_{ul}/k_B$$\ga $510\,K) nor the CII and OI fine structure
lines ($\rm E_{ul}/k_B$$\sim $92\,K (CII) and $\sim $230\,K (OI)), are
appreciably excited.  Nevertheless we must now include them since such
regions  can now  be  much  warmer (section  3.3).   The  CII line  in
particular  can remain  a  powerful coolant  even within  FUV-shielded
regions  where  CII  abundances  are  maintained  only  by  CRs.   The
continuum term $\rm \Lambda _{g-d}$ denotes cooling due to the thermal
interaction of the gas with  colder concomitant dust (see Appendix C). 
 For CO  line cooling  we use  the {\it observed}  CO SLEDs,  and the
  luminosity fractions per component  obtained from our decomposition.
  These are lower limits if  significant line power remains beyond the
  CO J=13--12 transition.

 Here we  must note that  while T$_{\rm kin}$$\ga $T$_{\rm  dust}$ for
 concomitant dust and gas is found  by a number of detailed PDR models
 (e.g.  Hollenbach \& Tielens 1999; Papadopoulos et al.  2011), all of
 them  use  approximations regarding  the  radiative  transfer of  the
 re-radiated  FIR/IR continuum  from the  (FUV/optically)-heated dust.
 The validity  of these approximations may  break down for  the ISM in
 (U)LIRGs where significant dust  optical depths prevail out to FIR/IR
 wavelengths.  In  such environments a strong  IR radiation background
 within  molecular clouds,  understimated by  current PDR  models, may
 warm the  dust above $\rm T_{kin}$,  and make $\rm  \Lambda _{g-d}$ a
 net  gas-heating  term in  Equation  1.   This  is possible  even  in
 ordinary PDRs  where T$_{\rm  dust}$ can dip  below T$_{\rm  kin}$ by
 $\sim  $(5-15)K in  FUV-shielded regions  (see Hollenbach  \& Tielens
 1999  their  Figure  16).

 Nevertheless, if $\rm T_{dust}$ values significantly larger than $\rm
 T_{kin}$ were to  prevail for large dust and  molecular gas masses in
 the  high  FIR/IR optical  depth  ISM  of  (U)LIRGs, it  would  yield
 strongly surpressed  high-J CO lines  and even in  absorption against
 the corresponding  dust continuum.  This  is not observed in  the FTS
 high-J CO line  spectra of (U)LIRGs, and NGC\,6240  in particular has
 the highest  line/continuum ratios observed among them,   even up
   to      J=13--12.       For      a      column      density      of
   $\sim$$10^{24}$--$10^{25}$\,cm$^{-2}$  (typical  in  ULIRGs),  dust
   optical depths of $\sim 0.1$  would prevail at the rest frequencies
   of such  high-J CO lines  suggesting that $\rm T_{dust}$  cannot be
   much  greater than  $\rm T_{kin}$.   However, establishing  this in
   detail would  require more detailed  modeling of the  combined dust
   plus line  radiative transfer throughout the galaxy  in tandem with
   more robust estimates of the  total column density and dust optical
   depth, which are beyond the scope of this paper.  Finally the high
 CR  energy densities  and strong  turbulence in  the ISM  of (U)LIRGs
 directly  heat  the  gas  but  not  the  dust  (unlike  photoelectric
 heating).    This  will  raise  $\rm  T_{kin}$  above  the  $\rm
   T_{dust}$ of the concomitant dust (e.g.  Papadopoulos et al.  2011,
   2012a).  For turbulence such  an inequality has been recently shown
   for  the very  turbulent molecular  clouds in  the  Galactic Center
   (often used  as local analogs  of ULIRG-type ISM  conditions) where
   dust  with  $\rm  T_{dust}$$\sim$(15-20)\,K  (Pierce-Price  et  al.
   2001; Rodr\'iguez-Fern\'andez et al.  2004; Grac\'ia-Mar\'in et al.
   2011) coexists with dense  gas ($\ga 10^{4}$\,cm$^{-3}$) gas having
   $\rm  T_{kin}$$\sim  $(60-100)\,K  (Rodr\'iguez-Fern\'andez et  al.
   2004; Ao et al.  2013 and references therein).

\subsection{Average photoelectric gas heating}

  In order  to compute $\rm  \Gamma _{pe}$ (see  C\,1) we need  the $\rm
  G^{(FUV)} _{\circ}$ intensity (in Habing units) of the {\it average}
  ISRF in  the ISM  of the two  merger/starbursts.  An  effective $\rm
  G^{(eff)} _{\circ}$ can be obtained from the $\rm T_{dust} $ of each
  ISM component by using:

\begin{equation}
\rm T_{dust} = 55\left(\frac{1\,\mu m}{a}\right)^{0.06}
 \left(\frac{G^{(eff)}_{\circ}}{10^4}\right)^{1/6}\,K
\end{equation}

\noindent
which is  the average  value for graphite  and silicate grains  for an
emissivity  law of  $\alpha $=2.   The formula  is insensitive  to the
 grain  size, which we set  to a=1\,$\mu $m. In terms
of $\rm  G^{(eff)} _{\circ}$ it is:

\begin{equation}
\rm G^{(eff)}_{\circ} \sim 5645\left(\frac{T_{dust}}{50\,K}\right)^{6},
\end{equation}

\noindent
    For  $\rm  T_{\rm   dust}$=(10--15)\,K,  $\rm  G^{(eff)}_{\circ}$$
    \sim$0.4--4,  typical for  the dust  in  the Galaxy.   We can  now
    obtain an  approximate estimate  of $\rm T_{dust}$  for the  dust mass
    content of  the three  gas components that  make up  the molecular
    SLEDs of NGC\,6240 and Arp\,193 by setting a gas/dust ratio of 100
    and $\rm T_{dust}$=(1/f)$\rm T_{kin}$  (f$\ga $1)  per ISM  component.  We
    then find the corresponding total dust SED, and adjust f's so that
    it matches the observed one  in the far-IR/submm range (see Figure
    19).    The   estimated   $\rm T_{dust}$   is  then   used   to   find
   $\rm G^{(eff)}_{\circ}$ per ISM component and its $\rm \Gamma _{pe}$.

 This method yields  a $\rm G^{(eff)}_{\circ}$ per  ISM component that
 is $\rm  G^{(eff)}_{\circ}$$>$$\rm G^{(FUV)} _{\circ}$, thus  its use
 in Equation C\,1  yields an upper limit for  $\rm\Gamma _{pe}$.  This
 is   because   in   the   very   high-extinction   ISM   of   compact
 merger/starbursts the average FUV/optical radiation field is strongly
 attenuated (see also 4.2.1), leaving  only IR radiation to heat their
 concomitant  dust.   Thus,  unlike   ordinary  ISM  environments,  in
 merger/starbursts the observed dust SED and equivalent $\rm T_{dust}$
 can  have  a   strong  contribution  from  IR   heating,  making  the
 corresponding $\rm G^{(eff)}_{\circ}$ (from Equation 3) mostly a $\rm
 G^{(IR)}_{\circ}$,  and   a  strict  upper  limit   of  the  strongly
 attenuated FUV radiation field.  Moreover  even if a correction could
 be made and  the true average $\rm G^{(FUV)}  _{\circ}$ computed, the
 latter would  still be higher  than that  within the CO-rich  part of
 molecular clouds  where the molecular  SLEDs emerge.  This  is simply
 because  FUV-heated warm  dust from  their HI  and CO-poor/H$_2$-rich
 outer  cloud layers  also  contributes to  the  total dust  emission,
 biasing  the  global dust  SED  (and  its effective  $\rm  T_{dust}$)
 towards  these warmer  outer GMC  regions. This  would then  make the
 average $\rm G^{(FUV)}  _{\circ}$ (from Equation 3)  higher than that
 actually   prevailing    inside   the   (CO/HCN)-rich    regions   of
 molecular~clouds.

\subsection{(FUV photon)-deficient  thermal states in Arp\,193 and NGC\,6240 }

We  can  now estimate  the  Y=$\rm  \Gamma _{pe}/\Lambda  _{line,g-d}$
factor  for each  gas  component making  up  the CO  SLED, where  $\rm
\Lambda  _{line,g-d}$   is  the  total  cooling   power  expressed  in
Equation~1.   We label  states with  Y$<$1 as  (FUV photon)-deficient,
indicating other  dominant gas heating mechanisms.   For Arp\,193 $\rm
Y_A$$\sim  $0.04--0.5  only for  the  low-density  component with  the
highest   temperatures   ($\rm   T_{kin}$$\sim   $(300--400)\,K,   and
containing   $\sim   $(3--8)\%    of   its   $\rm   M_{tot}(H_2)$$\sim
$8$\times$$10^{9}$\,M$_{\odot}$  (see  Table 3).   For  the other  two
cooler components $\rm Y_{B,C}$$\geq $1 (i.e. FUV photons can maintain
the thermal state). For these  most massive components of Arp\,193 the
Y  factors  could  be  larger  still  (thus  placing  them  even  more
comfortably  within the  domain of  FUV-maintained thermal  states) if
dust  optical  depth effects  rather  than  only dust-gas  temperature
differences  contribute  to the  small  difference  found between  the
observed dust SED and their SLED-deduced ones.

 In  NGC\,6240  however  large  fractions of  $\rm  M_{tot}(H_2)$  are
 clearly in (FUV photon)-deficient thermal states.  More specifically,
 as in Arp\,193, we find the warmest component (in this case component
 B)  to  have $\rm  Y_{B}$$<$1  ($\sim  $0.04--0.2).  However,  unlike
 Arp\,193, we obtain $\rm Y$$<$1 also for either (A) or (C) components
 (specific values  depending on the  adopted CII abundance and  set of
 solutions).   Together  with  component  (B),  the  fraction  of  the
 molecular gas  mass in such thermal  states then rises  to $\rm M_{B,
   A\,or\,C}/M_{tot}(H_2)$$\sim  $(40--70)\%.   There  are even  cases
 where  all three  components have  $\rm Y$$<$1  (e.g.  for  $\rm \chi
 _{CII}$=$10^{-6}$ and  the SLED  decomposition in the  0.5$\leq $$\rm
 K_{vir}$$\leq$20 domain),  placing the  entire molecular gas  mass in
 NGC\,6240  in such thermal  states.  This  is consistent  with recent
 results reported by  Meijerink et al (2013) where  the entire CO SLED
 is attributed  solely to shock-heated  gas.  In that work,  the large
 line/continuum  ratio  of the  observed  CO  SLED  of NGC\,6240  (see
 Figures  1, 4)  is  a direct  result  of a  large $\rm  T_{kin}$/$\rm
 T_{dust}$  ratio,   itself  a  signature   of  globally  shock-heated
 molecular gas.  The ultimate  power source of such galaxy-wide shocks
 can  be readily  found in  the time-evolving  gravitational potential
 expected within  compact mergers, a power source  unavailable for the
 molecular gas in isolated~spirals.

  Finally, returning  to issue  of IR-heated  dust as  a potential
  molecular  gas heating  source we  note that  for states  with Y$<$1
  typically  $\rm |\Lambda  _{g-d}|/\Gamma  _{pe}$$\sim $(0.01--0.20).
  Thus even  reversing the sign of  $\rm \Lambda _{g-d}$  (so that the
  entire $\rm  |\Lambda _{g-d}|$  becomes a net  heating term)  is not
  enough to elevate Y up to  $\ga $1 values.  Moreover such a reversal
  would imply improbably high $\rm T_{dust}$ given that in states with
  Y$<$1, $\rm T_{kin}$ (as  constrained solely by the molecular SLEDs)
  is  already  high  ($\sim  $(65-300)\,K).   In  practice  the  small
  numerical  coefficient  of $\rm  \Lambda  _{g-d}$  ensures that  gas
  heating by a  warmer dust cannot compete  with $\rm \Gamma _{pe}$
  in typical SF environments (e.g.  for $\rm G^{(FUV)} _{\circ}$=100,
  $\rm \epsilon_{ph}$=0.01  and $\rm n(H_2)$=$10^{4}$\,cm$^{-3}$: $\rm
  |\Lambda      _{g-d}|/\Gamma_{pe}$$\sim     $$7\times$$10^{-5}$$|\rm
  (f-1)/f|T^{3/2}   _{kin}$),  which  remains   $<$1  even   for  $\rm
  T_{kin}$=100\,K  unless  f$\la$0.075,  i.e.   dust 13x  warmer  than
  the~gas).

\subsubsection{Strong  surpression of the average  FUV radiation fields in merger/starbursts}

 The Y  factors estimated  in the previous  section will  be generally
 lower  still  if  the  propagation   of  FUV  light  in  the  heavily
 dust-obscured SF  regions of merger/starbursts is  more realistically
 treated.  Using  the formalism by  Wolfire et al.  1990  for galactic
 centers, the average FUV radiation field incident on molecular clouds
 randomly mixed with the forming stars has an intensity:

\begin{equation}
\rm G^{(FUV)} _{\circ} \sim 3\times 10^2 \lambda_{*}(pc)\left(\frac{L_{IR}}{10^{10}\,L_{\odot}}\right)
\left(\frac{R_{SB}}{100\,pc}\right)^{-3} \left[1-e^{-(R_{SB}/\lambda_{*})}\right], 
\end{equation}

\noindent
 where  $\lambda _{*}$  is  the mean  distance  FUV photons  propagate
 before being absorbed.  The size of the starburst responsible for the
 observed  $\rm L_{IR}$  is  denoted by  $\rm  R_{SB}$ and  can be  as
 compact as $\sim $(200-300)\,pc (see  our section 3.4; Sakamoto et al
 2008).  For  the dust-enshrouded  ISM of merger/starbursts,  with its
 high     {\it     average}      densities     of     $\rm     \langle
 n(H_2)\rangle$$\ga $$10^3$\,cm$^{-3}$ clumpy PDR models give $\lambda
 _{*}$$\la  $1\,pc  (Meixner  \&   Tielens  1993).  For  Arp\,193  and
 NGC\,6240         where         $\rm        L_{IR}$$\sim$(2-4)$\times
 $10$^{11}$\,L$_{\odot}$    Equation   4   yields:    $\rm   G^{(FUV)}
 _{\circ}$$\sim $220--1500,  which is  $\sim $3-4 times  and up  to an
 order of  magnitude lower than  the values derived in  4.1.  Moreover
 Equation 4  yields an estimate of  the average FUV  field incident on
 molecular  cloud {\it surfaces},  which will  be lower  deeper inside
 their (CO/HCN)-rich  inner regions.  The fundamental role  of the FUV
 radiation in gas  heating in PDRs but also  the limitations placed by
 its  strong  dust  absorption  can  be  further  highlighted  by  the
 archetypal  PDR  in  Orion's   Bar.   Its  thickness  is  only  $\sim
 $(0.2-0.5)\,pc (Tielens \& Hollenbach  1985), leaving the bulk of the
 Orion  molecular  cloud cold  (Sakamoto  et  al.   1994).  The  large
 surpressions of the average FUV  radiation fields expected in the ISM
 of  dusty   mergers  further  accentuate   the  need  for   (non  FUV
 photon)-driven gas heating mechanisms in such galaxies.

\section{The gas heating mechanisms in LIRGs: towards a complete diagnostics}

 Similar investigations  of the thermal/dynamical states  of the dense
 molecular  gas  have  already  been  made  for  the  Galactic  Center
 (Bradford et al.  2005; Ao et  al.  2013) and the nucleus of NGC\,253
 (Bradford et al.   2003; Hailey-Dunsheath et al.   2008).  These were
 the first to establish CRs  and/or turbulence as dominant gas heating
 mechanisms in these regions using  energy balance criteria similar to
 ours in Section 4.  Another  criterion they used was the PDR-expected
 warm  gas mass  fraction, finding  a much  larger one  than could  be
 maintained by  FUV photons from  PDRs (e.g.  Bradford et  al.  2003).
 This  can be  so even  as the  molecular line  ratios could  still be
 fitted by  PDR ensembles.   Indeed {\it  molecular line  ratios alone
   cannot  discern the  dominant molecular  gas heating  mechanisms in
   LIRGs.}  This  was shown in Mrk\,231  where hot and dense  PDRs can
 reproduce its CO  line ratios up to J=13--12 but  cannot contain much
 molecular  gas mass  without their  warm dust  $''$outshining$''$ the
 observed dust  SED, and an  X-ray dominated region is  necessary (van
 der Werf  et al.   2010).  Such $''$normalization$''$  tests, whether
 using the warm molecular gas mass fraction, its concomitant dust mass
 and  SED, the  dust IR-brightness  (Rangwala  et al.   2011), or  the
 observed high-J  CO line  luminosities (Bradford  et al.   2003), are
 crucial for  discerning whether PDR  ensembles can truly  account for
 the molecular line emission in LIRGs or~not.

\subsection{New gas heating mechanisms in merger/starbursts: a general argument }

  A  $''$normalization$''$  criterion  based  on  the  warm  gas  $\rm
  f_{PDR}$=$\rm M_{PDR}/M_{total}$(HI+H$_2$) mass fraction expected in
  PDRs  can be  readily used  for  the ISM  of merger/starbursts.   An
  approximate $\rm f_{PDR}$ value in  FUV-irradiated GMCs can be found
  by  computing the  mass of  the outer  atomic HI  layer marking  the
  HI$\rightarrow $H$_2$ transition as it is comparable to the warm PDR
  H$_2$ layer extending further inwards.  Thus the PDR-residing column
  density N(HI+H$_2$)$\sim  $2$\times $$\rm N_{tr}(HI)$.  For  a given
  $\rm G^{(FUV)}  _{\circ}$, metallicity  Z (Z=1, Solar),  and average
  gas  density n,  this column  density can  be computed  in units  of
  optical extinction as (see Pelupessy et al. 2006 for details):

\begin{equation}
\rm A^{(tr)} _{v} = 1.086 \xi^{-1} _{FUV} \ln\left[1+\frac{G^{(FUV)} _{\circ} k_{\circ}}{n R_f}\Phi\right]
\end{equation}

\noindent
with   $\rm   k_{\circ}$=4$\times  $$10^{-11}$\,s$^{-1}$   the   H$_2$
dissociation  rate (for  $\rm G^{(FUV)}  _{\circ}$=1), $\rm  R_f$$\sim
$3$\times  $$10^{-17}$\,cm$^{-3}$\,s$^{-1}$  its   formation  rate  on
grains       (for      typical       CNM      HI),       and      $\rm
\Phi$=6.6$\times $$10^{-6}$$\sqrt{\pi}$$\rm Z^{1/2}$$\rm \xi_{FUV}$ is
the H$_2$ self-shielding function  over the HI/H$_2$ transition layer,
with  $\rm \xi  _{FUV}$=$\rm \sigma  _{FUV}/\sigma_{V}$$\sim $2-3  the
dust cross section  ratio for FUV and optical  light.  The PDR-related
gas mass fraction per GMC then~is:

\begin{equation}
\rm f_{PDR}\sim 2\times \left[1-\left(1-\frac{4A^{(tr)} _{v}}{3\langle A_v \rangle}\right)^{3}\right],
\end{equation}

\noindent
assuming spherical and uniform GMCs that do not cross-shield and hence
each receives the full FUV  radiation field.  These simple assumptions
make the  computed $\rm  f_{PDR} (HI+H_2)$ a  {\it maximum}  since the
denser  substructures   existing  within  GMCs,  and   the  inevitable
cloud-cloud cross-shielding, will lower its actual value as denser gas
deeper  inside   will  have  even   thinner  PDR  layers,   and  cloud
cross-shielding  will reduce  the actual~$\rm  G^{(FUV)}_{\circ}$ (see
4.2.1).

The mean  optical extinction  $\rm \langle A_v  \rangle$ per GMC  is a
function  of  the  ambient  ISM  conditions, mainly  the  average  gas
pressure  (dominated  by  supersonic  non-thermal gas  motions).   The
empirical   Galactic  (average   density)-(size)   relation  n$\propto
$(2R)$^{-1}$, along with its  normalization in terms of cloud boundary
pressure $\rm P_{e}$, yields (see Pelupessy et al.~2006):

\begin{equation}
\rm \langle A_v \rangle \sim 0.22\,Z\,\left(\frac{n_{\circ}}{100\,cm^{-3}}\right)
\left(\frac{P_e/k_B}{10^4 cm^{-3} K}\right)^{1/2},
\end{equation}

\noindent
where   $\rm  n_{\circ}$$\sim   $1500\,cm$^{-3}$.   For   average  GMC
densities  of  $\sim  $100\,cm$^{-3}$, Solar  metalicities,  $\rm  \xi
_{FUV}$=2,  and   $\rm  G^{(FUV)}  _{\circ}$=5-10  (for   ordinary  SF
environments) Equation  5 yields $\rm A^{(tr)}  _{v}$$\sim $0.50-0.75,
while for typical  ISM pressures in galactic  disks $\rm P_e/k_B$$\sim
$(1-2)$\times $10$^{4}$\,K\,cm$^{-3}$: $\rm  \langle A_v\rangle $$\sim
$3.3-4.7.  Under  such conditions  the HI+(warm  H$_2$) PDR  gas phase
contains $\rm f_{PDR}$$\sim $0.74--1 (from  Equation 6), and thus most
HI and H$_2$ gas  is indeed in PDRs as is  often stated (Hollenbach \&
Tielens  1999).  In  merger/starbursts  however  the highly  turbulent
molecular   gas   can   reach    pressures   of   $\rm   P_e/k_B$$\sim
$($10^{7}$--$10^8$)\,cm$^{-3}$, conditions recently inferred even in a
distant ULIRG at z$\sim $2.3 (Swinbank et al.  2011).  In such an ISM:
$\rm \langle  A_v\rangle $$_{\rm  ULIRG}$$\sim $100--330,  creating an
environment where FUV photons do not penetrate deep.  Indeed, even for
a  high  $\rm  G^{(FUV)}  _{\circ}$=$10^4$ and  average  densities  of
n=$10^4$\,cm$^{-3}$  (typical  for merger/starbursts):  $\rm  A^{(tr)}
_v$$\sim $1.4--1.9, and thus $\rm f_{PDR}$$\sim $0.03--0.15, with only
half of that as molecular gas ($\rm f_{PDR, H_2}$$\sim $0.015--0.075).
In reality  {\it average} FUV  radiation fields irradiating  the dense
molecular    gas    in    ULIRGs     will    have    $\rm    G^{(FUV)}
_{\circ}$$\sim  $$10^2$--$10^3$  (4.2.1),   for  which  $\rm  A^{(tr)}
_v$$\sim    $0.15--0.77    (for   n=$10^4$\,cm$^{-3}$),    and    $\rm
f_{PDR}$(HI+H$_2$)$\sim $0.0035--0.06.

 Thus  {\it  for the  high  average  densities  and pressures  of  the
   molecular gas in  merger/starbursts, PDRs will be  confined in very
   thin  layers  per  molecular  cloud  with  $\rm  f_{PDR,  H_2}$$\la
   $few\%,}  essentially undoing  the  effects  of stronger  radiation
 fields  (which  act  to   $''$thicken$''$  PDR  layers).   Hence  any
 molecular SLED  decomposition that yields  much larger warm  gas mass
 fractions cannot  be attributed  to PDR ensembles.   We note  that we
 assumed  metal-rich  ISM  (Z$\sim  $1),  which  may  not  be  so  for
 merger/starbursts at high redshifts.  The expected $\rm f_{PDR, H_2}$
 must  then be  computed for  their average  metalicities, and  can be
 substantially  larger  for  Z$<$1.    In  metal-poor  environments  a
 particularly luminous  C$^{+}$ line  can be  expected, with  a larger
 $\rm L(C^{+})/L_{IR}$ from the much $''$thicker$''$ CO-poor envelopes
 of PDRs around molecular clouds.

\subsection{Turbulence and CRs as new global molecular gas heating mechanisms}

In  compact  merger/starbursts  high  SF rate  densities  boost  their
average   CR  energy   density  by   $\sim  $($10^2$--$10^{3}$)$\times
$Galactic  while highly  supersonic turbulence  (fueled by  the merger
process) is typical.   In such environments it has  already been shown
that  CR   and  turbulent  heating   ($\rm  \Gamma  _{CR}$   and  $\rm
\Gamma_{turb}$)  alone  can  maintain  warm gas  states  even  in  the
complete absence of photoelectric  heating (Papadopoulos et al.  2010,
2012a).  In  reality they  can easily surpass  $\rm \Gamma  _{pe}$ for
large amounts of  molecular gas mass even as all  three processes take
place.

For  CRs:  $\rm   \Gamma  _{CR}/\Gamma  _{pe}$$\sim$[0.038--0.075]$\rm
(\zeta     _{CR}/\zeta_{CR,Gal})$$\rm    [G^{(FUV)}_{\circ}]^{-1}$$\rm
exp(A_v/1.086)$   (for    photolectric   efficiency    $\rm   \epsilon
_{ph}$=(1-2)\%,  and using  the  expressions in  Appendix  C).  For  a
constant  average ratio  $\rm  G^{(FUV)}_{\circ}$/$\rm \zeta_{CR}$  at
cloud boundaries  ($\rm G^{(FUV)}_{\circ}$  and $\rm  \zeta_{CR}$ both
being proportional  to the local SFR  density), we find that  for $\rm
A_v$=3-4,  $\rm  \Gamma  _{CR}/\Gamma  _{pe}$$\sim  $0.6-3,  i.e.   CR
heating becomes  comparable and overtakes photolectric  heating as the
ambient FUV  field becomes  strongly attenuated.   In the  Galaxy such
optical depths still include most of  a typical GMC's mass, but in the
high-pressure ISM of mergers these include only small mass fractions (see
5.1).

For    turbulence:    $\rm     \Gamma    _{turb}/\Gamma    _{pe}$$\sim
$(0.15--0.30)$\times $$\rm  \left[(P_e/P_{\circ})^{3/4}G^{-1} _{\circ}
  L^{1/2}  _{pc}\right]$  (using  expressions  in  Appendix  C).   For
Galactic  GMCs  with  boundary  pressures:  $\rm  P_e/k_B$$\sim  $$\rm
P_{\circ}/k_B$=10$^{4}$\,K\,cm$^{-3}$     and     sizes    of     $\rm
L_{pc}$=(10-20)\,pc (turbulence  driven at  the largest  scales): $\rm
\Gamma  _{turb}/\Gamma _{pe}$$\sim$(0.47--1.34)$\rm  G^{-1} _{\circ}$,
which  remains $<$1  for $\rm  G^{(FUV)}_{\circ}$$\sim $$10^2$--$10^3$
typical  in   starbursts.   However  for  the   high-pressure  ISM  of
merger/starbursts:                 $\rm                P_{e}/k_B$$\sim
$($10^7$--$10^8$)\,K\,cm$^{-3}$   and   $\rm   \Gamma   _{turb}/\Gamma
_{pe}$$\sim$(85--1340)$\rm   [G^{(FUV)}  _{\circ}]^{-1}$.    Turbulent
heating can  then overtake photolectric  heating, except in  the outer
GMC layers near O,B stars  where $\rm G^{(FUV)} _{\circ}$ remains high
(we considered no attenuation of the average $\rm G^{(FUV)} _{\circ}$,
making the computed $\rm \Gamma _{turb}/\Gamma _{pe}$ a~minimum).  For
a   $''$bottom$''$-driven   turbulence  in   merger/starbursts   (i.e.
injected at small scales and high-density regions, Papadopoulos et al.
2012a), turbulent heating may be strong  even for the densest and most
FUV-shielded  molecular   gas,  quite   unlike  Galactic   GMCs  where
turbulence in such~regions has dissipated.

\subsection{Future  prospects in the age of ALMA and JVLA}

  The  divergence  of  the  IR-normalized   high-J  CO  SLEDs  of  two
  IR-luminous starbursts  with otherwise  similar low-J CO  SLEDs, and
  $\rm L_{IR}/L^{'}  _{CO, 1-0}$,  $\rm L^{'} _{CO,  1-0}/L^{'} _{HCN,
    1-0}$ ratios  highlights the uncertainties of  assuming the high-J
  CO line excitation in the  absence of appropriate observations as it
  has  been the  practice  at  high redshifts  (e.g.   Tacconi et  al.
  2006).    As    Arp\,193   and   NGC\,6240    demonstrate,   similar
  $''$proxies$''$  of  SF  $''$efficiencies$''$  and  dense  gas  mass
  fractions can still correspond to different dense gas conditions and
  thus high-J CO SLEDs.  Moreover a strong decoupling of molecular gas
  and   dust  temperatures   with   $\rm  T_{kin}$$>$$\rm   T_{dust}$,
  maintained  deep  inside  FUV-shielded regions,  is  expected  where
  turbulent and/or  CR heating  dominate.  These  in turn  {\it induce
    very  different initial  conditions  for  star formation,}  making
  molecular line diagnostic of FUV-shielded dense gas crucial for star
  formation theories.

Molecular line diagnostic of (non  FUV)-photon gas heating will become
readily accessible with  ALMA (Papadopoulos 2010; Bayet  et al.  2011;
Meijerink et al.   2011), with a recent such study  done for the dense
molecular  gas in  the  Galactic Center  (Ao et  al.   2013).  In  the
absence of  very high-J  CO SLEDs,  which for  the local  Universe are
accessible only from Space, multi-J CO {\it and} $^{13}$CO lines up to
J=6--5, 7--6, and large dipole moment molecular lines are adequate for
revealing  non-photon  gas  heating.  Extreme  thermal  and  dynamical
states in the dense gas  of merger/starbursts can be powerfully probed
using imaging of line emission  from high-dipole molecules (e.g.  HCN)
{\it and} one of its rare isotopologues (e.g.  H$^{13}$CN).  ALMA with
its  large sensitivity,  correlator  versatility  and wide  bandwidths
(allowing  multi-line   observations),  can  routinely   conduct  such
investigations  in   star-forming  galaxies  in   the  coming~decades.
Finally,  with the  determination  of $\rm  M_{total}$(H$_2$) being  a
necessary step for finding what fraction  of it is warm and dense (and
whether PDRs  can account  for it),  the JVLA with  its access  to the
low-J CO SLEDs at high redshifts remains indispensible.

On  the   theoretical  front,  discrete-component   decompositions  of
molecular  SLEDs like  those  used  in this  work  and throughout  the
literature (e.g. Rangwala et al.  2011;  Panuzzo et al.  2010; van der
Werf et  al. 2010) must  be replaced  by continous $\rm  (n, T_{kin})$
distributions  that   represent  more  realistic  renderings   of  the
conditions inside  molecular clouds.   This can  then yield  much more
powerful,       physically-motivated,      $''$continous$''$-component
decompositions of molecular SLEDs  of galaxies with fewer degeneracies
than the  discrete decompositions,  while containing  more information
about the underlying physical conditions of turbulent molecular clouds
(see  also Appendix  B).  The  recent shock-model  of the  CO SLED  of
NGC\,6240  by  Meijerink  et  al.   2013 is  a  step  towards  such  a
$''$continous$''$-component CO  SLED modeling based (in  this case) on
the underlying physics of shocked  gas.  The large molecular line data
sets that  ALMA will deliver for  the extragalactic ISM will  then set
much  more  powerful constraints  on  such  SLED  models, and  on  the
physical conditions  of the  molecular gas in  merger/starbursts where
the most decisive tests for SF theories may lie.

\subsection{Conclusions}

We present a large set of  molecular line observations of Arp\,193 and
NGC\,6240,  two  prominent   merger/starbursts,  using  the  SPIRE/FTS
instrument aboard the Herschel Space  Observatory (HSO), the IRAM 30-m
telescope at  Pico Veleta (Spain) and  the 12-m APEX telescope  in the
Atacama Desert (Chile).   These two LIRGs, selected  from our Herschel
Comprehensive (U)LIRG Emission Survey (HerCULES) have similar low-J CO
SLEDs,  $\rm \epsilon_{SF,  co} $=$\rm  L_{FIR}/L_{co, 1-0}$  and $\rm
r_{HCN/CO}$=$\rm L^{'}_{\rm HCN,1-0}$/$\rm L^{'} _{\rm CO,1-0}$ ratios
(proxies  of SF  efficiency  and  dense gas  mass  fraction) are  good
benchmarks for studying the dense  gas in mergers where this component
may~dominate.   Our  analysis  of  the molecular  SLEDs  yields  the
following~results:

\begin{itemize}

\item  The two  IR-normalized CO  SLEDs markedly  deviate from  J=4--3
  onwards  with NGC\,6240  having  a highly  excited  SLED with  large
  line/continuum  ratios   up  to   J=13--12  while  Arp\,193   has  a
  significantly lower global high-J  CO excitation (and line/continuum
  ratios) despite  being one of  the three most intense  starbursts in
  the local Universe.

\item Only  $\sim $(5--15)\% of the  H$_2$ gas mass in  Arp\,193 is at
  n$\ga    $$10^{4}$\,cm$^{-3}$    (the   primary    star    formation
  $''$fuel$''$), while  in NGC\,6240 this rises  to $\sim $(70--90)\%,
  more  typical for  a merger/starburst.   Terminal SF  feedback, with
  Arp\,193      $''$caught$''$      during      a      short-timescale
  gas-(dispersal/consumption) maximum  of its  duty cycle,  can induce
  such a~disparity.

\item We deduce $\rm [CO/^{13}CO]$$\ga $150 (Arp\,193) and up to $\sim
  $300-500 (NGC\,6240),  much higher than in the  Galaxy.  A top-heavy
  galactic IMF sustained over long timescales in merger/starbursts can
  produce  such  high  $\rm  [CO/^{13}CO]$ abundances.   By  measuring
  several  isotope  ratios  of  atoms   like  C,  N,  and  S  using  a
  multiplicity  of  corresponding  isotopologue  molecules  and  their
  rotational  transitions ALMA  can now  powerfully probe  the  IMF in
  galaxies  where  very  high  dust  extinctions  render  the  use  of
  starlight for such a task nearly impossible.

\item  In  both  galaxies  the gas  component  responsible  for  their
  luminous  heavy rotor  SLEDs can  account also  for their  high-J CO
  SLEDs  from J=5--4,  6--5 up  to J=13--12.   In NGC\,6240  most the
  H$_2$  gas mass  is in  states irreducible  to self-gravitating  and
  photoelectrically heated gas, and this also the case for the warmest
  gas phase in~Arp\,193.

\end{itemize}

\noindent

Finally we give a general argument  on why the FUV radiation from PDRs
is   unlikely   to   encompass   large   molecular   gas   masses   in
merger/starbursts and maintain the  extraordinary thermal states often
found for their dense gas.  Turbulent and/or CR heating can readily do
so, unhindered by  the large dust extinctions,  and without destroying
the complex heavy rotor molecules like  HCN as FUV radiation from PDRs
does.  Such heating mechanisms will  then alter the initial conditions
of star formation away from those  in the Galaxy and isolated spirals,
making   starbursts/mergers  critical   testbeds  of   star  formation
theories.  ALMA and  the JVLA, with their  exceptional sensitivity can
play  a leading  role in  determining the  initial conditions  of star
formation in dust-obscured galaxies across cosmic epoch.

\acknowledgements 

The  authors  gratefully   acknowledge  financial  support  under  the
"DeMoGas"  project.  The  project "DeMoGas"  is implemented  under the
"ARISTEIA" Action of the "OPERATIONAL PROGRAMME EDUCATION AND LIFELONG
LEARNING". The authors also acknowledge support by  the  European
Social  Fund (ESF)  and National  Resources.
Padelis  P.   Papadopoulos acknowledges  support
from an Ernest  Rutherford Fellowship (ERF) from STFC.   We would like
to thank the referee for  a strong and constructive criticism that was
critical  in  sharpening  some  of  our  orginal  arguments.   Padelis
Papadopoulos  is much  indebted  to Dr  Francesca  Matteucci for  many
discussions on the capability of chemical evolution models of galaxies
in unraveling  their IMF history.  Finally  Padelis Papadopoulos would
like  to take  this opportunity  to thank  his old  PhD  advisor Ernie
Seaquist for  those early wings, and  dedicates this last  work to his
little son  $\Lambda \epsilon \omega \nu  \iota \delta \alpha$--X$\rho
\eta \sigma \tau o$.

\newpage

\appendix

\section{The LVG radiative transfer model and typical degeneracies}

 For  each gas  component  a  3-dimensional parameter  grid  with
  regularly spaced  temperature $\rm  T_{kin}$, density  n($\rm H_2$),
  and  fractional  abundance  versus  velocity  gradient  $\rm  X_{\rm
    mol}/(dV/dR)$ is  used as  an input to  a Large  Velocity Gradient
  (LVG) radiative transfer  code.  $\rm X_{\rm mol}$  is the molecular
  abundance  ratio  with   respect  to  H$_2$.   In   this  work  $\rm
  X_{HCN}$=2$\times   $$10^{-8}$   (Greve    et   al.    2009),   $\rm
  X_{HCO^+}$=8$\times    $$10^{-9}$    (Jansen   1995),    and    $\rm
  X_{CS}$=1$\times  $$10^{-9}$ (Greve  et al.   2009).  The  parameter
  grid  spans $\rm  T_{kin}$=($\rm  10^{0.5}$--$\rm 10^3$)\,K,  n($\rm
  H_2$)=(10$^{2}$--10$^{8}$)cm$^{-3}$, and (dV/dR)=($\rm 10^{0}$--$\rm
  10^{3}$)\,km\,s$^{-1}$\,pc$^{-1}$.    We   sample  it   using   with
  logarithmic steps  of 0.1  and generate the  model grids  with RADEX
  (van der Tak et al. 2007).

For each  individual model a $\chi  ^2$ value is calculated  using the
ratios of  line brightness temperatures  obtained from the  models and
the    observations.     We    then    compute    $\chi^{2}$=$\Sigma_i
(1/\sigma_i)^{2}[R_{{\rm  obs}(i)}  -  R_{{\rm  model}(i)}]^2$,  where
$R_{\rm  obs}$  is  the  ratio  of  the  measured  line  luminosities,
$\sigma_{\rm i}$  the error  of the measured  line ratio,  and $R_{\rm
  model}$ the ratio of the  line brightness temperatures calculated by
the  LVG  model.  We  then  obtain  the Bayesian  probability  density
function (pdf) for a given set of observational data with respect to a
model $p$~using:

\begin{equation}
 P(p\|x) = \frac{P(p) P(x\|p)}{P(x)}, 
\end{equation}

\noindent
where $P(p)$  is the prior  probability of  the model. We  assume flat
priors ($P(p)$)  for n($\rm H_2$),  (dV/dR), and $\rm T_{kin}$  with $
P(p)  =  1 $,  and  set  $P (p)$=0  for  solutions  outside the  prior
criteria.  We calculate the probability for each observation $P(x\|p)$
of obtaining the observed data  with model $p$, which follows Gaussian
distribution,~using

\begin{equation}
P(x\|p) = \Sigma_i [\exp(-\chi_i^2 (x)/2)]/P(x),
\end{equation}

\noindent
where  $x$  is the  measurement  set,  $P(x)$ the  normalization,  and
$\chi^2(x)$ calculated  for each  model.  We  set line  optical depths
$\tau $$<$100 as required by the  RADEX manual.  Finally we search two
solution   classes   classified   by    the   gas   dynamical   state,
self-gravitating:   0.5$<$$\rm  K_{vir}$$<$2,   and  unbound   states:
0.5$<$$\rm K_{vir}$$<$20 (see~3.1.1).

An  obvious  degeneracy inherent  in  all  our solutions,  and  indeed
present  in all  LVG radiative  transfer models,  is that  between the
asssumed molecular  abundance $\rm X_{mol}$  and (dV/dR) since  an LVG
model  always  uses  the   $\rm  X_{mol}/(dV/dR)$  combination.   Only
radiative transfer models that solve  for the chemical networks while
also computing  the gas thermal  state (since it strongly  affects the
reaction rates)  can avoid such degeneracies.   Since 1985 (Hollenbach
\& Tielens 1985) all such models have been developed for PDRs and only
much later expanded to include X-ray Dominated Regions (XDRs) expected
near AGN (Meijerink \& Spaans  2005), and CR-dominated Regions (CRDRs)
(Papadopoulos et  al.  2011). None  yet employs an  LVG approximation,
necessary  for modeling  line emission  from macroturbulent  molecular
clouds.     We   are    currently    developing   a    multi-component
(PDR/XDR/CRDR)-LVG code to be used in the~future.

Another  degeneracy common  among  LVG  models, is  that  of the  $\rm
K_{vir}$ parameter.  Unless  one sets a probability  prior to restrict
the  $\rm [T_{kin},  n(H_2),  (dV/dR)]$ grid  to self-gravitating  gas
states ($\rm  K_{vir}$$\sim $1), solutions  with $\rm K_{vir}$$>$1
are nearly always possible for a given set of CO/HCN SLEDs besides the
self-gravitating solutions.  This particular degeneracy direct impacts
the  corresponding $\rm  X_{HCN}$  and $\rm  X_{CO}$ factors  (compare
Figures 12,  13 with 14,  15).  We nevertheless consider  also unbound
states, even for  the dense gas, as these are  possible in the extreme
ISM environments of merger/starbursts (see 3.1.1).  In our HCN/CO SLED
decomposition  we list  both  types  of solutions  (see  Table 3,  and
Figures  16, 17).   Finally a  n($\rm H_2$)--$\rm  T_{kin}$ degeneracy
exists  in  all  our (HCN/HCO$^{+}$)-constrained  LVG  solutions  (see
Figures 5, 6, 9, 10).

\section{The fitting procedure and  future developments }

 In  supersonically  turbulent  molecular   clouds  a  wide  range  of
 densities  and  temperatures  {\it   is}  expected,  and  the  n($\rm
 H_2$)--$\rm T_{kin}$ parameter space defined  their HCN SLEDs will be
 $''$populated$''$ by  them.  In our  modeling this is used  to obtain
 the  CO SLEDs  as a  superposition of  states drawn  from the  n($\rm
 H_2$)--$\rm T_{kin}$ space as constrained  by the luminous global HCN
 SLEDs  (for NGC\,6240  also the  CS  and HCO$^{+}$  SLEDs).  The  ISM
 physics  behind  this  choice  is outlined  in~3.3.   Dynamical  mass
 constraints can yield additional  restrictions towards the cold/dense
 $''$corner$''$ of  this n($\rm  H_2$)--$\rm T_{kin}$  parameter space
 (see 3.2, 3.3).   Future higher-J HCN and  HCO$^{+}$ measurements and
 of their rarer isotopologues (e.g.   H$^{13}$CN) can also do this for
 its (high-$\rm T_{kin}$)/(low-n) domain.

Our fitting  process starts from the  (high-$\rm T_{kin}$)/(low-n) end
of  the parameter  space  shown in  Figures 5,  9,  producing all  the
corresponding  CO SLEDs  within  the regions  of maximum  probability.
Once a template SLED whose  shape matches the largest possible segment
of the high-J CO SLED is  found, its mass scaling factor is determined
so the observed and the model SLED are brought to the same scale.  The
resulting scaled-up SLED template is then subtracted from the observed
one,  and the  process is  repeated for  obtaining the  next component
needed to match the remaining global  CO SLED.  We use a discrete grid
to produce the aforementioned sets of template CO SLEDs, and given its
dense  sampling we  do not  expect the  main features  of the  CO SLED
decomposition (n(H$_2$),  $\rm T_{kin}$, relative  mass constributions
of the  components) to  depend on the  grid characteristics.   This is
currently the practice  in all such studies (e.g. van  der Werf et al.
2010),  but it  does contain  a  degree of  arbitariness.  Moreover  a
discrete  set of  components  is a  poor,  unphysical, description  of
supersonically  turbulent molecular  clouds.  A  physically motivated,
continuous mass  weighting function $\rm w(n,T_{kin})$,  would be much
more realistic.  Currently only $\rm dM(n)/dn$ can be obtained in this
fashion  from simulations  of  isothermal  turbulent molecular  clouds
(Padoan \& Nordlund 2002).  Such  an approach, apart from reducing the
arbitariness and  the number of free  parameters of discrete-component
decompositions of  molecular SLED, will  allow more information  to be
extracted  for the  underlying conditions  of the  turbulent gas  from
such modeling.

\section{The heating and cooling functions for molecular gas}

The heating and  cooling functions used to compute  the energy balance
(Equation 1)  of the LVG-derived gas  states making up the  CO and HCN
SLEDs of NGC\,6240  and Arp\,193 are given below  (see Papadopoulos et
al.  2011 for details on  these functions).  For photoelectric heating
the corresponding expression is:

\begin{equation}
\rm \Gamma _{pe}=2\times 10^{-24}\left[\epsilon _{ph} n(H_2) G^{(FUV)} _{\circ}\right]
ergs\, cm^{-3}\, s^{-1}
\end{equation}

\noindent
where $\rm \epsilon _{ph}$ is the photoelectric heating efficiency, given
by

\begin{equation}
\rm \epsilon _{ph} = \frac{4.9\times10^{-2}}{1+4\times10^{-3}\gamma^{0.73}}+
\frac{3.65\times10^{-2}\left(T_{kin}/10^4\right)^{0.7}}{1+2\times10^{-4}\gamma},
\end{equation}

\noindent
and  $\gamma$=$\rm (G^{(FUV)} _{\circ}  T^{1/2} _{kin})/n_e$  is a  factor that
determines   the   balance    between   photonization   and   electron
recombination.   We  obtain  the  electron  density  by  setting  $\rm
n_{e}/[2n(H_2)]$ equal to the assumed abundance of C\,II (i.e. ionized
carbon as the sole source of free electrons in molecular clouds).
The cooling due to gas-dust interaction (since $\rm T_{kin}$$\geq $$\rm T_{dust}$
for concomitant gas and dust phases) is given by

\begin{equation}
\rm \Lambda _{g-d}=1.4\times 
10^{-32}\left[n(H_2)\right]^2\left(\frac{f-1}{f}\right) T^{3/2} _{kin}\,
 ergs\,cm^{-3}\,s^{-1},
\end{equation}

\noindent
where $\rm  T_{kin}$=$\rm f T_{dust}$ (f$\geq$1)  for the temperatures
of the  molecular gas  and dust mass  reservoir.  CO  rotational lines
provide substantial  cooling, especially for the dense  gas phase deep
inside GMCs.   To compute $\rm \Lambda  _{CO}$ we use  the observed CO
SLEDs  from  J=1--0 up  to  J=13--12.   Thus  for  each  phase
k=1,2... (whose sum makes up the observed CO SLEDs) it is:

\begin{equation}
\rm \Lambda ^{(k)} _{CO} =\frac{\sum_{J=0}^{12} \rho ^{(k)} _{J+1,J}
L_{J+1,J}}{\Delta V_{k}}=
\mu m_{H_2} n_k(H_2) \left[\frac{\sum_{J=0}^{12} \rho ^{(k)} _{J+1,J}
L_{J+1,J}}{M_{k}(H_2)}\right], 
\end{equation}

\noindent
where   $\rm  \rho   ^{(k)}   _{J+1,J}$=$\rm  L_{J+1,J}$(k-phase)/$\rm
L_{J+1,J}$  is  the  CO  line  luminosity  fraction  corresponding  to
(k)-phase, $\mu  $=1.34 the mean  molecular weight, $\rm  m_{H_2}$ the
H$_2$ molecule mass, and  $\rm n_K(H_2)$, $\rm M_{k}(H_2)$ the density
and mass of the corresponding  (k) component (obtained via our CO SLED
decomposition).   Inserting the various  quantities and  converting to
astrophysical units yields:

\begin{equation}
\rm \Lambda _{CO}=8.77\times 10^{-33} n(H_2) 
\left[\frac{\sum_{J=0}^{12} \rho  _{J+1,J}
L_{J+1,J}(L_{\odot})}{M(H_2)/(10^9\,M_{\odot})}\right]
 ergs\,cm^{-3}\,s^{-1},
\end{equation}

\noindent
where we have  omitted subscript (k) for simplicity  as the meaning of
computing $\rm \Lambda _{CO}$ per component is now clear.

The  C\,II  line  at  158\,$\mu  $m can  remain  powerful  coolant  of
molecular gas even in the absence of far-UV radiation, as C\,II can be
maintained by  CRs deep  inside molecular  clouds (Papadopoulos  et al
2011).  The corresponding cooling power is given by

\begin{equation}
\rm \Lambda _{CII\,158\mu m}=7.9\times 10^{-23} \chi_{CII} [n(H_2)]^2 
e^{-92.2/T_{kin}}\, ergs\, cm^{-3}\, s^{-1}
\end{equation}

\noindent
  where $\rm \chi_{CII} $=$\rm [CII/H]$  is the C\,II abundance.  In a
  similar fashion the cooling due to the O\,I line fine structure line
  at 63\,$\mu$m is given by

\begin{equation}
\rm \Lambda _{OI\,63\mu m}=2.27\times 10^{-24}\chi_{O} [n(H_2)]^2 T^{0.32} _{kin}
e^{-228/T_{kin}}\, ergs\, cm^{-3}\, s^{-1},
\end{equation}

\noindent
where  $\rm  \chi_{OI}$=$\rm  [O/H]$  is  the  Oxygen  abundance  (see
Papadopoulos et al.  2011 for  a derivation of these two expressions).
We assume  $\rm \chi_{O}$=2.9$\times  $$10^{-4}$ for the  abundance of
Oxygen not locked in CO. For CII we set three abundance values namely:
$\rm \chi  _{CII}$=3$\times $$10^{-4}$ (all carbon  ionized), and $\rm
\chi  _{CII}$=$10^{-5}$,  $10^{-6}$   corresponding  to  CR-maintained
abundances  inside  (far-UV)-shielded  gas regions,  and  compute  its
cooling (and Y factors) accordingly.

 Finally we include  the two lowest H$_2$  rotational lines: S(0):$\rm
 J_u$-$\rm J_l$=2--0, S(1):$\rm  J_u$-$\rm J_l$=3--1, the only ones
 that will be  excited  for $\rm  T_{kin}$$\sim $(100--1000)\,K.   The
 cooling function then is:

\begin{equation}
\rm \Lambda _{H_2} = 2.06\times 10^{-24} \left(\frac{n(H_2)}{1+r_{op}}\right)
\left[1+\frac{1}{5}
e^{510K/T_{kin}}\left(1+\frac{n_{0}}{n(H_2)}\right)\right]^{-1}(1+R_{10}) \, ergs\,
cm^{-3}\, s^{-1}, 
\end{equation}

\noindent
with  $\rm   n_{0}$$\sim  $54\,cm$^{-3}$   the  critical   density  of
S(0):2--0, $\rm r_{op}$ the ortho-H$_2$/para-H$_2$ ratio (=1-3), and

\begin{equation}
\rm R_{10}=26.8 r_{op} \left[\frac{1+(1/5)
e^{510K/T_{kin}}\left(1+\frac{n_{0}}{n(H_2)}\right)}{
1+(3/7)e^{845K/T_{kin}}\left(1+\frac{n_{1}}{n(H_2)}\right)}\right],
\end{equation}

\noindent
expresses  the relative  cooling contributions  of the  two rotational
lines considered here  (with $\rm n_{1}$$\sim $10$^{3}$\,cm$^{-3}$ the
critical density of the S(1):3--1 line).

 CRs and turbulence can also  deposit large amounts of energy into the
 molecular gas,  unaffected by  extinction. For CRs  the corresponding
 heating function is:

\begin{equation}
\rm \Gamma _{CR} =1.5\times 10^{-27} n(H_2) \left(\zeta_{CR}/\zeta _{CR,Gal}\right)\, ergs\,cm^{-3}\,s^{-1},
\end{equation}

\noindent
where we used  the expressions found in Papadopoulos  et al. 2011, for
fully  molecular gas.  A Galactic  CR  ionization rate  of $\rm  \zeta
_{CR,Gal}$=5$\times  $$10^{-17}$\,s$^{-1}$   has  been  assumed.   For
turbulent heating we use the work of Pan \& Padoan 2009 as it has been
reformulated by Papadopoulos et al.  2012a eventually yielding:

\begin{equation}
\rm \Gamma _{turb} = 3.3\times 10^{-27} n(H_2) \sigma^3 _{\circ, n}\left(\frac{P_e}{P_{\circ}}\right)^{3/4}
L^{1/2} _{pc}\,ergs\,cm^{-3}\,s^{-1},
\end{equation}

\noindent
where $\rm \sigma _{\circ, n}$=1.2\,km\,s$^{-1}$ and $\rm P_{\circ}/k_B$=$10^4$\,K\,cm$^{-3}$. 

\newpage

\begin{deluxetable}{ll|cc|cc}
\tablecolumns{6}
\rotate
\tablewidth{0pc}
\tablecaption{CO, $^{13}$CO and CI velocity-intergated line flux densities and luminosities for NGC\,6240 and Arp\,193}
\tablehead{
\colhead{Line} & \colhead{$\lambda $ ($\nu $)} & \colhead{$\rm S_{line}$ (NGC\,6240)} &
\colhead{$\rm L_{line}$\tablenotemark{a} (NGC\,6240)} & \colhead{$\rm S_{line}$ (Arp\,193)} &
\colhead{$\rm L_{line}$\tablenotemark{a} (Arp\,193)} \\
  & \,\,\,\,\, [$\mu $m]\, (GHz)  & [Jy\,km\,s$^{-1}$] & [$\rm L_{\odot}$] & [Jy\,km\,s$^{-1}$] & [$\rm L_{\odot}$]}  
\startdata
CO J=1--0\tablenotemark{b} & 2600.757 (115.271) & 322$\pm $29   & 4.2$\times $$10^{5}$  & 194$\pm $16   & 2.2$\times$$10^{5}$\\
CO J=2--1\tablenotemark{b} & 1300.404 (230.538) & 1492$\pm $253 & 3.9$\times $$10^{6}$  & 850$\pm $130  & 2.0$\times$$10^{6}$\\
CO J=3--2\tablenotemark{b} & 866.963  (345.796) & 3205$\pm $642 & 1.2$\times $$10^{7}$  & 1294$\pm $171 & 4.5$\times$$10^{6}$\\
CO J=4--3                  & 650.252  (461.041) & 4634$\pm $371 & 2.4$\times $$10^{7}$  & 1603$\pm $266 & 7.4$\times$$10^{6}$\\
CO J=5--4                  & 520.231  (576.268) & 5636$\pm $149 & 3.6$\times $$10^{7}$  & 1366$\pm $143 & 7.8$\times$$10^{6}$\\
CO J=6--5                  & 433.556  (691.473) & 5913$\pm $82  & 4.6$\times $$10^{7}$  & 1369$\pm $62  & 9.4$\times$$10^{6}$\\
CO J=7--6                  & 371.650  (806.652) & 6009$\pm $60  & 5.5$\times $$10^{7}$  & 1178$\pm $38  & 9.5$\times$$10^{6}$\\
CO J=8--7                  & 325.225  (921.799) & 5833$\pm $89  & 6.0$\times $$10^{7}$  & 1036$\pm $46  & 9.5$\times$$10^{6}$\\
CO J=9--8                  & 289.120  (1036.912)& 4769$\pm $82  & 5.6$\times $$10^{7}$  & 844$\pm $78   & 8.7$\times$$10^{6}$\\
CO J=10--9                 & 260.240  (1151.985)& 4162$\pm $67  & 5.4$\times $$10^{7}$  & 662$\pm $26   & 7.6$\times$$10^{6}$\\
CO J=11--10                & 236.613  (1267.014)& 3161$\pm $74  & 4.5$\times $$10^{7}$  & 371$\pm $56   & 4.7$\times$$10^{6}$\\
CO J=12--11                & 216.927  (1381.995)& 2592$\pm $60  & 4.0$\times $$10^{7}$  & 282$\pm $32   & 3.9$\times$$10^{6}$\\
CO J=13--12                & 200.272  (1496.923)& 2081$\pm $60  & 3.5$\times $$10^{7}$  & 126$\pm $35   & 1.9$\times$$10^{6}$\\
$^{13}$CO J=1--0\tablenotemark{b} & 2720.406 (110.201) & 6.5$\pm $1.9  & 8.0$\times $$10^{3}$ & \nodata      & \nodata\\
$^{13}$CO J=2--1\tablenotemark{b} & 1360.228 (220.398) & 26$\pm $4     & 6.4$\times $$10^{4}$ & 26$\pm $4  & 5.7$\times$$10^{5}$ \\
$^{13}$CO J=3--2\tablenotemark{c} & 906.846  (330.588) & $\la $79      & $\la$2.9$\times $$10^{5}$ & \nodata & \nodata\\ 
$^{13}$CO J=6--5                  & 453.498  (661.067) & 112$\pm $67   & 8.3$\times $$10^{5}$   & \nodata & \nodata\\    
$\rm [CI]$ $^{3}$$P_1\rightarrow $$^3$$P_0$&609.136 (492.161)&1751$\pm $224& 9.7$\times $10$^{6}$& 636$\pm $146&3.1$\times $10$^{6}$\\
$\rm [CI]$ $^{3}$$P_2\rightarrow $$^3$$P_1$ &370.415 (809.342)& 3332$\pm $52& 3.0$\times $10$^{7}$& 1660$\pm $43&1.3$\times $10$^{7}$\\ 
\enddata
\tablenotetext{a}{For $\rm z_{co}$=0.0245, $\rm  D_L(z)$=105.5\,Mpc (NGC\,6240) and $\rm z_{co}$=0.0231, $\rm D_L(z)$=99.3\,Mpc
  (Arp\,193).}
\tablenotetext{b}{from Papadopoulos et al. 2012a}
\tablenotetext{c}{from Greve et al. 2009, 3$\sigma $ upper limit}
\end{deluxetable}


\begin{deluxetable}{lc|cc|cc}
\tablecolumns{6}
\tablewidth{0pc}
\tablecaption{HCN, HCO$^{+}$ and CS data}
\tablehead{
\colhead{Line} & \colhead{$\rm \nu _{rest} $} & \colhead{$\rm S_{line}$ (NGC\,6240)} &
\colhead{References\tablenotemark{a}} & \colhead{$\rm S_{line}$ (Arp\,193)} &
\colhead{References\tablenotemark{a}} \\
  & \,\,\, (GHz)  & [Jy\,km\,s$^{-1}$] &    & [Jy\,km\,s$^{-1}$] &   }
\startdata
HCN J=1--0       & 88.631 &  14$\pm $2  & 1,2,3,4       & 6.2$\pm $0.9         & 4\\
HCN J=2--1       & 177.261&  44$\pm $7  & 3             & \nodata              &  \\
HCN J=3--2       & 265.886&  74$\pm $7  & 3, 4, 5, x    & 12.3$\pm $2.4        & 4\\
HCN J=4--3       & 354.505&  41$\pm $6  & x             & $\la $10 (3$\sigma$) & 6\\
HCO$^{+}$ J=1--0  & 89.188 &  21$\pm $3  & 2             & \nodata              & \\
HCO$^{+}$ J=3--2  & 267.557&  141$\pm$21\tablenotemark{b}&  x & \nodata              & \\
HCO$^{+}$ J=4--3  & 356.734&  74$\pm$9   &  2, x         & \nodata              & \\
CS J=2--1        &  97.980&  7.5$\pm$1.5 &  x           & $\la $2.1 (3$\sigma$)& x\\
CS J=3--2        & 146.969&  9$\pm$2 &  x           & $\la $3 (3$\sigma$)  & x\\
CS J=7--6 & 342.882& $\la $32 (3$\sigma$)  & x & \nodata & \\ \enddata
\tablenotetext{a}{1=Nakanishi  et  al.  2005,  2=Greve  et  al.  2009,
  3=Krips et al. 2008, 4=Gracia-Carpio  et al. 2008, 5=Israel (private
  communication),  6=Papadopoulos 2007,  x=this work.  All values  are
  averages of  those reported  in the  mentioned literature  (see also
  2.3). } \tablenotetext{b}{ Our HCO$^{+}$ J=3--2 flux for NGC 6240 is
  higher than that reported by  Gracia-Carpio et al.  (2008) using the
  IRAM  30\,m  telescope  (the  only other  such  measurement  in  the
  literature).   We   adopt  our  value   as  some  of   the  267\,GHz
  observations  with the  IRAM 30\,m  have been  affected by  pointing
  offsets (the 30-m beam at such frequencies is $\sim $9$''$).}

\end{deluxetable}

\newpage

\begin{deluxetable}{llllcl}
\tablecolumns{6}
\rotate
\tabletypesize{\small}
\tablewidth{0pc}
\tablecaption{CO SLED decompositions for NGC\,6240 and Arp\,193}
\tablehead{
\colhead{Galaxy\,\,\,\,($\rm K_{vir}$)\tablenotemark{a}} & \colhead{Component (A)\tablenotemark{b}} & \colhead{Component (B)\tablenotemark{b}} &
\colhead{component (C)\tablenotemark{b}} & \colhead{$\rm M_{tot}(H_2)/M_{dyn}$\tablenotemark{c}} & \colhead{$\rm X^{(eff)} _{CO}$\tablenotemark{d}}} 
\startdata
NGC\,6240 (0.5--2) & [4.1, 1.6, 0.9, 77\%, 11.7/$\sqrt{\alpha}$] & [3.8, 2.5, 1.2, 8\%, 1.9/$\sqrt{\alpha}$] & 
[2.7, 2.6, 2, 15\%, 0.8/$\sqrt{\alpha}$] & 2.1/$\sqrt{\alpha}$$^{(*)}$ & 3.2/$\sqrt{\alpha}$\\
\,\,$''$\,\,$''$\,\,\,\,\,\,\,\,\,\,\,\,\,\,\,\,\,\,(0.5--2)  & [4.6, 1.5, 1.2, 82\%, 18.9/$\sqrt{\alpha}$] &
 [3.8, 2.5, 1.2, 5\%, 1.9/$\sqrt{\alpha}$] & [2.6, 2.5, 2, 13\%, 0.8/$\sqrt{\alpha}$] & 3.4/$\sqrt{\alpha}$$^{(*)}$ & 5.3/$\sqrt{\alpha}$\\  
\,\,$''$\,\,$''$\,\,\,\,\,\,\,\,\,\,\,\,\,\,\,\,\,\,(0.5--2) & [3.8, 1.8, 0.8, 58\%, 5.9/$\sqrt{\alpha}$] & [3.8, 2.5, 1.2, 11\%, 1.9/$\sqrt{\alpha}$] &
[2.7, 2.4, 2.0, 31\%, 0.9/$\sqrt{\alpha}$] & 1.4/$\sqrt{\alpha}$ & 2.2/$\sqrt{\alpha}$\\ 
\,\,$''$\,\,$''$\,\,\,\,\,\,\,\,\,\,\,\,\,\,\,\,\,\,(0.5--20) & [4.4, 1.5, 0.8, 88\%, 23.8/$\sqrt{\alpha}$] & [4.3, 2.6, 8.7, 1\%, 2.1/$\sqrt{\alpha}$] &
[2.9, 2.0, 3.0, 11\%, 1.0/$\sqrt{\alpha}$] & 5.5/$\sqrt{\alpha}$$^{(*)}$ & 8.6/$\sqrt{\alpha}$\\ 
\,\,$''$\,\,$''$\,\,\,\,\,\,\,\,\,\,\,\,\,\,\,\,\,\,(0.5--20) &  [5.0, 1.5, 4.9, 67\%, 7.5/$\sqrt{\alpha}$]  & [4.3, 2.6, 8.7, 2\%, 2.1/$\sqrt{\alpha}$] &
[3.0, 2.0, 1.5, 30\%, 1.4/$\sqrt{\alpha}$] & 1.8/$\sqrt{\alpha}$ & 2.8/$\sqrt{\alpha}$  \\
\,\,$''$\,\,$''$\,\,\,\,\,\,\,\,\,\,\,\,\,\,\,\,\,\,(0.5--20) &  [3.8, 1.7, 0.8, 70\%, 7.4/$\sqrt{\alpha}$] & [3.8, 2.5, 1.2, 7\%, 1.9/$\sqrt{\alpha}$] &
[3.0, 1.8, 3.0, 23\%, 1.1/$\sqrt{\alpha}$] & 2.4/$\sqrt{\alpha}$$^{(*)}$  & 3.8/$\sqrt{\alpha}$  \\
\hline
Arp\,193\,\,\,\,\, (0.5--2) & [3.5, 2.5, 1.7, 8\%, 1.3/$\sqrt{\alpha}$] & [4.0, 1.7, 2.0, 15\%, 3.7/$\sqrt{\alpha}$] & [2.9, 1.7, 1.7, 77\%, 1.7/$\sqrt{\alpha}$] &  
0.38/$\sqrt{\alpha}$ & 1.7/$\sqrt{\alpha}$\\
\,\,$''$\,\,$''$ \,\,\,\,\,\,\,\,\,\,\,\,\,\,\,  (0.5--20) & [3.9, 2.6, 14, 3\%, 1.2/$\sqrt{\alpha}$] & [4.4, 1.5, 3.9, 4\%, 4.8/$\sqrt{\alpha}$] &
[3.0, 1.5, 3.1, 93\%, 1.6/$\sqrt{\alpha}$] & 0.38/$\sqrt{\alpha}$ & 1.7/$\sqrt{\alpha}$\\
\enddata
\tablenotetext{a}{The galaxy and the  $\rm K_{vir}$  range of the LVG radiative transfer solutions.}
\tablenotetext{b}{The gas component: [log(n), log($\rm T_{kin}$), $\rm K_{vir}$, $\rm M_{comp}/M_{tot}(H_2)$, $\rm X^{(comp)}_{CO}$] where
$\rm X^{(comp)}_{CO}$ (in units of $\rm X_l$=$\rm M_{\odot}(K\,km\,s^{-1}\,pc^2)^{-1}$)\\ \hspace*{0.5cm} is computed from Equation 1 
in Papadopoulos et al. 2012b (where $\alpha$$\sim $0.6--2.4).}
\tablenotetext{c}{The ratio of the total gas mass to the mass contained within the CO-bright regions, estimated using
 $\rm M_{dyn}$=1.3$\rm M_{dyn,vir}$ (see 3.3).\\
\hspace*{0.5cm} The values marked with (*) are those for which $\rm M_{tot}(H_2)/M_{dyn}$  remains significantly above unity 
even for $\alpha$=2.4, and thus\\
 \hspace*{0.5cm}  excludes the corresponding CO SLED decompositions. }
\tablenotetext{d}{The effective $\rm X^{(eff)} _{CO}$=$\rm (M_A+M_B+M_C)/L^{'} _{CO,1-0}$ factor of the SLED decomposition.}
\end{deluxetable}

\newpage

\begin{figure}
\epsscale{1.3}
\plottwo{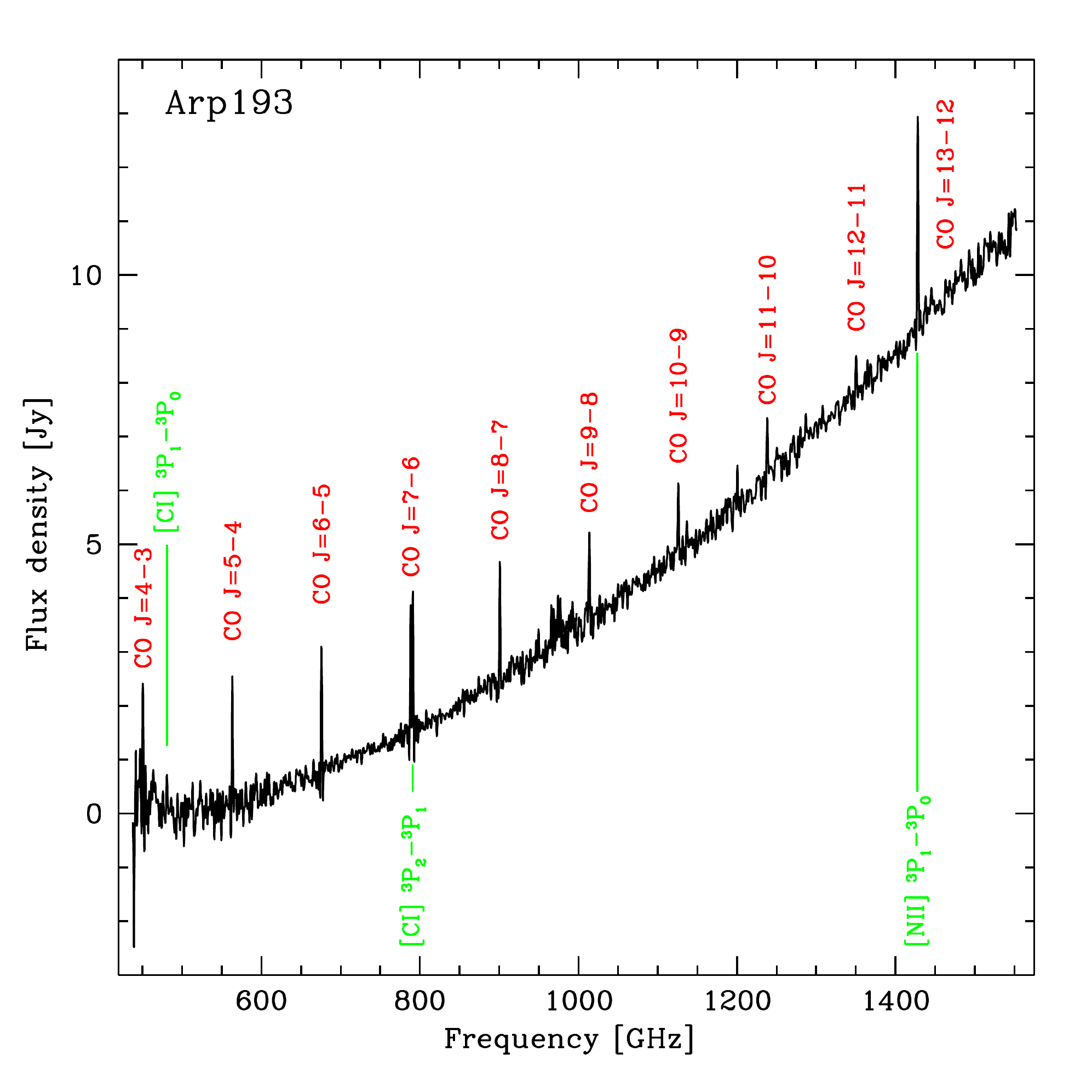}{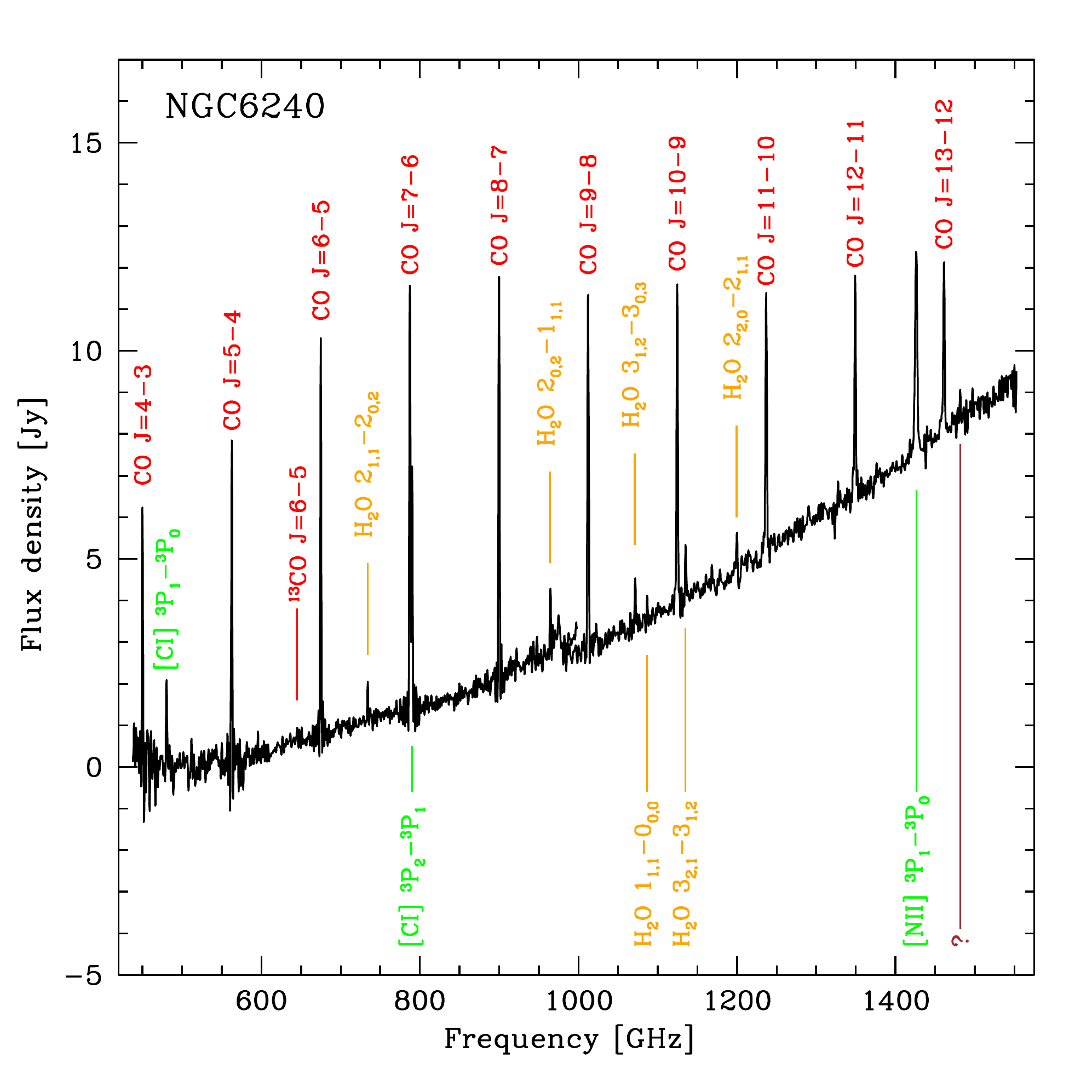}
\caption{The  full SPIRE/FTS  spectra of  Arp\,193 and  NGC\,6240. The
  detected lines marked  are: CO J=4--3 up to J=13--12  (red), the two
  fine structure lines of [C\,I] $^{3}$$P _1\rightarrow $$^3$$P_0$ and
  $  ^{3}$$P_2\rightarrow  $$^3$$P_1$  and that  of  [N\,II]  (green).
  There is also an unidentified line (brown).  The exceptionally large
  line-continuum  contrast  in  NGC\,6240 betrays  the  large  thermal
  decoupling between molecular gas and  dust reservoirs (and thus dust
  SED  and CO  SLED) with  $\rm T_{kin}$$>$$\rm  T_{dust}$, while  its
  well-excited  high-J  CO lines  contrast  the  much weaker  ones  in
  Arp\,193  even as  the latter  is one  of the  three most  prominent
  merger/starbursts in the local Universe.}
\end{figure}

\newpage

\begin{figure}
\epsscale{1.00}
\plotone{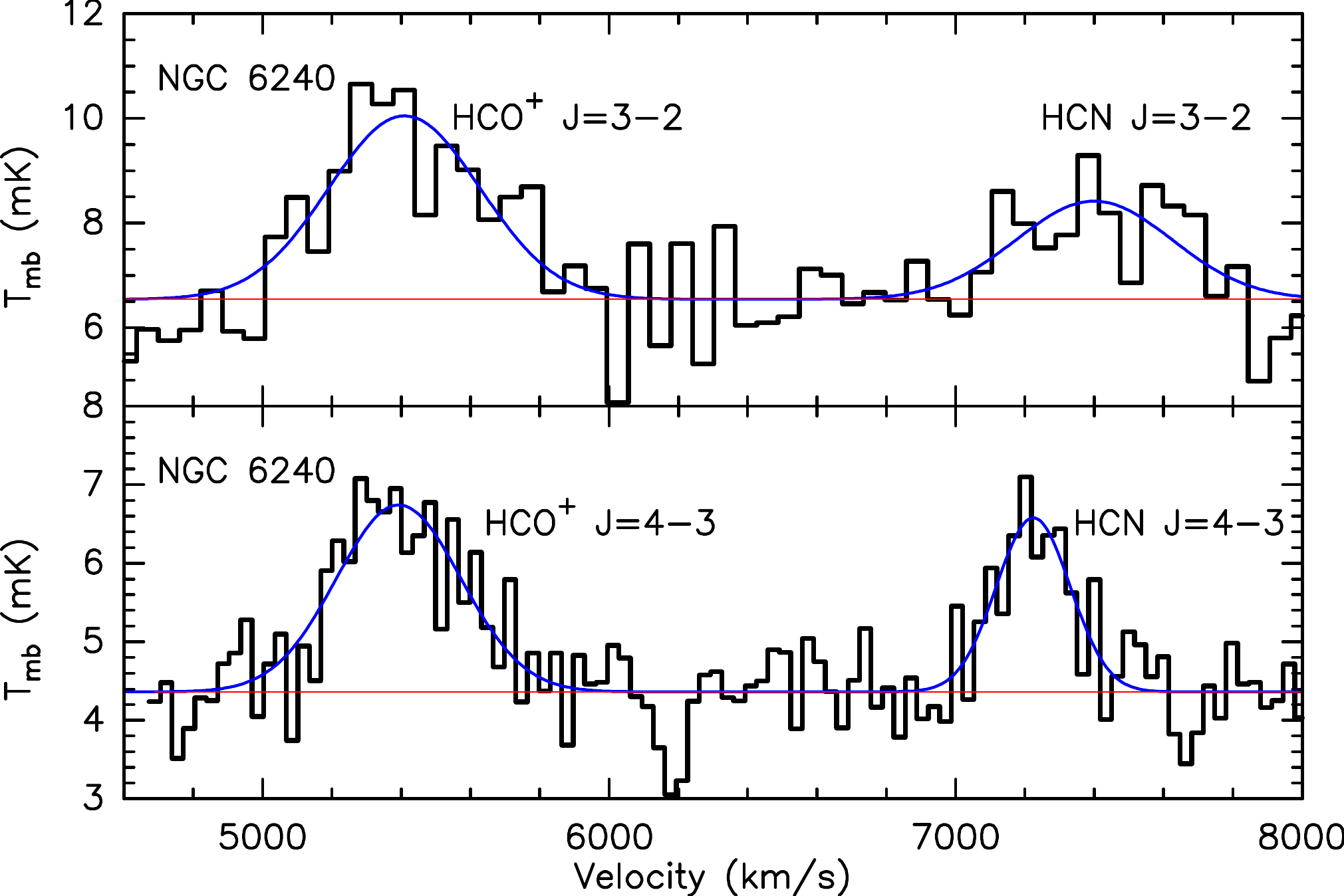}
\caption{The HCN  and HCO$^{+}$ J=3--2, 4--3 spectra  of NGC\,6240
  obtained with  APEX, smoothed to resolutions  of $\rm \Delta
  V_{ch}$=33\,km\,s$^{-1}$ for J=4--3, and $\rm \Delta V_{ch}$=62\,km\,s$^{-1}$
  for J=3--2 (the cz velocity scale is centered on the HCN transitions).}
\end{figure}

\clearpage

\begin{figure}
\epsscale{1.00}
\plotone{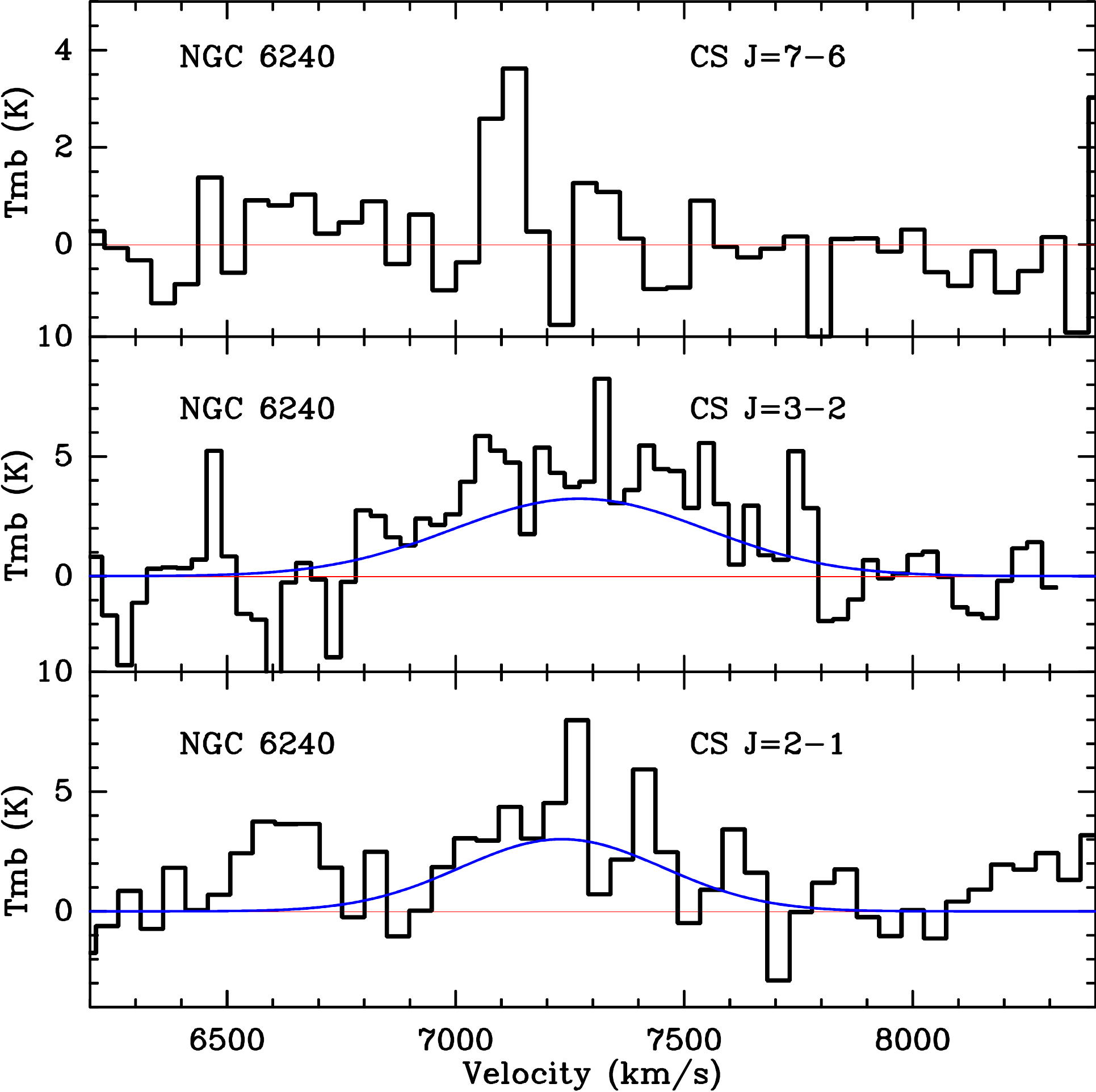}
\caption{The CS  J=2--1, 3--2 and  7--6 lines obtained with  the 30\,m
  telescope  (2--1, 3--2)  and  APEX (CS  J=7--6)  for NGC\,6240.  The
  spectra  are  smoothed  to  51\,km\,s$^{-1}$,  33\,km\,s$^{-1}$  and
  49\,km\,s$^{-1}$ for J=7--6, 3--2 and 2--1 respectively. }
\end{figure}

\newpage

\begin{figure}
\epsscale{1.3}
\plottwo{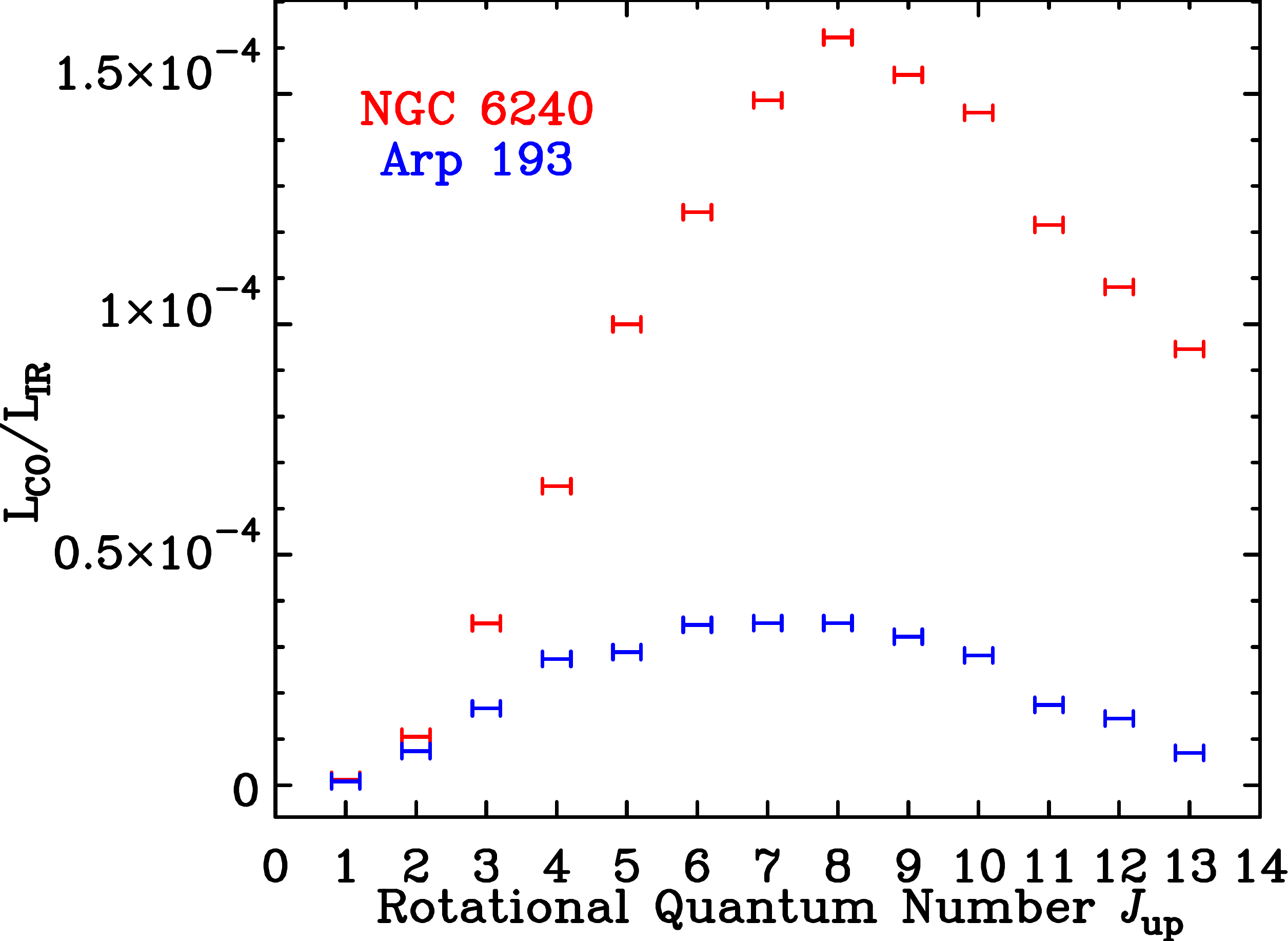}{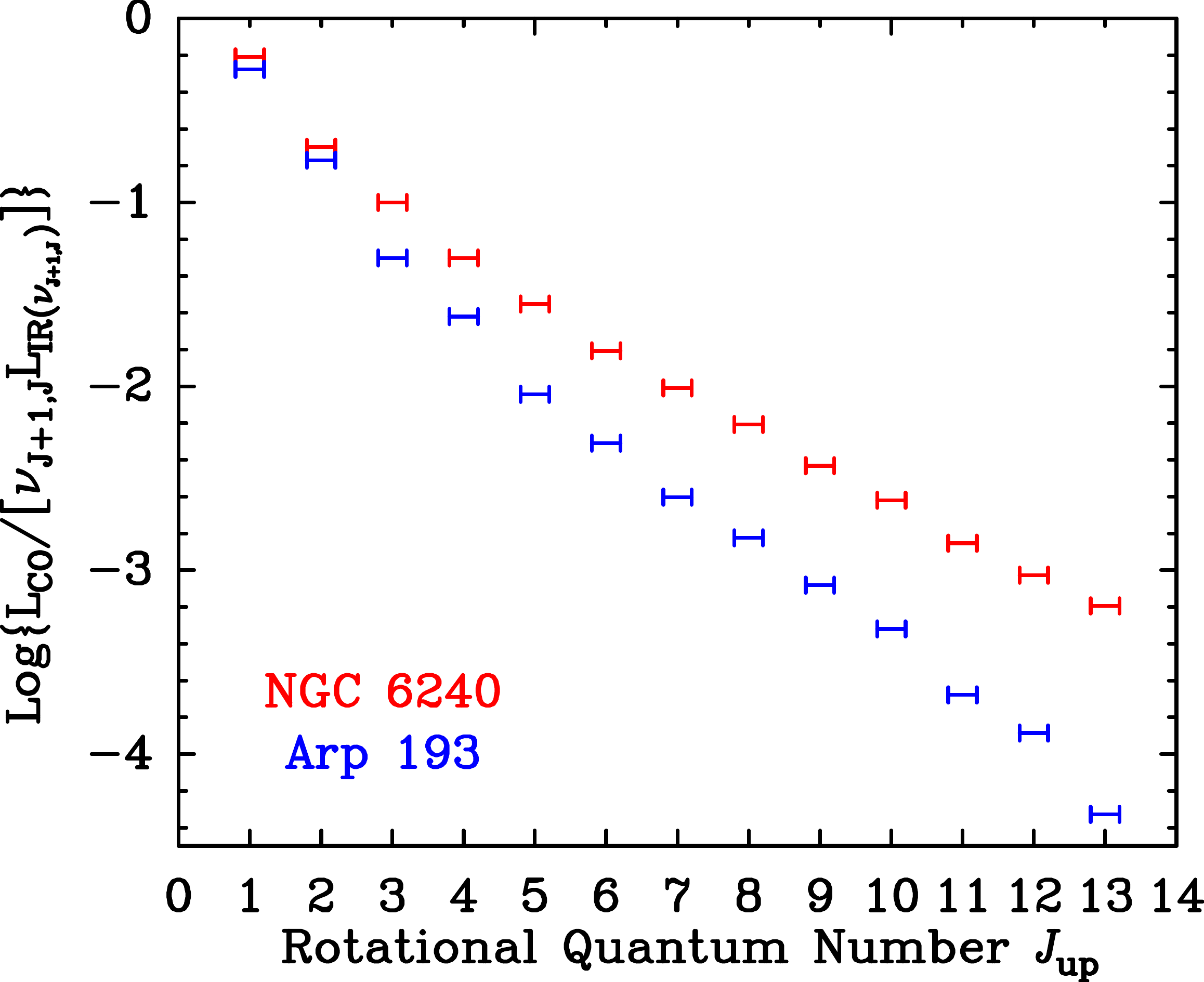}
\caption{The  dimensionless   CO  SLEDs  of   Arp\,193  and  NGC\,6240
  normalized  by  their  far-IR  luminosities  (top),  and  $\rm  [\nu
    _{J+1,J}L_{IR}(\nu  _{J+1,J})]$ of  the underlying  dust continuum
  (bottom)  (all  luminosities  in  $\rm L_{\odot}$)  revealing  the large
  divergence beyond the J=3--2 transition. }
\end{figure}

\newpage

\begin{figure}
\epsscale{1.5}
\plottwo{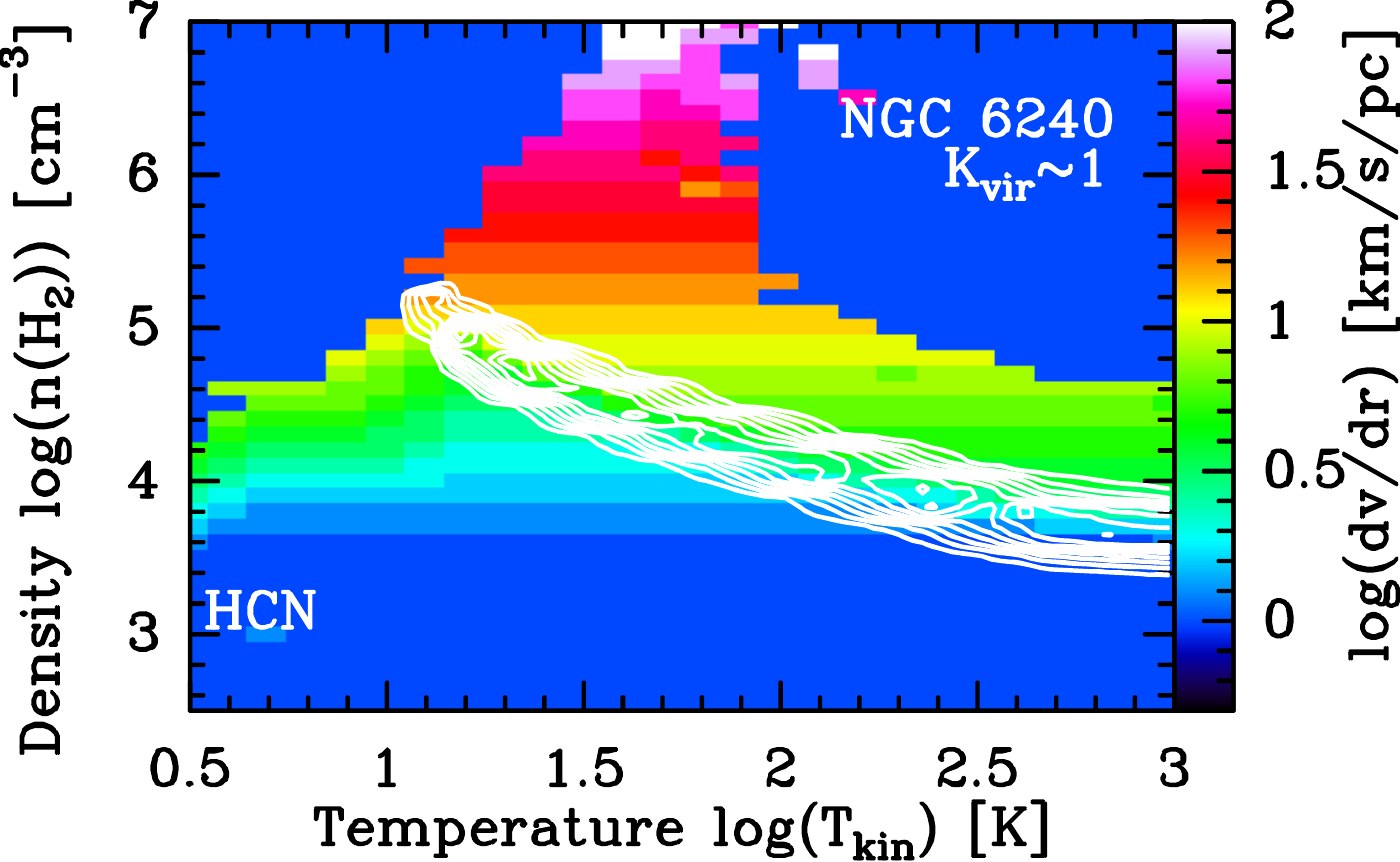}{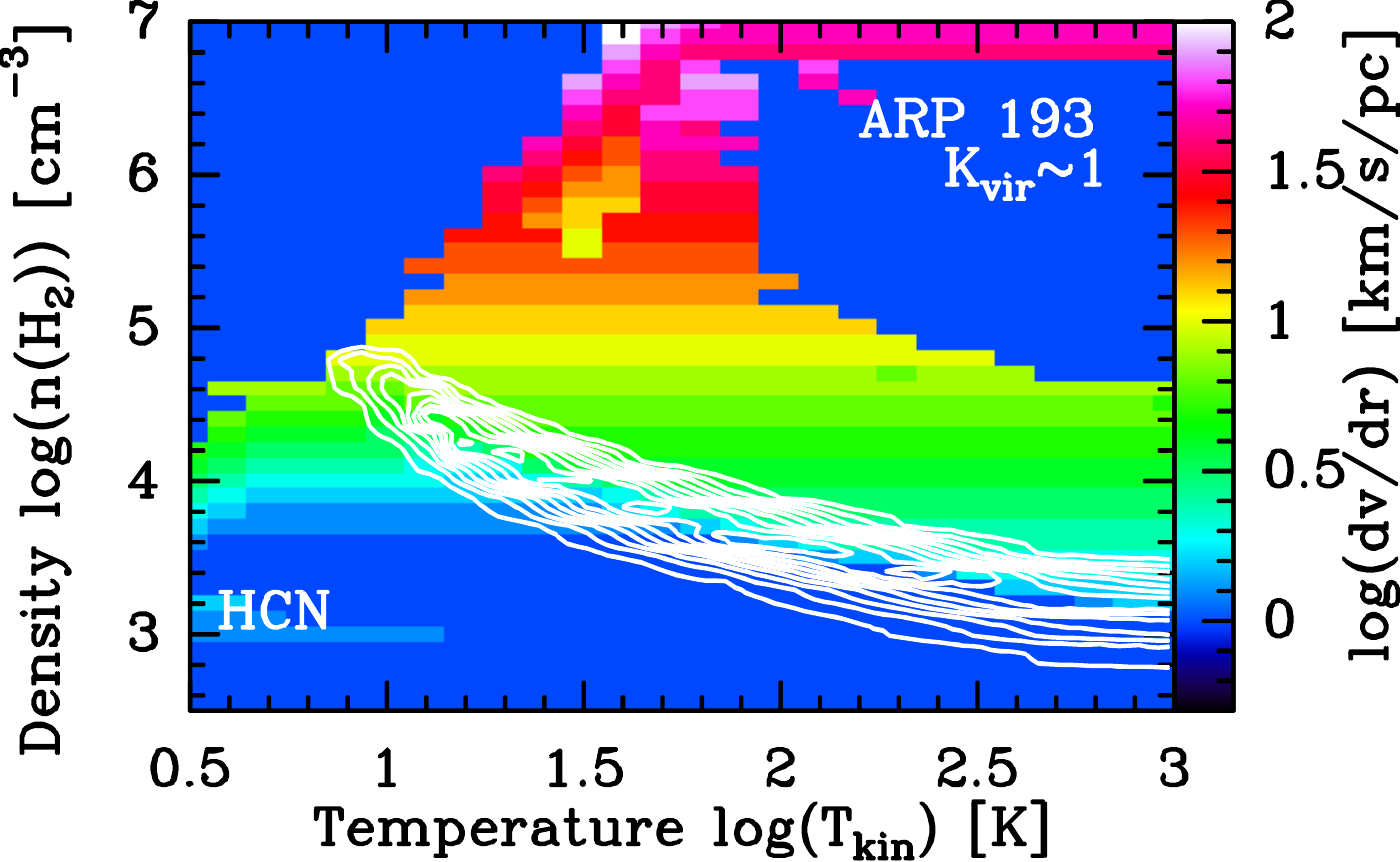}
\caption{Contours:  The two-dimensional  probability density  function
  (pdf) of the $\rm [n, T_{kin}]$ LVG parameters (in steps of 0.2), as
  constrained by the  HCN SLED in NGC\,6240 and  Arp\,193, Color:
    the range of (dV/dR) within the $\rm K_{vir}$=0.5--2 range (virial
    states) allowed  in the solution  search (3.1).  Details of
    the model and pdf analysis are in Appendix A.}  
\end{figure}

\newpage

\begin{figure}
\epsscale{0.8}
\plotone{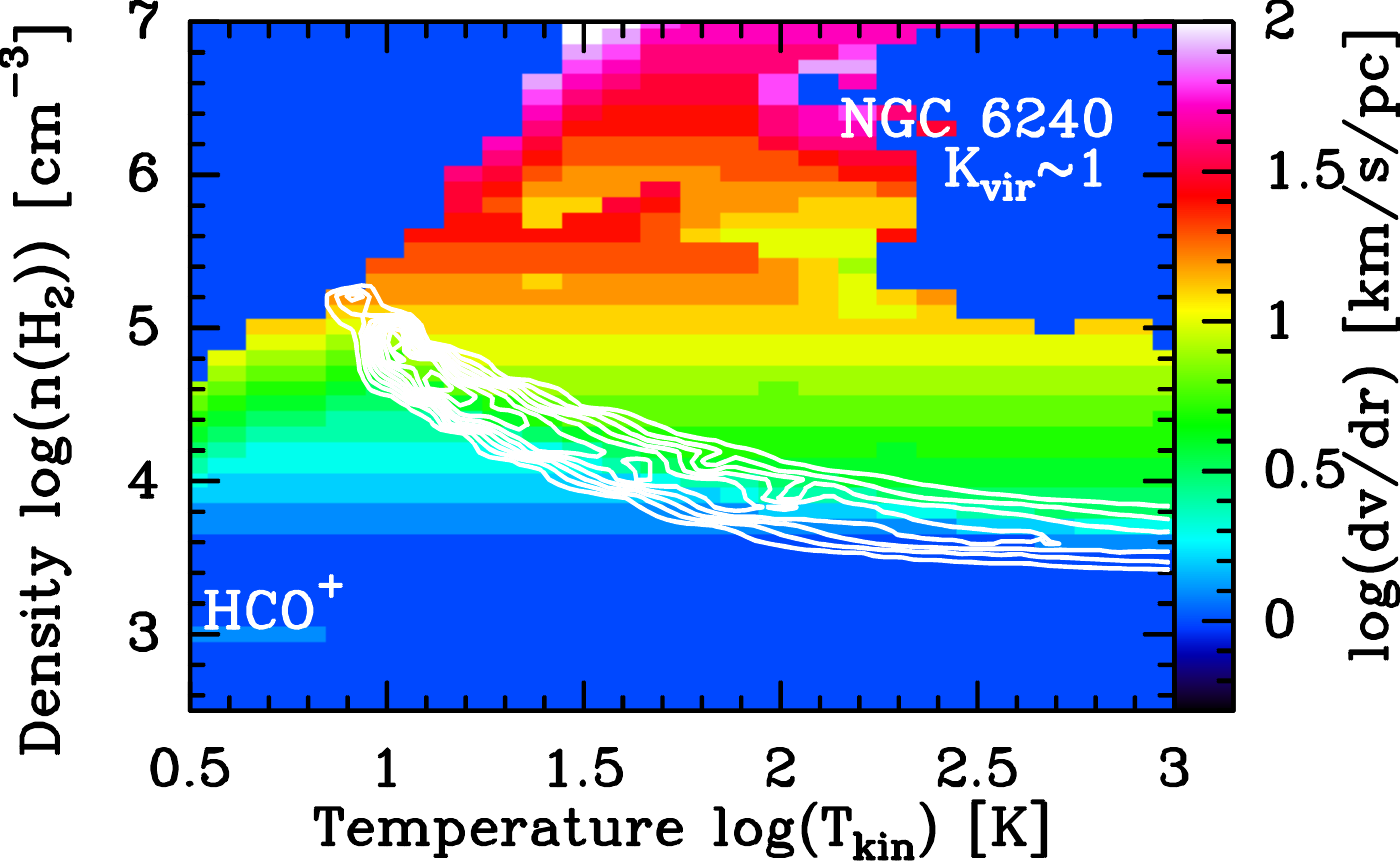}
\caption{Contours:  the two-dimensional  probability density  function
  (pdf) of the $\rm [n, T_{kin}]$ LVG parameters (in steps of 0.2), as
  constrained  by the  HCO$^{+}$  line ratios  measured in  NGC\,6240.
  Color:  the range  of (dV/dR)  within the  $\rm K_{vir}$=0.5--2
    range (virial states)  allowed in the solution  search (3.1).  
    Details of  the model and  pdf analysis are  in Appendix
    A.}
\end{figure}

\clearpage

\newpage

\begin{figure}
\epsscale{0.4}
\plotone{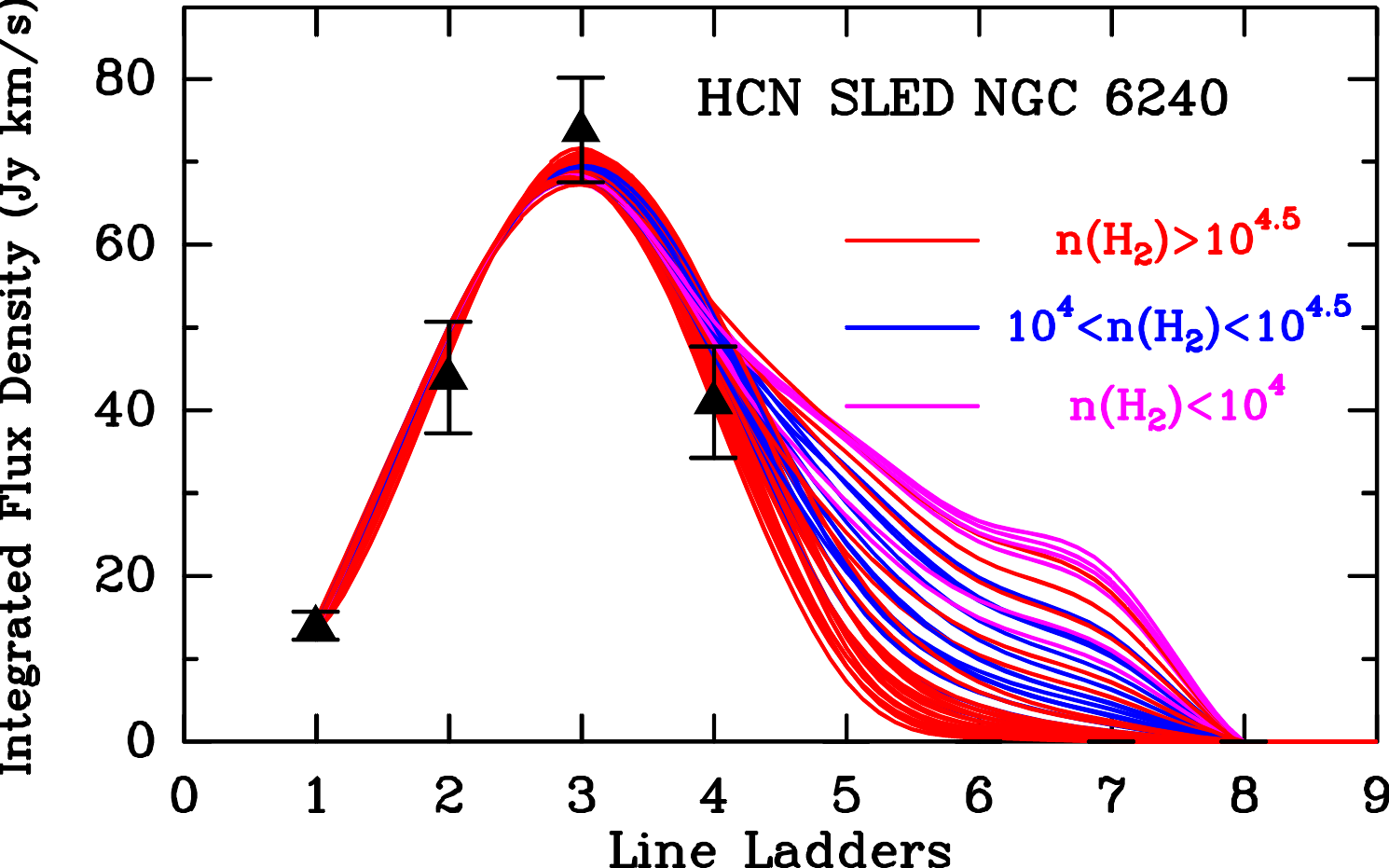}
\plotone{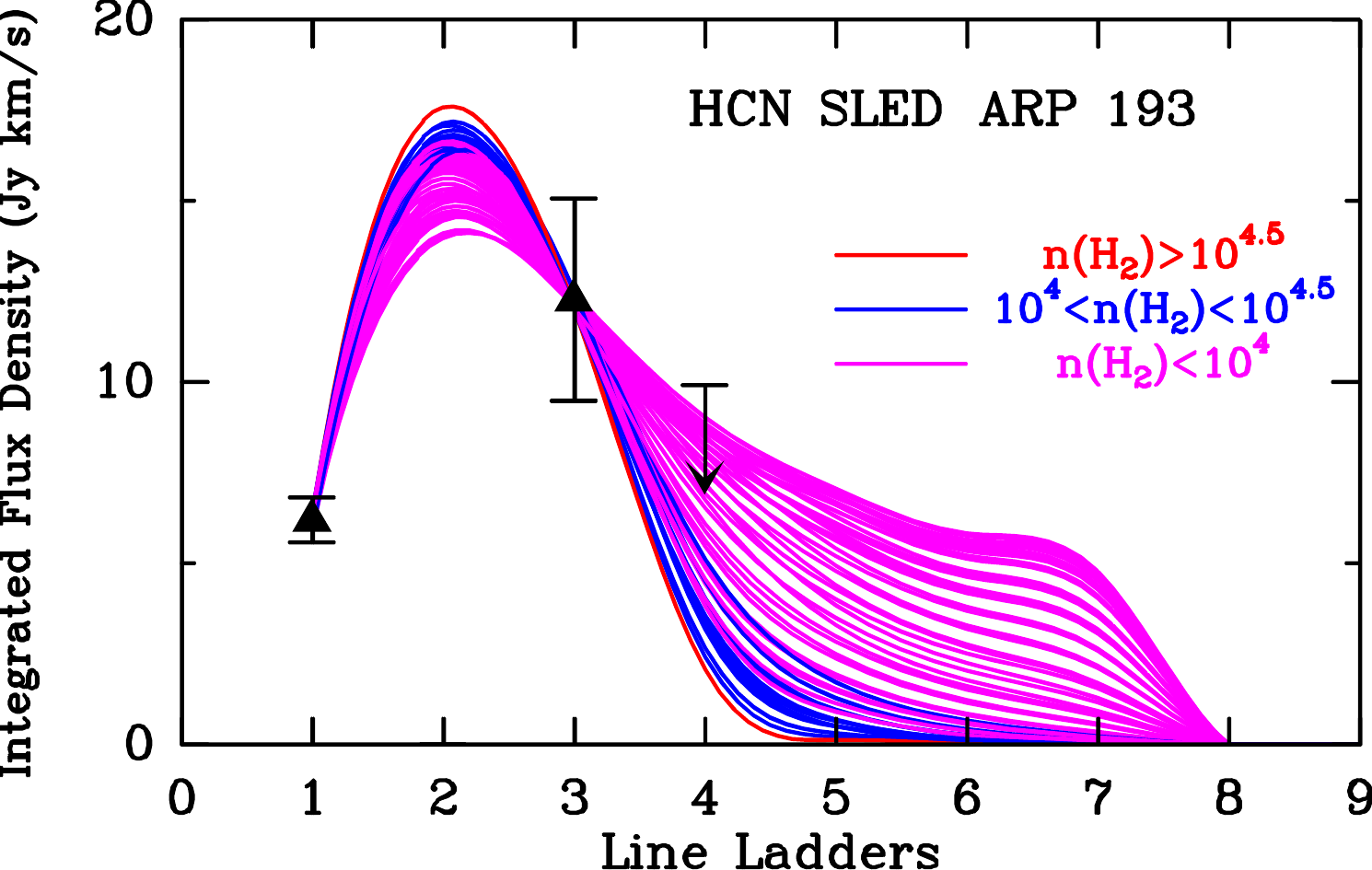}
\plotone{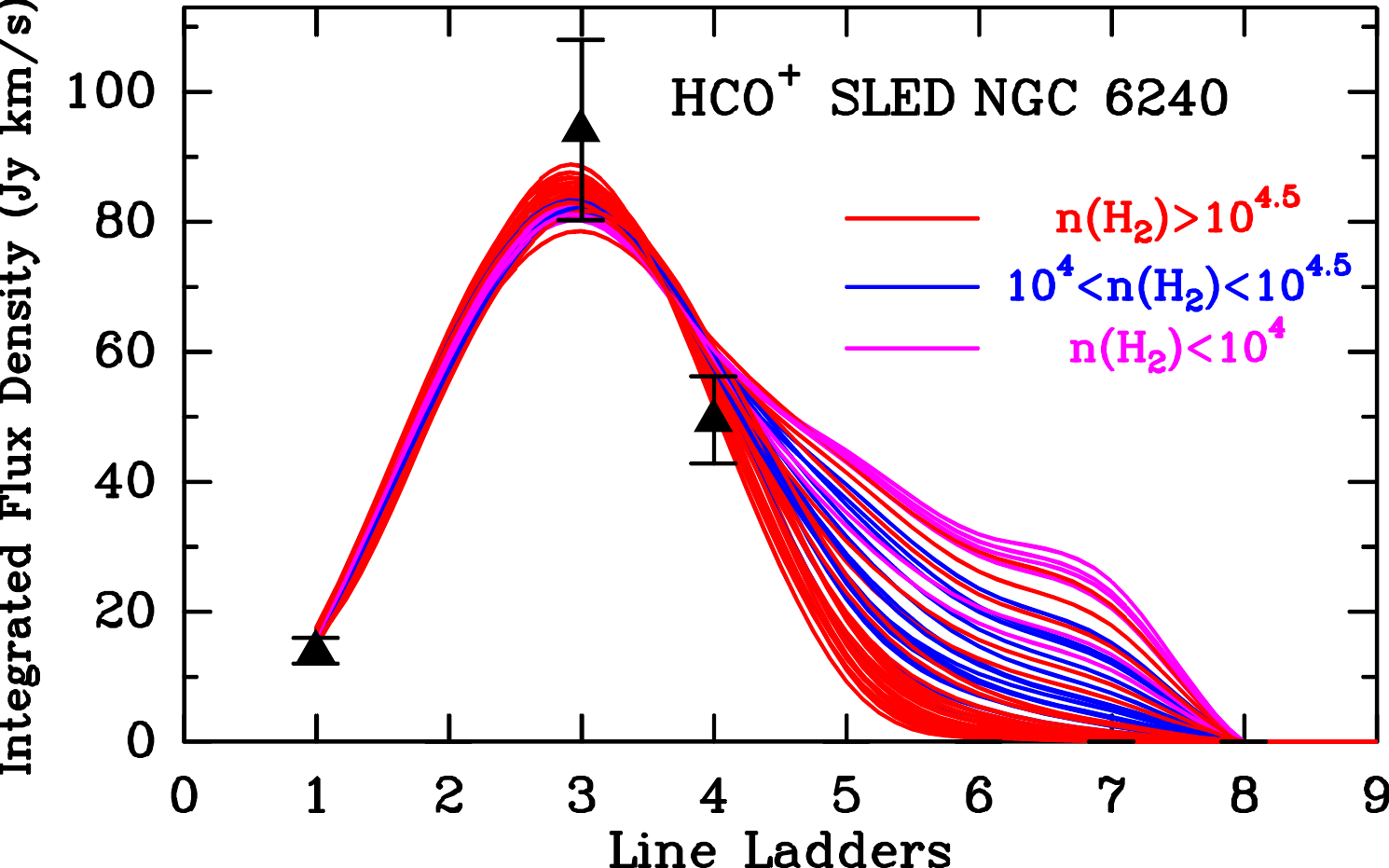}
\plotone{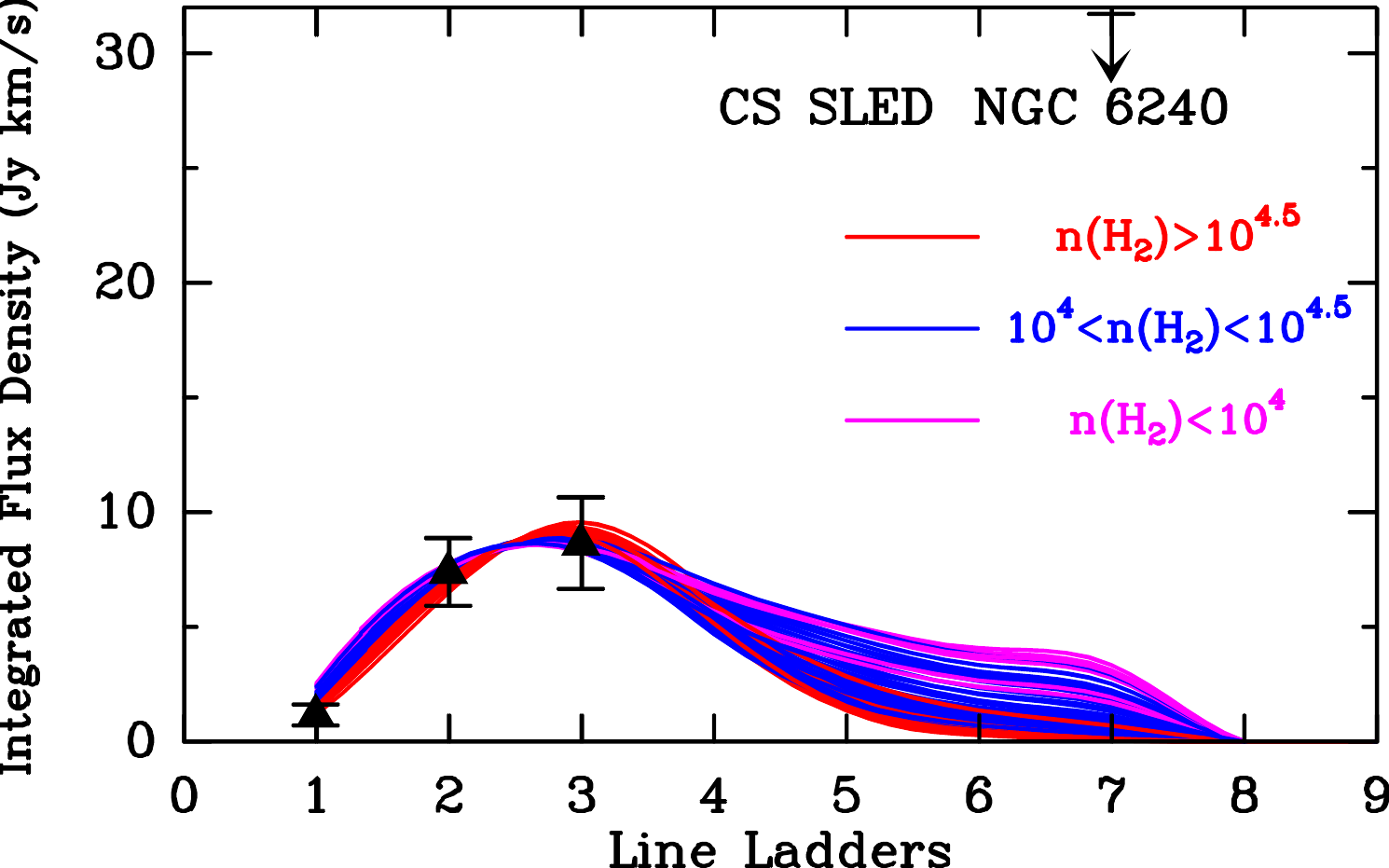}
\caption{The  molecular  SLEDs  of   the  $\rm  [n,  T_{kin},  dV/dR]$
  solutions shown in  Figure 5, parametrized by  their density ranges.
  There  is good  agreement with  all available  high-dipole molecular
  lines,  while  additional  HCN  and/or  the  HCO$^{+}$  J=5--4  line
  measurements  could   much  reduce   the  density and thus also  temperature
  degeneracies (see Figures 5, 6).}
\end{figure}

\clearpage

\newpage

\begin{figure}
\epsscale{0.8}
\plotone{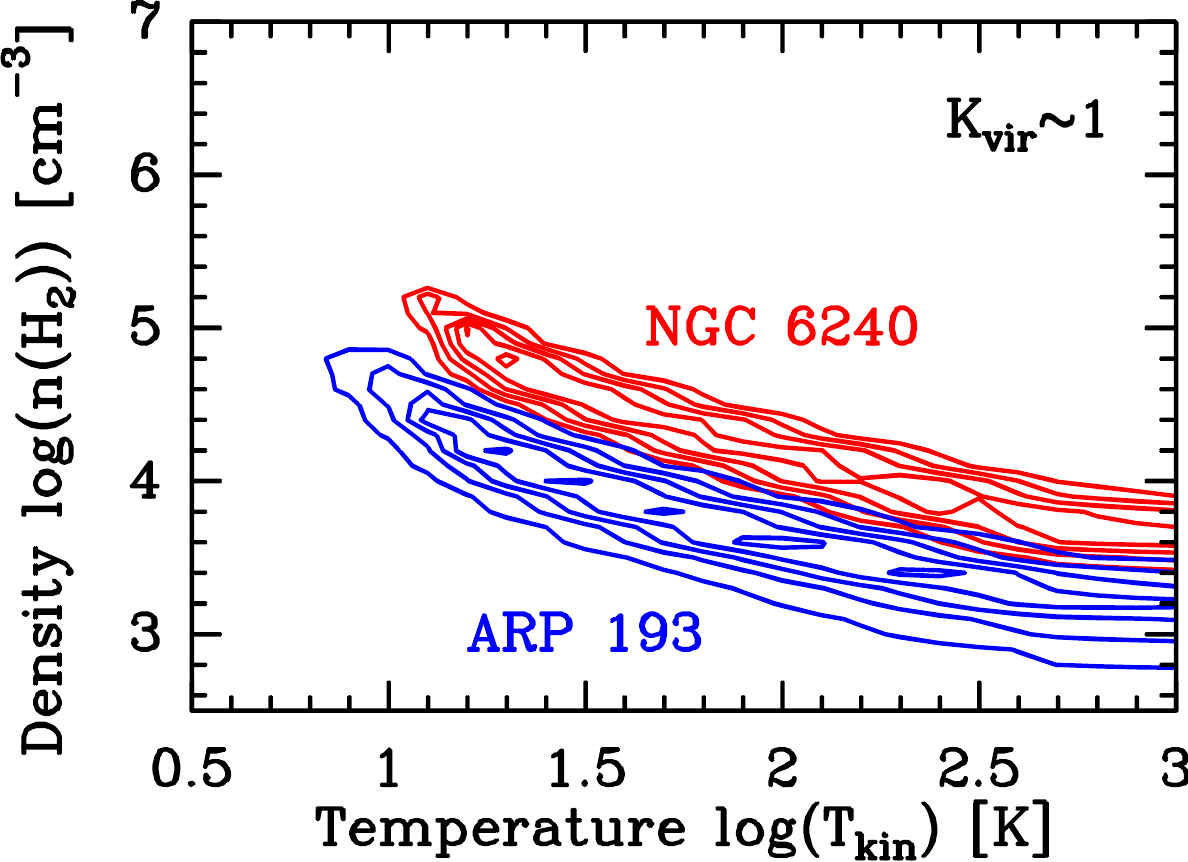}
\caption{The  overlaid  pdfs  of  the  $\rm  [n,  T_{kin}]$  solutions
  constrained by the HCN SLEDs of Arp\,193 and NGC\,6240, and for
 self-gravitating gas states (see also Figure 5).}
\end{figure}

\clearpage

\newpage

\begin{figure}
\epsscale{1.5}
\plottwo{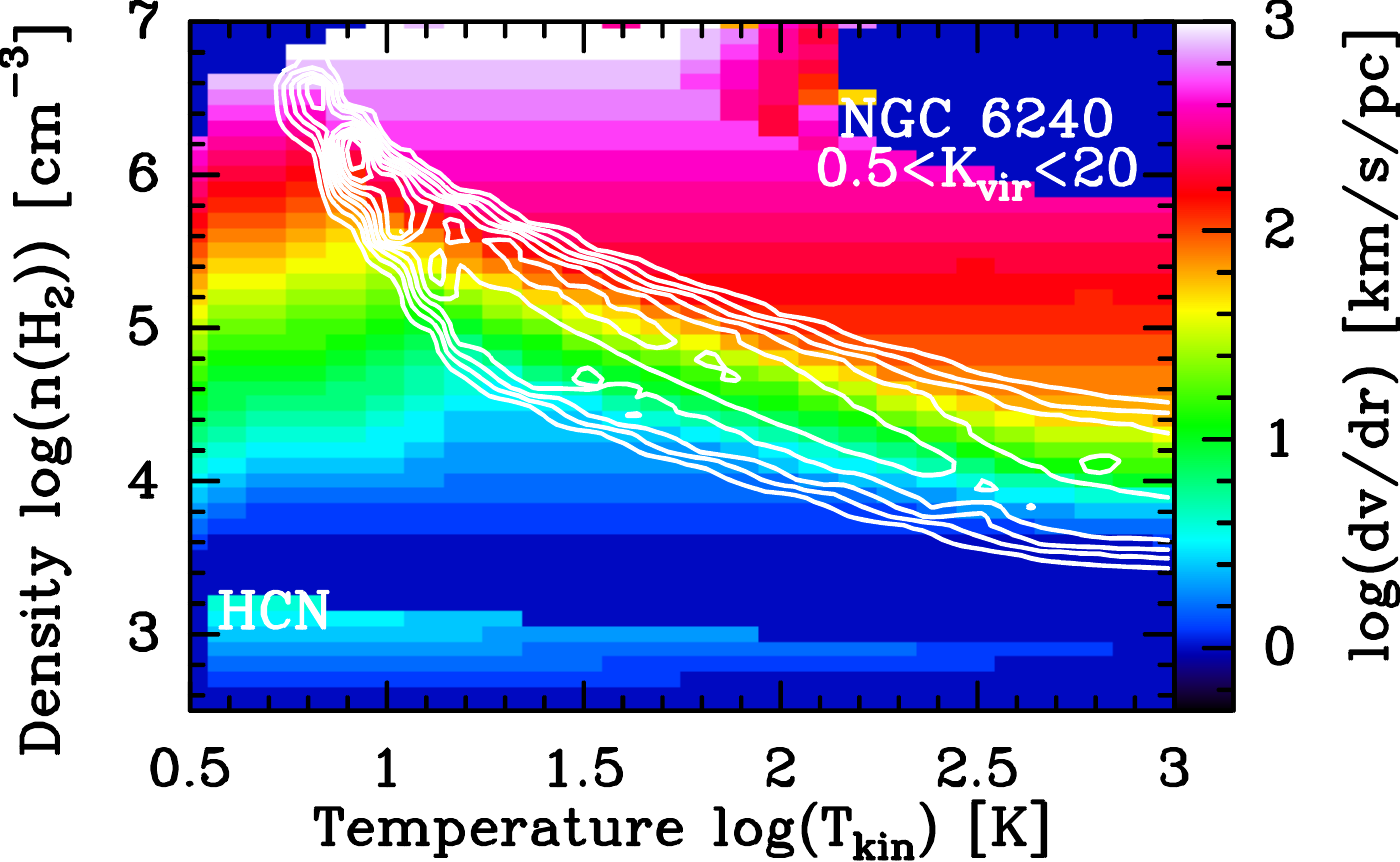}{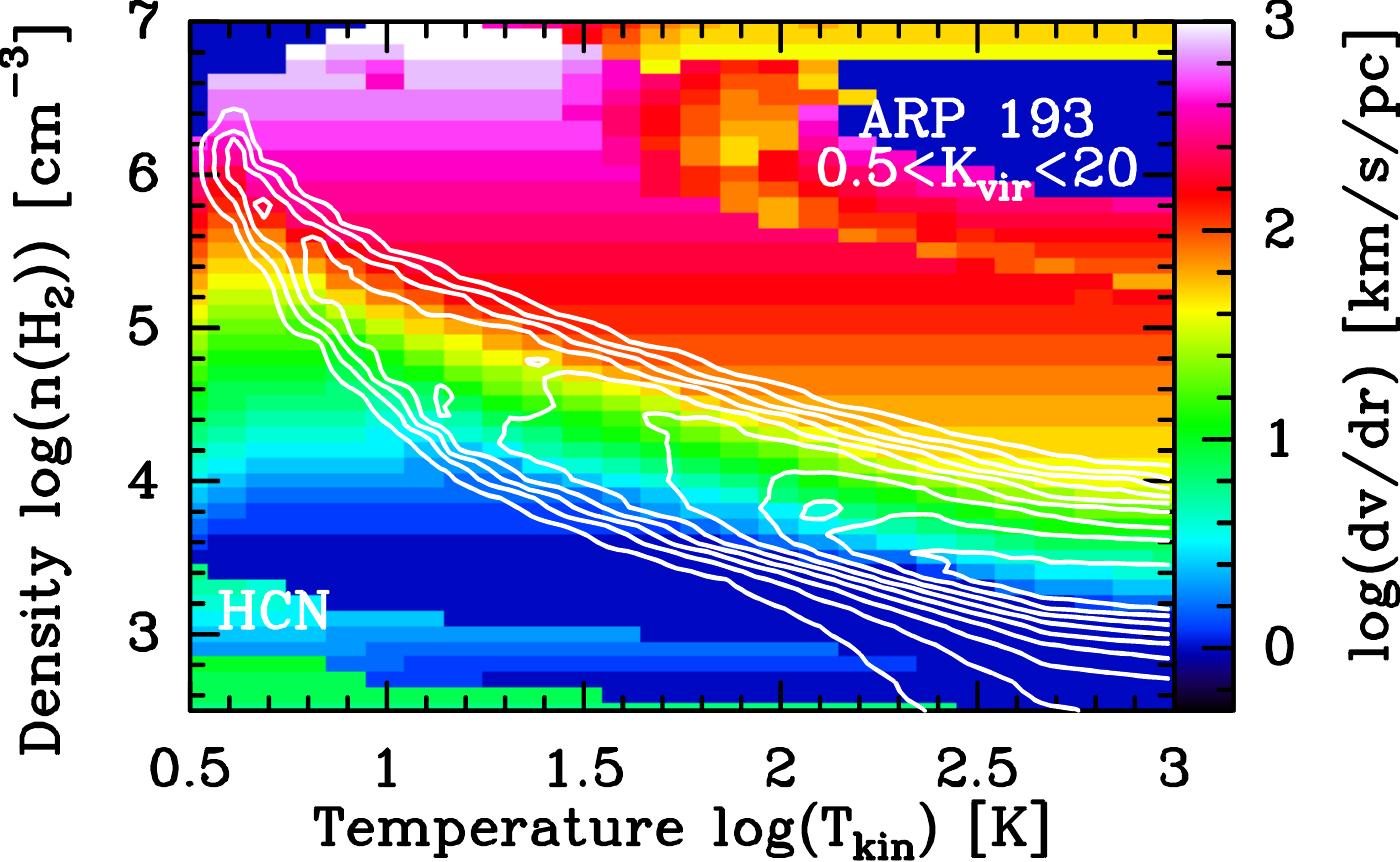}
\caption{Contours:  the two-dimensional  probability density  function
  (pdf) of the $\rm [n, T_{kin}]$ LVG parameters (in steps of 0.2), as
  constrained  by  the  HCN  line ratios  measured  in  NGC\,6240  and
  Arp\,193 but allowing also non-virial gas states (See Appendix A for
  details)   Color: the  dV/dR values of  the 0.5$<$$\rm
    K_{vir}$$<$20 range of the LVG solution search (see 3.1.1).}
\end{figure}

\clearpage

\newpage

\begin{figure}
\epsscale{0.8}
\plotone{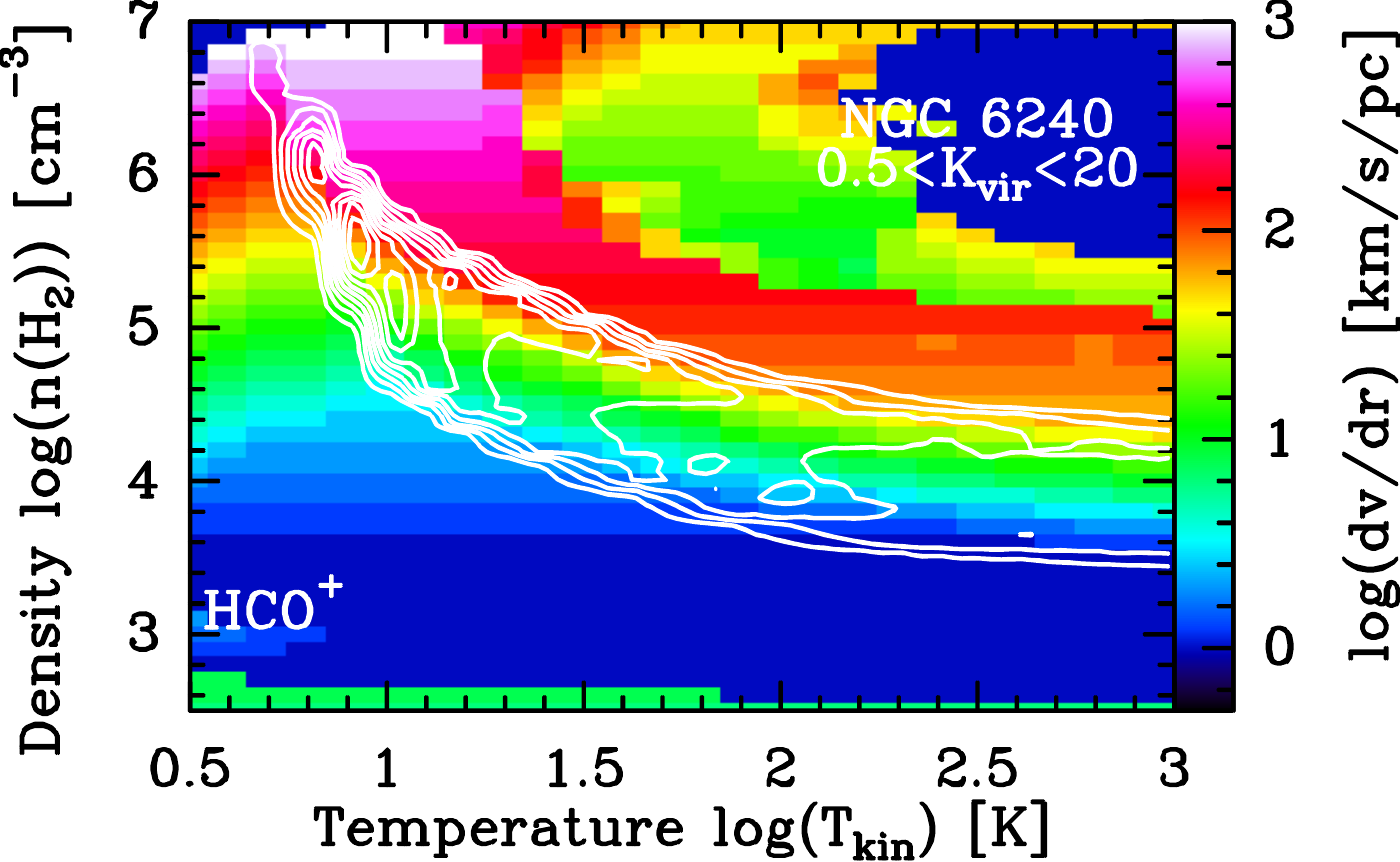}
\caption{Contours:  the two-dimensional  probability density  function
  (pdf) of the $\rm [n, T_{kin}]$ LVG parameters (in steps of 0.2), as
  constrained by the HCO$^{+}$ line  ratios measured in NGC\,6240 with
  the solution search  extended to unbound gas states  (See Appendix A
  for details).  Color: the  dV/dR values for  the 0.5$<$$\rm
    K_{vir}$$<$20 range of the LVG solution search (see 3.1.1).}
\end{figure}

\clearpage

\newpage

\begin{figure}
\epsscale{0.8}
\plotone{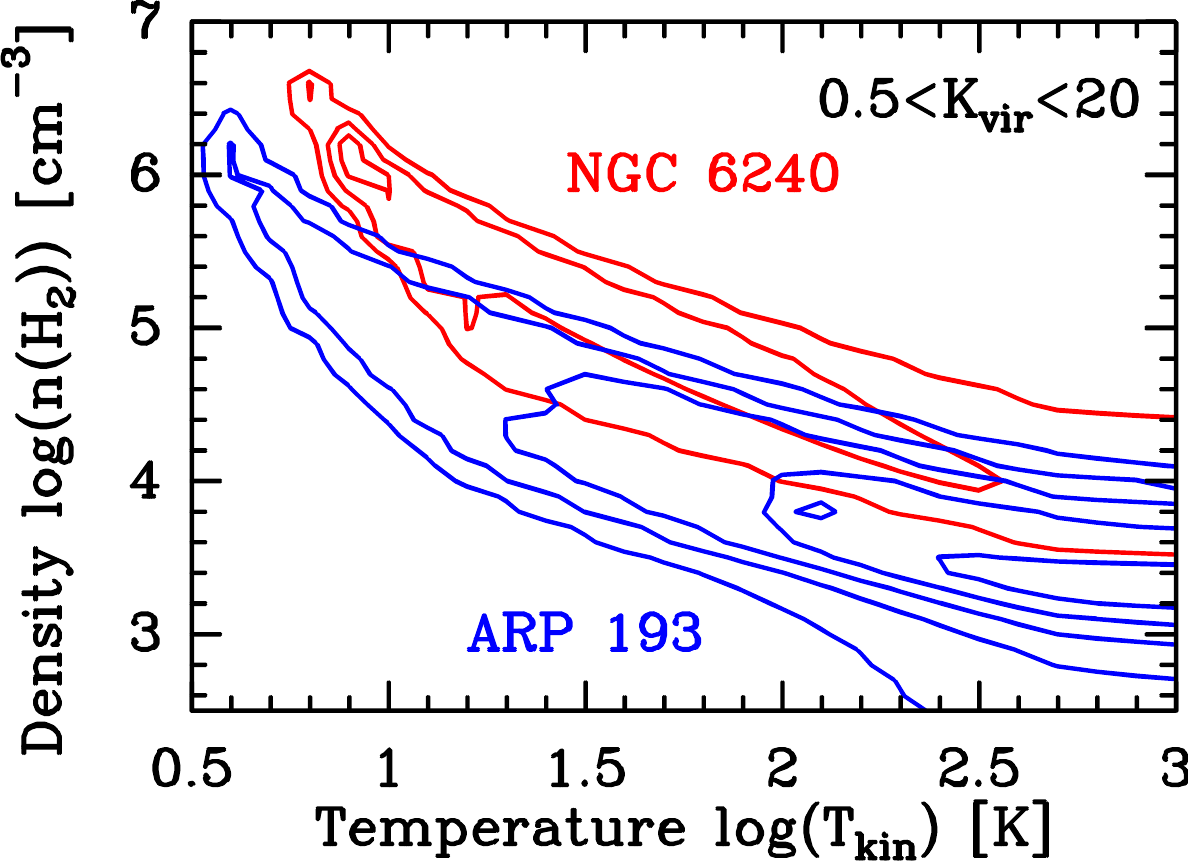}
\caption{The  overlaid  pdfs  of  the  $\rm  [n,  T_{kin}]$  solutions
  constrained by  the HCN ratios  measured in Arp\,193  and NGC\,6240,
  but  now with  the LVG  model search  extended to  include also
    unbound  states within  the  0.5$<$$\rm  K_{vir}$$<$20 range.}
\end{figure}

\clearpage

\newpage

\begin{figure}
\epsscale{0.6}
\plotone{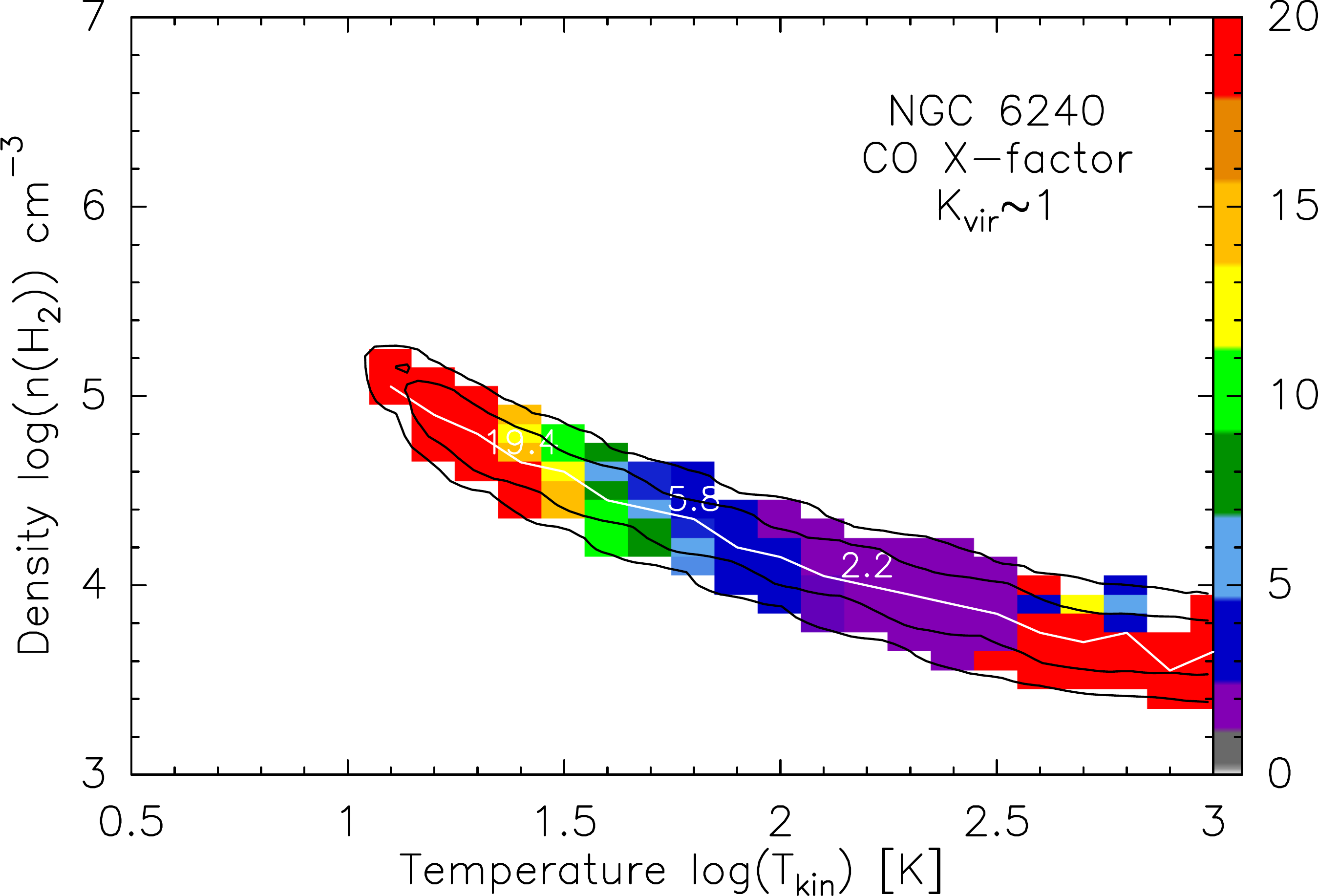}
\plotone{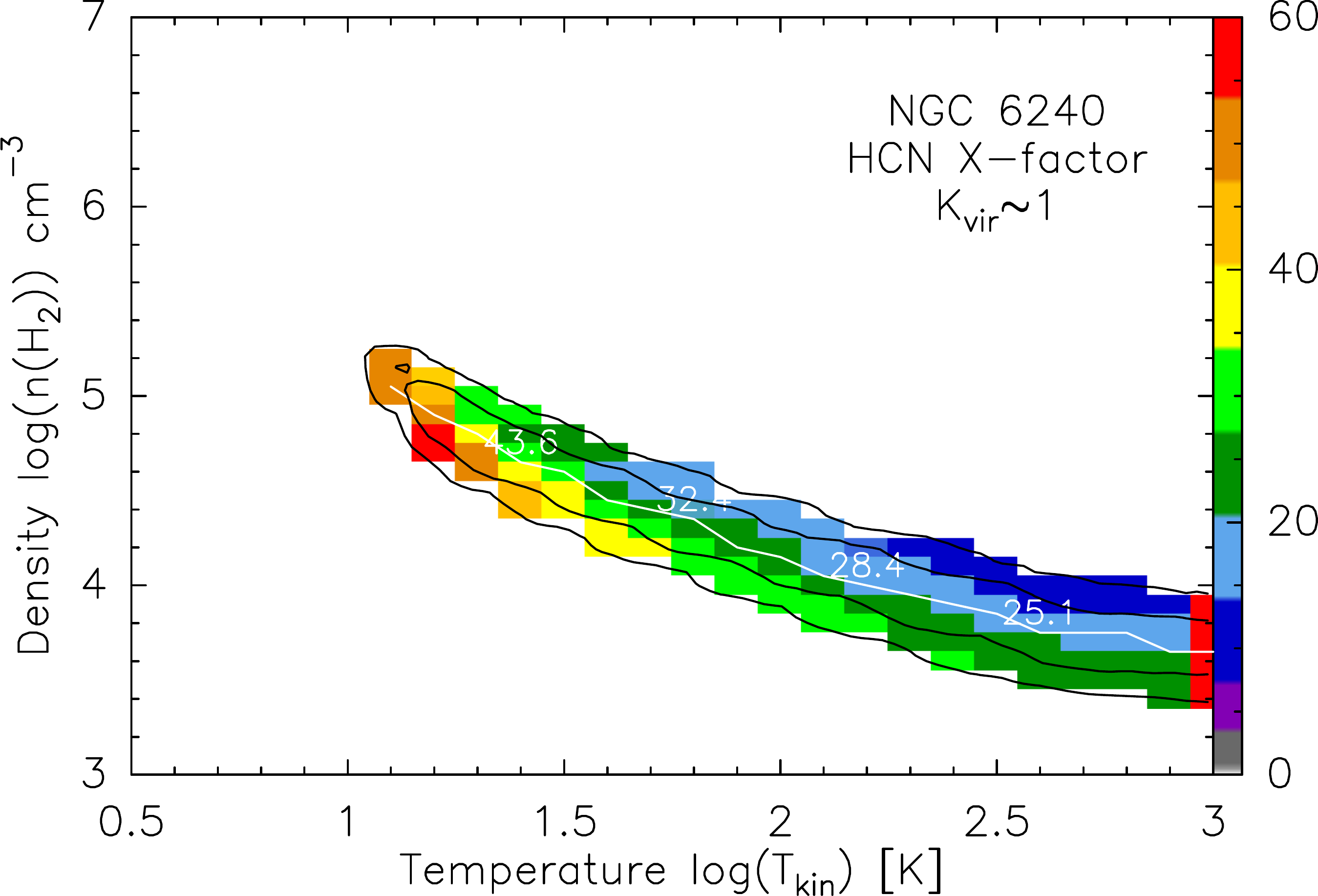}
\plotone{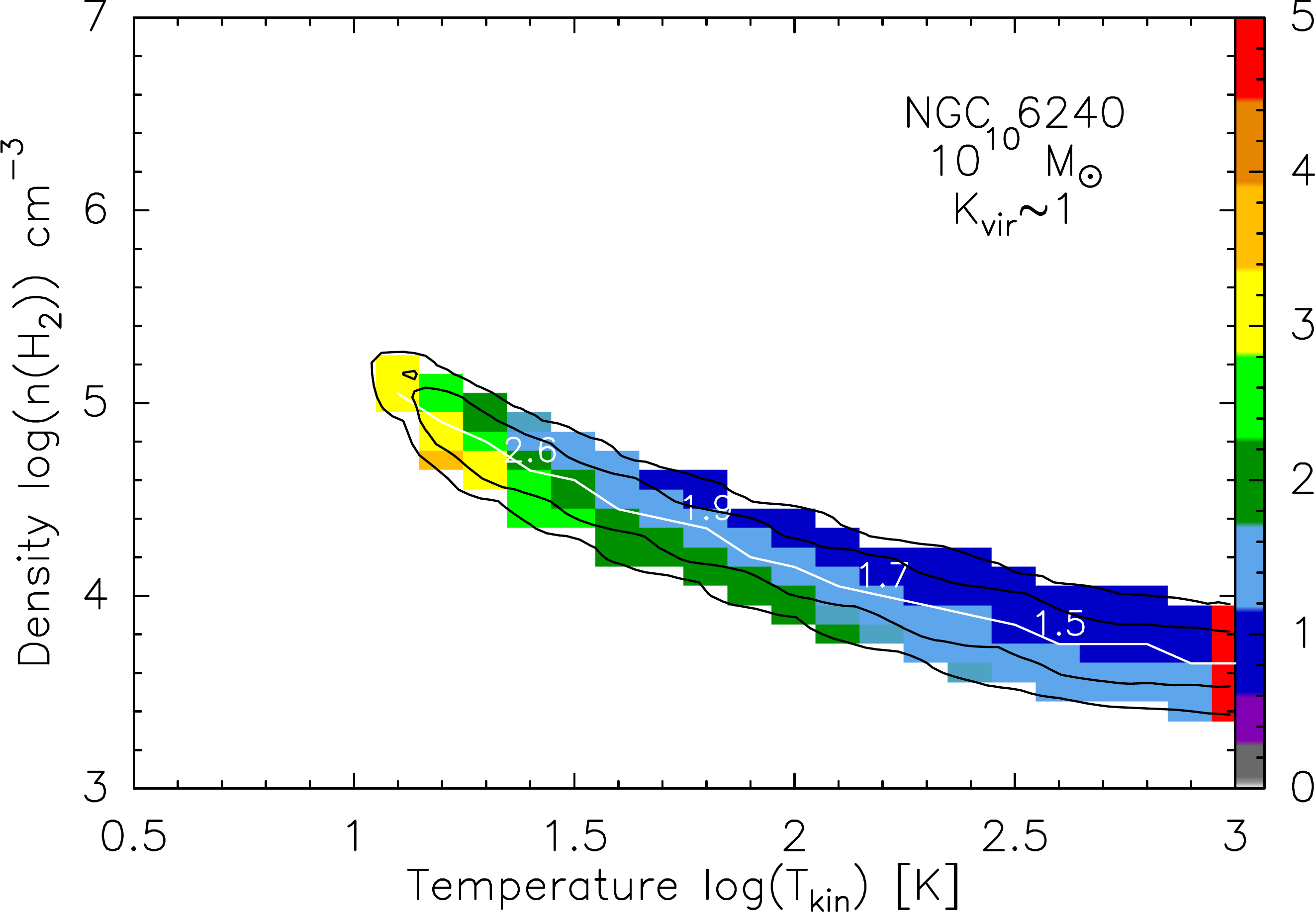}
\caption{Color:  The $\rm  X_{co}$,  $\rm X_{HCN}$  factors, and  $\rm
  M_{dense}(H_2)$    for    NGC\,6240    obtained    for    for    its
  (HCN/HCO$^{+}$)-constrained LVG solution  range and self-gravitating
  states (Figure 5). Contours: black lines mark the 0.1 and 0.5 values
  of  the  pdf.   The  white  line   traces  the  X  factor  (in  $\rm
  M_{\odot}(K\,km\,s^{-1}\,pc^2)^{-1}$)  and  the  corresponding  $\rm
  M_{dense}$  (in  $\rm 10^{10}\,M_{\odot}$)   computed  from  the  median
  density within the 0.5 contour at the corresponding $\rm T_{kin}$ in the
  x-axis.    The  few   very   large  values   (red   color)  at   the
  (low-n)/(high-T)   end  of   the  distribution   are  artifacts   of
  non-convergence of RADEX, and are not included in our analysis. }
\end{figure}

\clearpage

\newpage

\begin{figure}
\epsscale{0.6}
\plotone{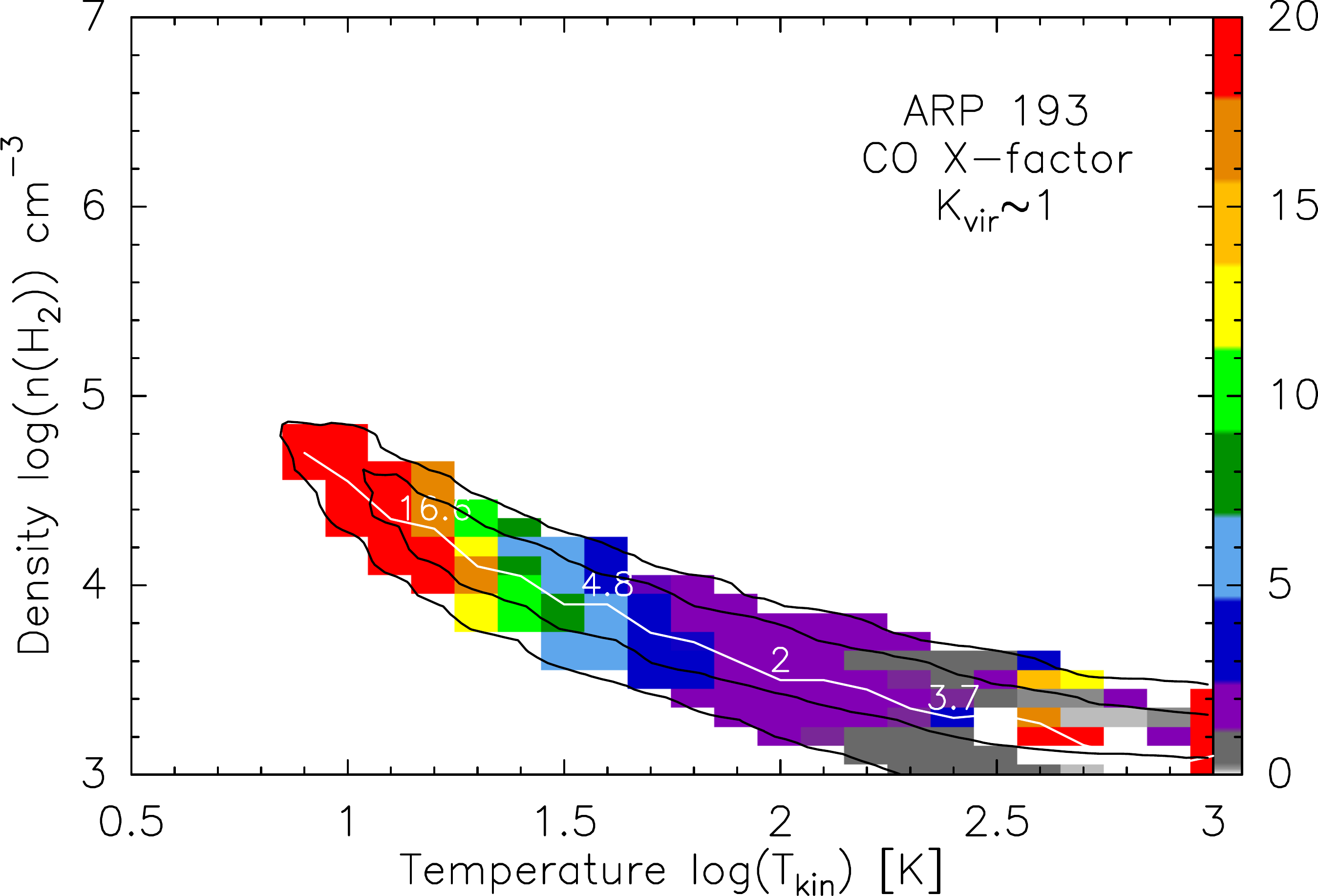}
\plotone{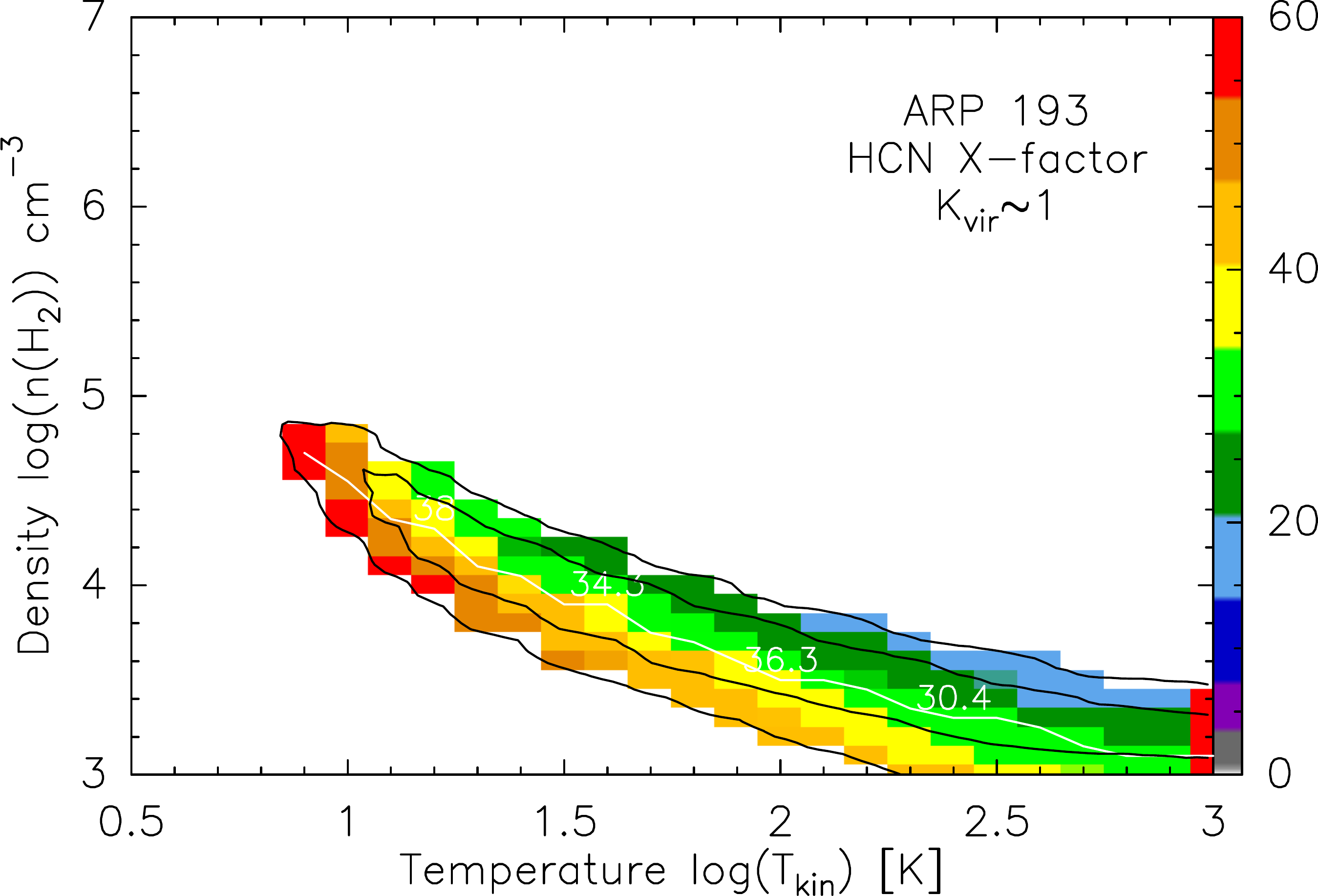}
\plotone{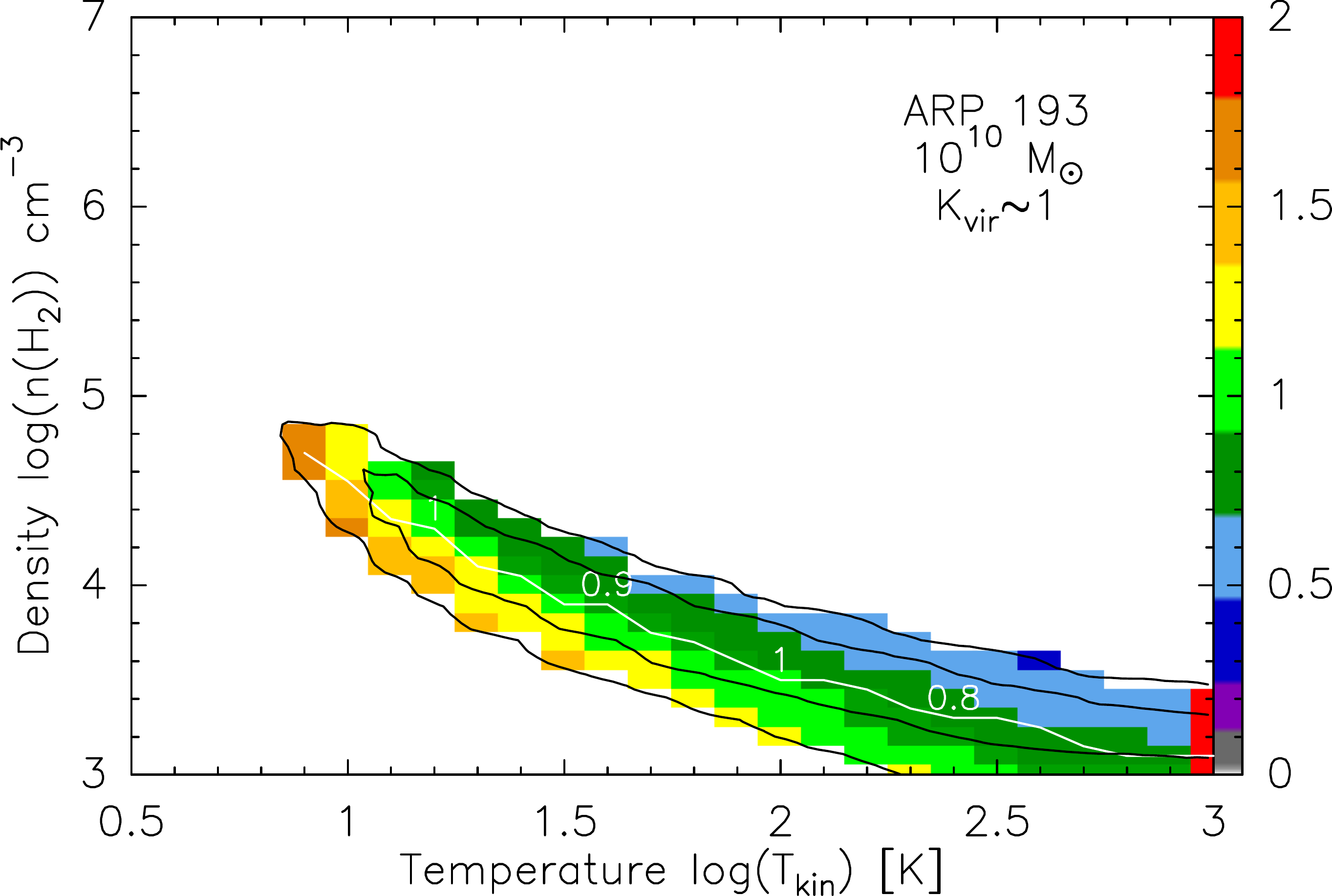}
\caption{Colors and contours as  in Figure 12, but for  Arp\,193 and using the
HCN-constrained parameter space shown in Figure  5. The X factor in $\rm
  M_{\odot}(K\,km\,s^{-1}\,pc^2)^{-1}$ and $\rm M_{dense}$ in $10^{10}$\,M$_{\odot}$. }
\end{figure}

\clearpage

\newpage

\begin{figure}
\epsscale{0.6}
\plotone{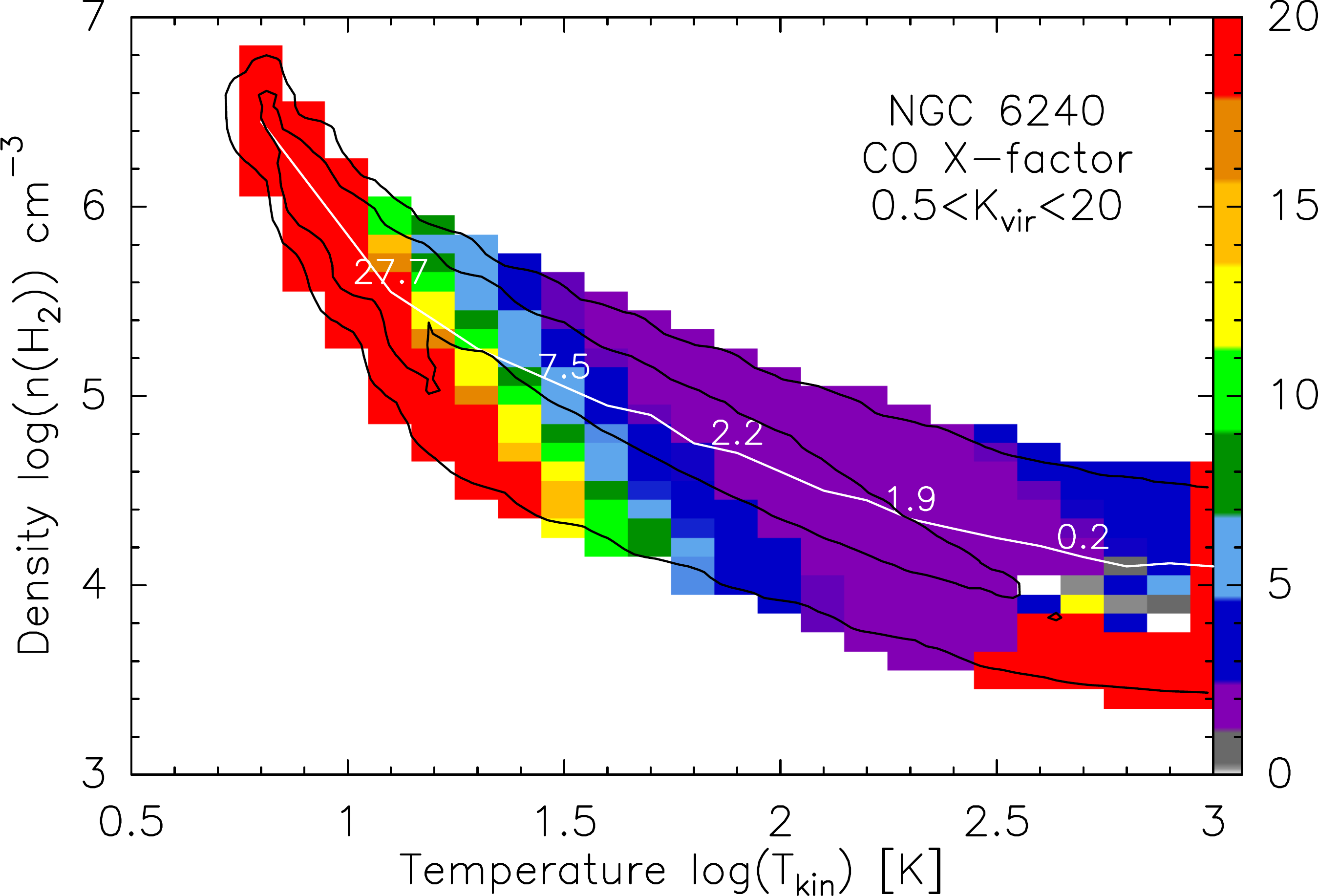}
\plotone{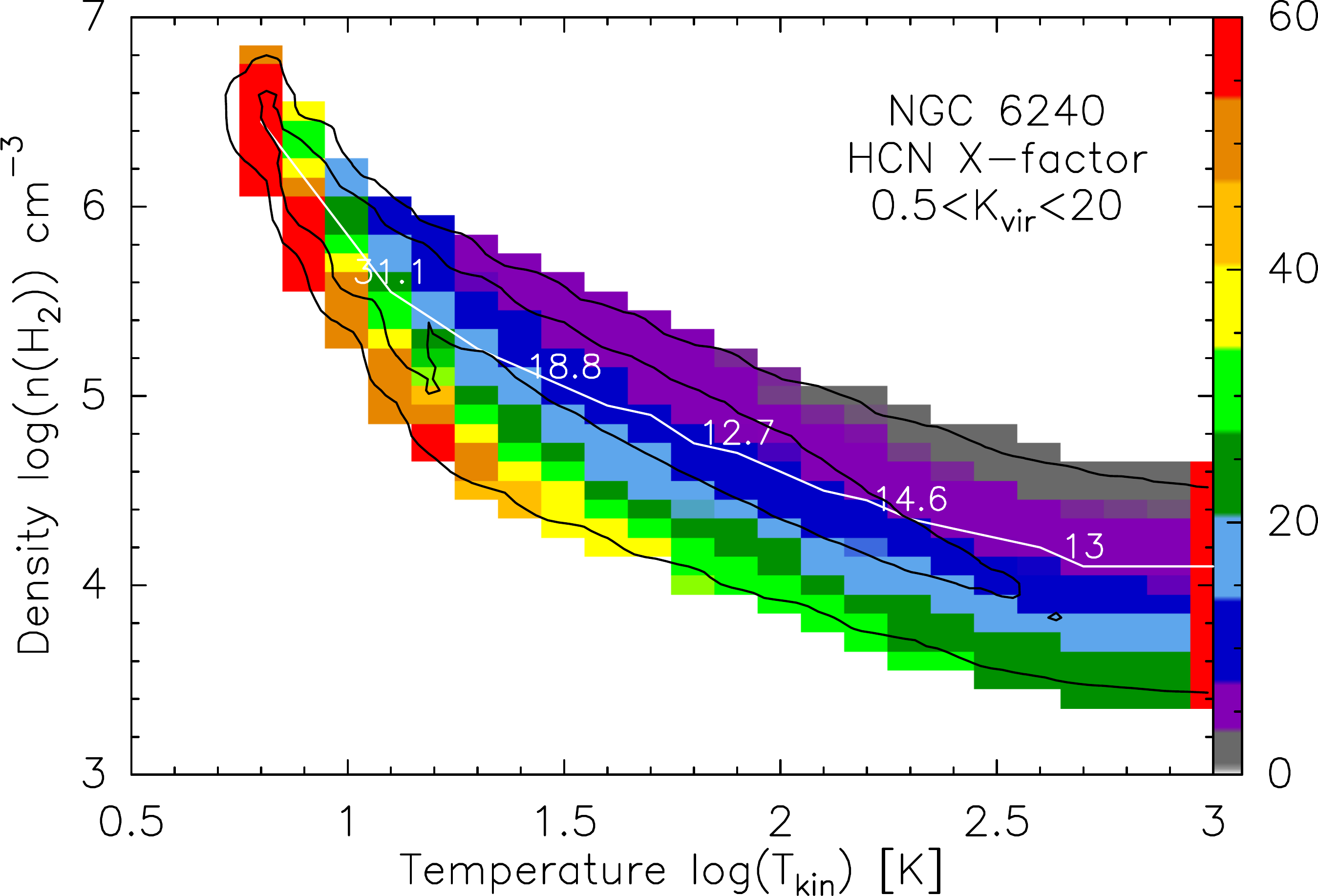}
\plotone{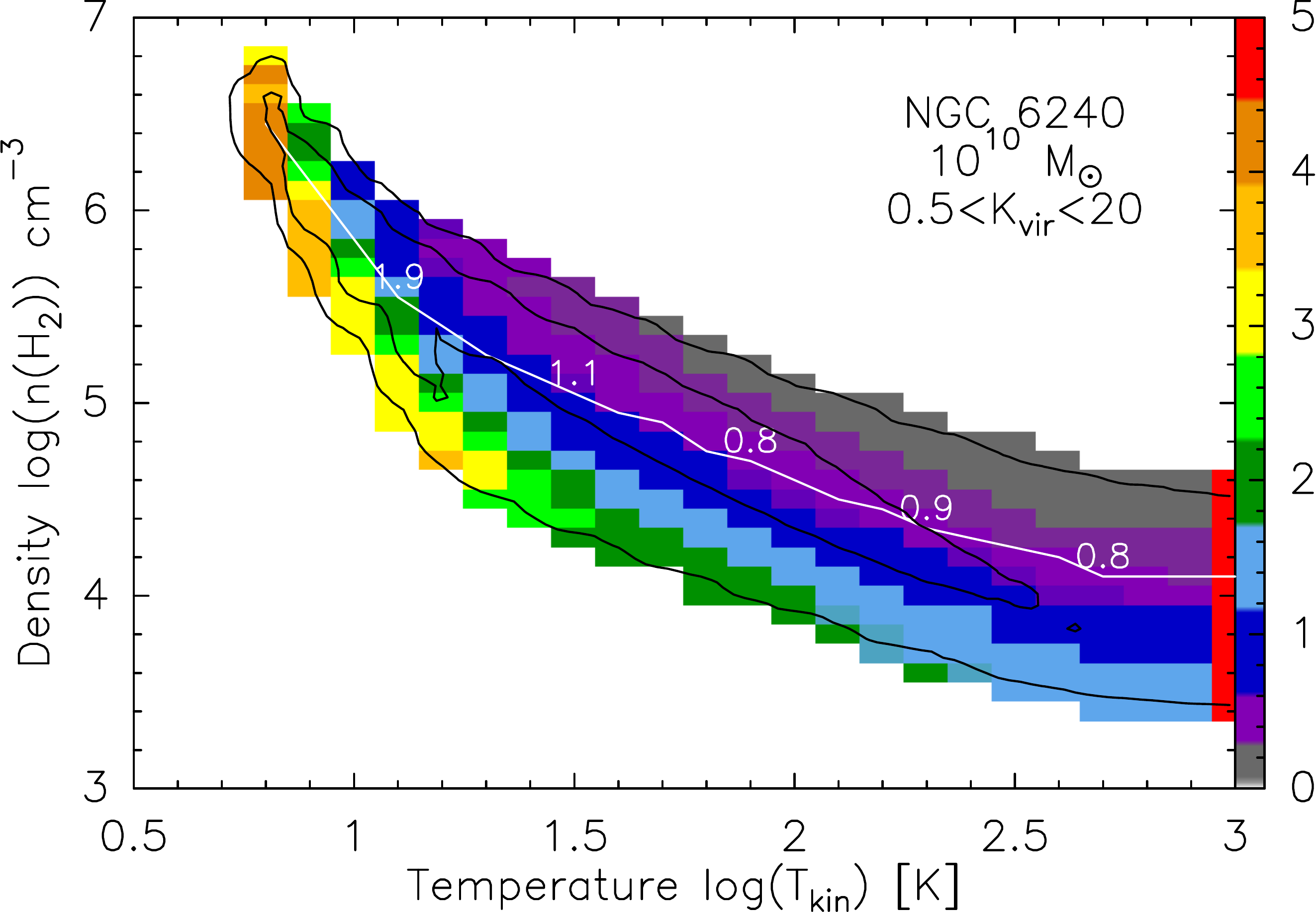}
\caption{Color:  The $\rm  X_{co}$,  $\rm X_{HCN}$  factors, and  $\rm
  M_{dense}$   obtained  from   its  (HCN/HCO$^{+}$)-constrained   LVG
  solution   range   that   now  includes   unbound   states   (Figure
  9). Contours:  black lines mark the  0.1 and 0.5 values  of the pdf.
  The    white    line    traces     the    X    factor    (in    $\rm
  M_{\odot}(K\,km\,s^{-1}\,pc^2)^{-1}$)  and  the  corresponding  $\rm
  M_{dense}$  (in  $\rm 10^{10}\,M_{\odot}$)   computed  from  the  median
  density within the 0.5 contour at the corresponding $\rm T_{kin}$ in the
  x-axis.    The  few   very   large  values   (red   color)  at   the
  (low-n)/(high-T)   end  of   the  distribution   are  artifacts   of
  non-convergence of RADEX, and are not included in our analysis.}
\end{figure}

\clearpage

\newpage

\begin{figure}
\epsscale{0.6}
\plotone{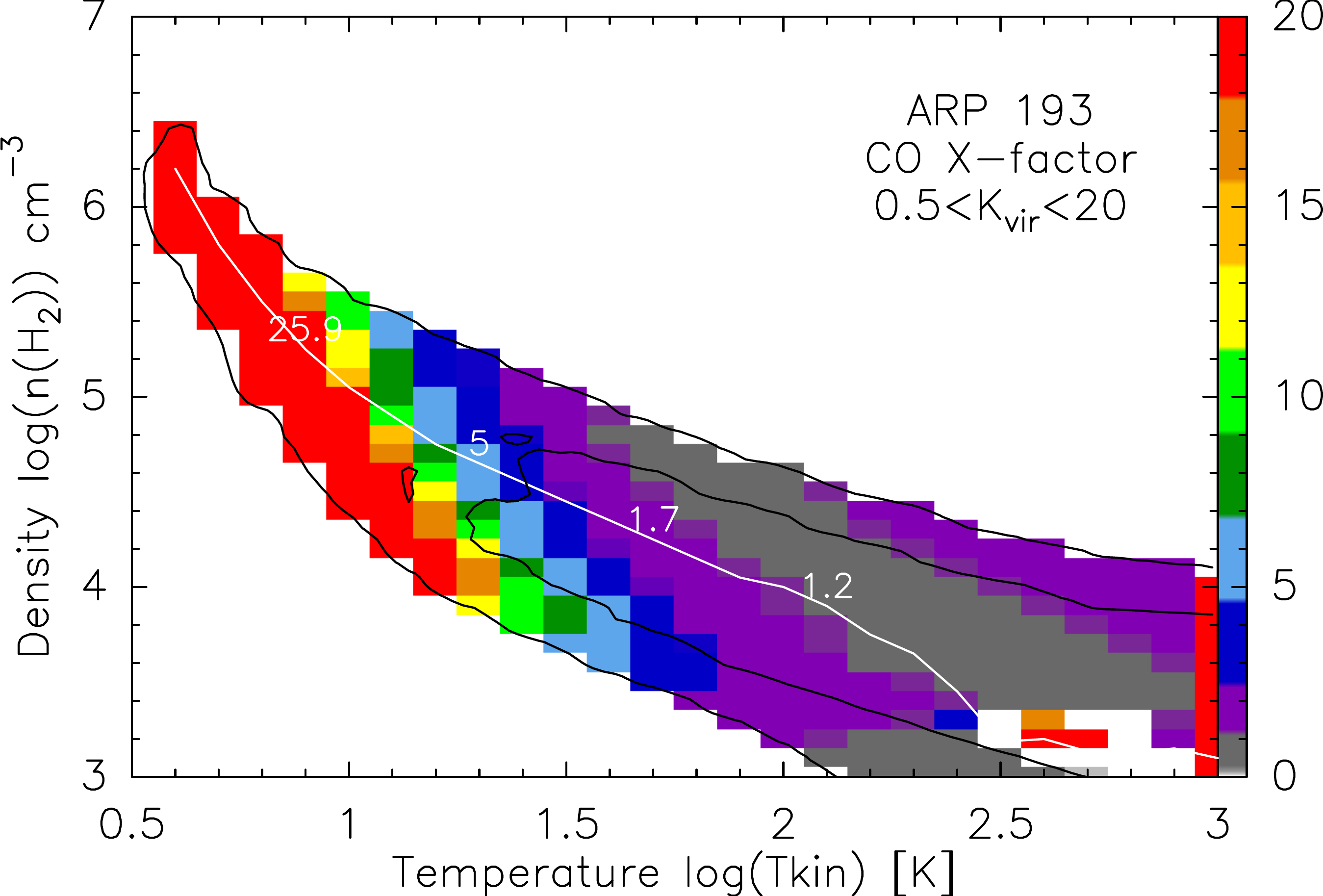}
\plotone{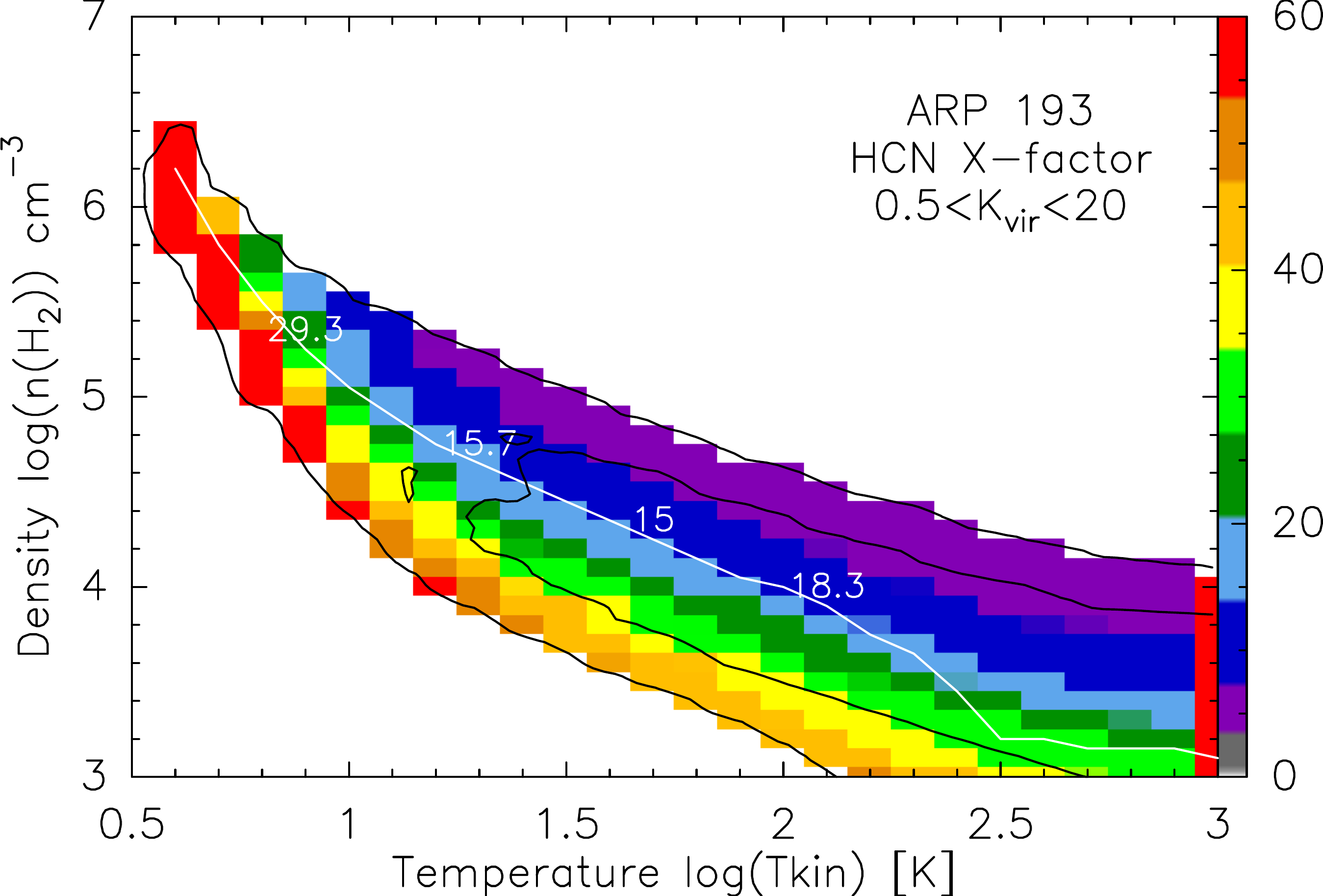}
\plotone{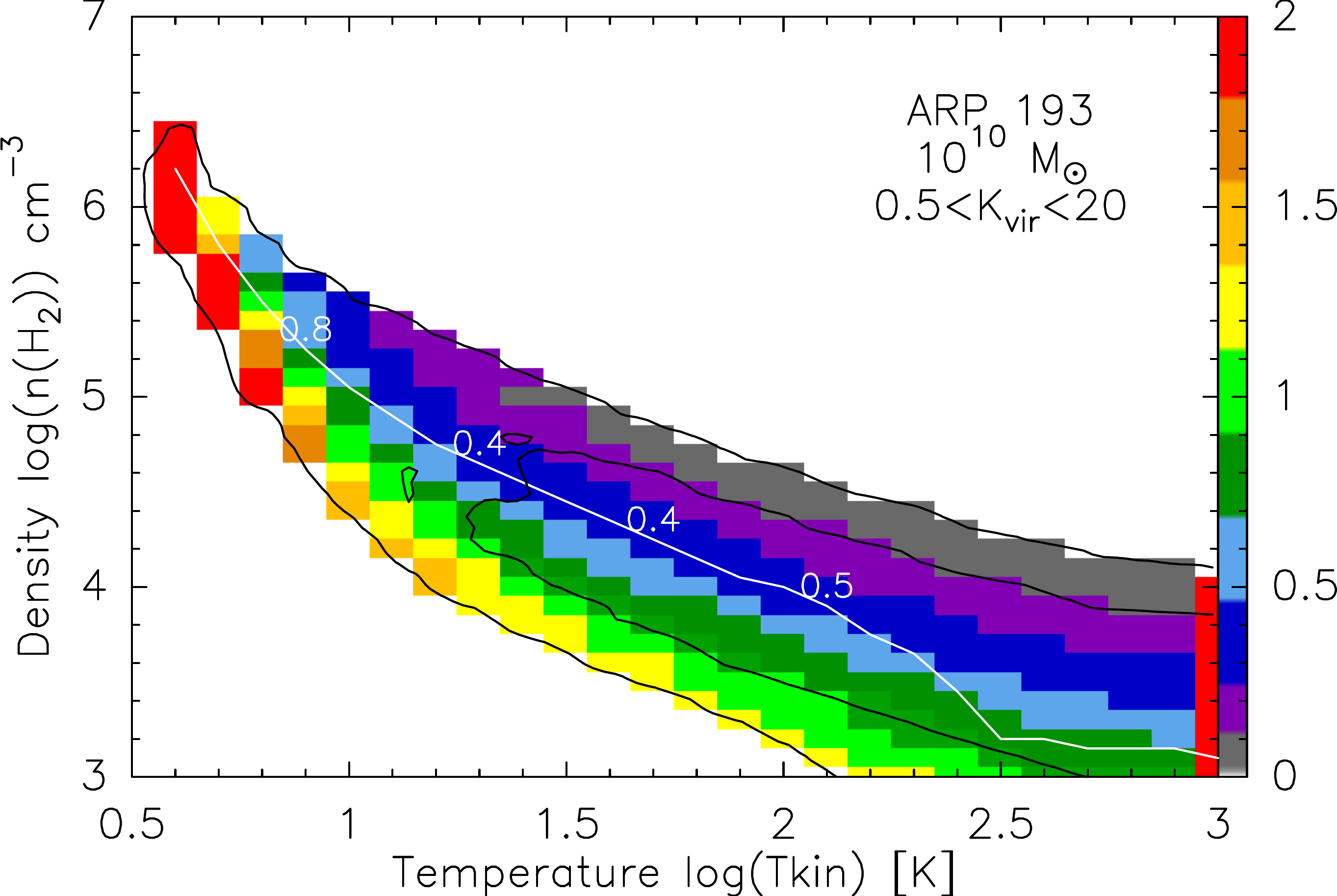}
\caption{Colors and  contours as  in Figure 14,  but for  Arp\,193 and
  using  the HCN-constrained  LVG  parameter space  that now  includes
  unbound  gas   states  (see  Figure   9).  The  X  factor   in  $\rm
  M_{\odot}(K\,km\,s^{-1}\,pc^2)^{-1}$   and    $\rm   M_{dense}$   in
  $10^{10}$\,M$_{\odot}$.}
\end{figure}

\clearpage
\newpage

\begin{figure}
\centering
\epsscale{1.10}
\plottwo{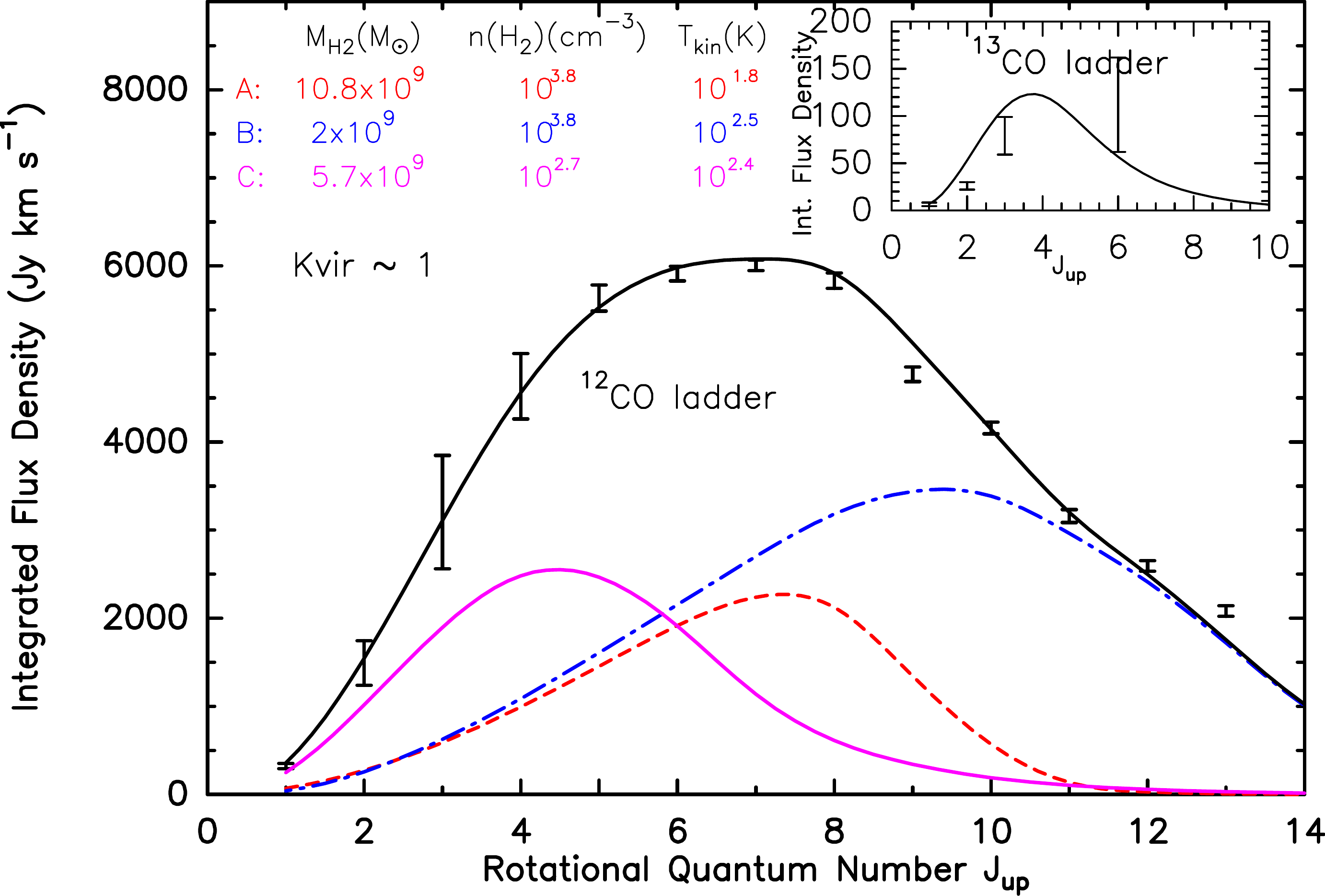}{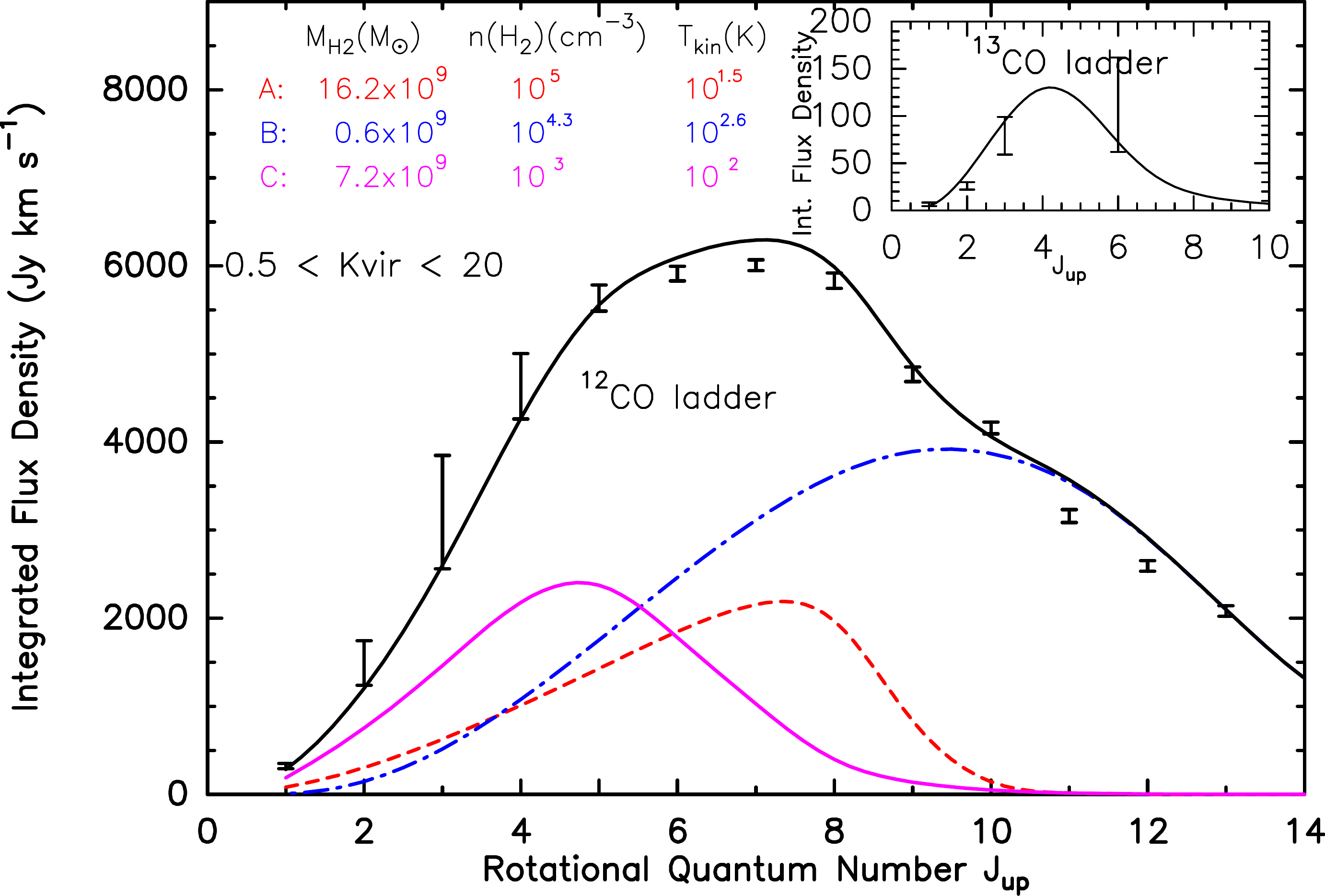}
\caption{NGC\,6240:  The two  CO SLED decompositions  allowed by
  the $\rm M_{dyn}$ constraint (see  Table 3, section 3.3).  The dense
  components (A) and  (B) (red, blue dotted lines) are  drawn from the
  LVG solution space compatible with  the HCN, HCO$^{+}$, and CS SLEDs
  of  this  system (see  section  3.1,  Figs 5,  6,  and  7), while  a
  lower-density component  (C) (pink) accounts  for the low-J  CO line
  emission.     The   $^{13}$CO    SLED    was    fitted   using    an
  r=[$^{12}$CO/$^{13}$CO]  abundance of  r=300 ($\rm  K_{vir}$$\sim $1
  decomposition)   and    r=500   (1$\la   $$\rm    K_{vir}$$\la   $20
  decomposition).}
\end{figure}

\begin{figure}
\centering
\epsscale{1.10}
\plottwo{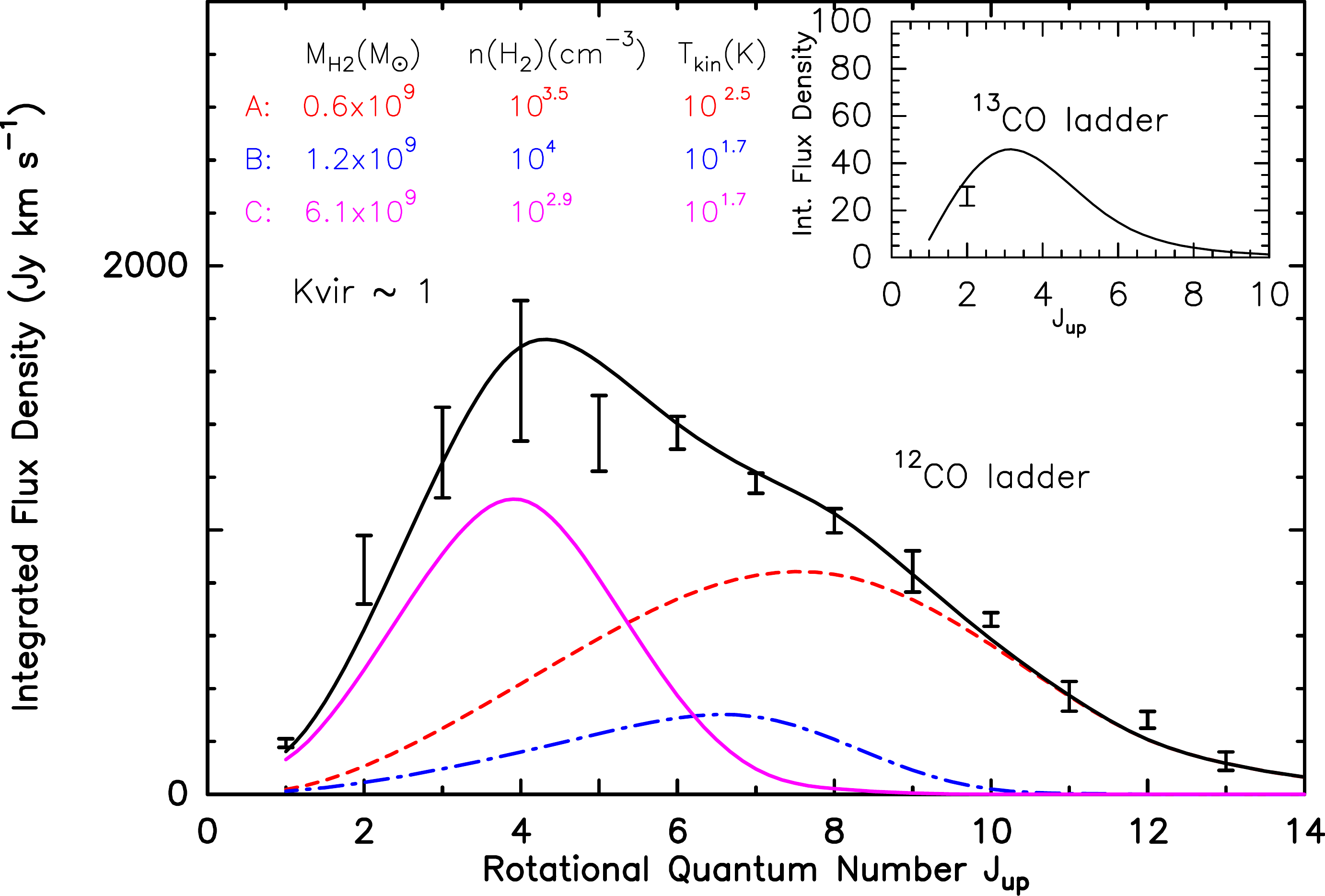}{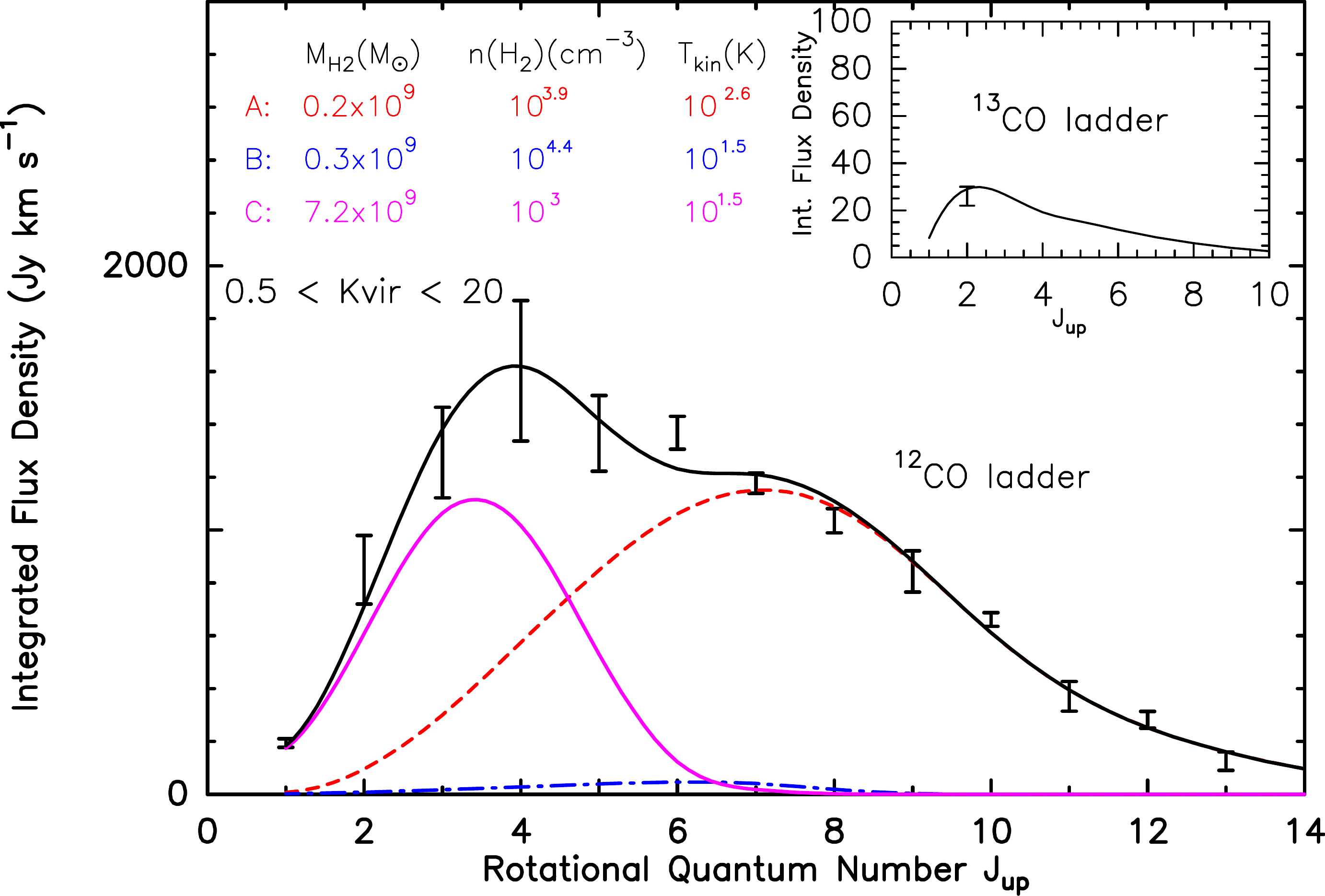}
\caption{Arp\,193: The two CO  SLED decompositions (see Table 3,
  and  section 3.3).   The dense  components  (A) and  (B) (red,  blue
  dotted lines) are drawn from  the LVG solution space compatible with
  the HCN  SLED of this  galaxy (see section  3.1, Figure 5),  while a
  lower-density component  (C) (pink) accounts  for the low-J  CO line
  emission.  The $^{13}$CO J=2--1 line  in both cases was fitted using
  r=[$^{12}$CO/$^{13}$CO]=150.}
\end{figure}

\clearpage
\newpage

\begin{figure}
\centering
\epsscale{1.0}
\plottwo{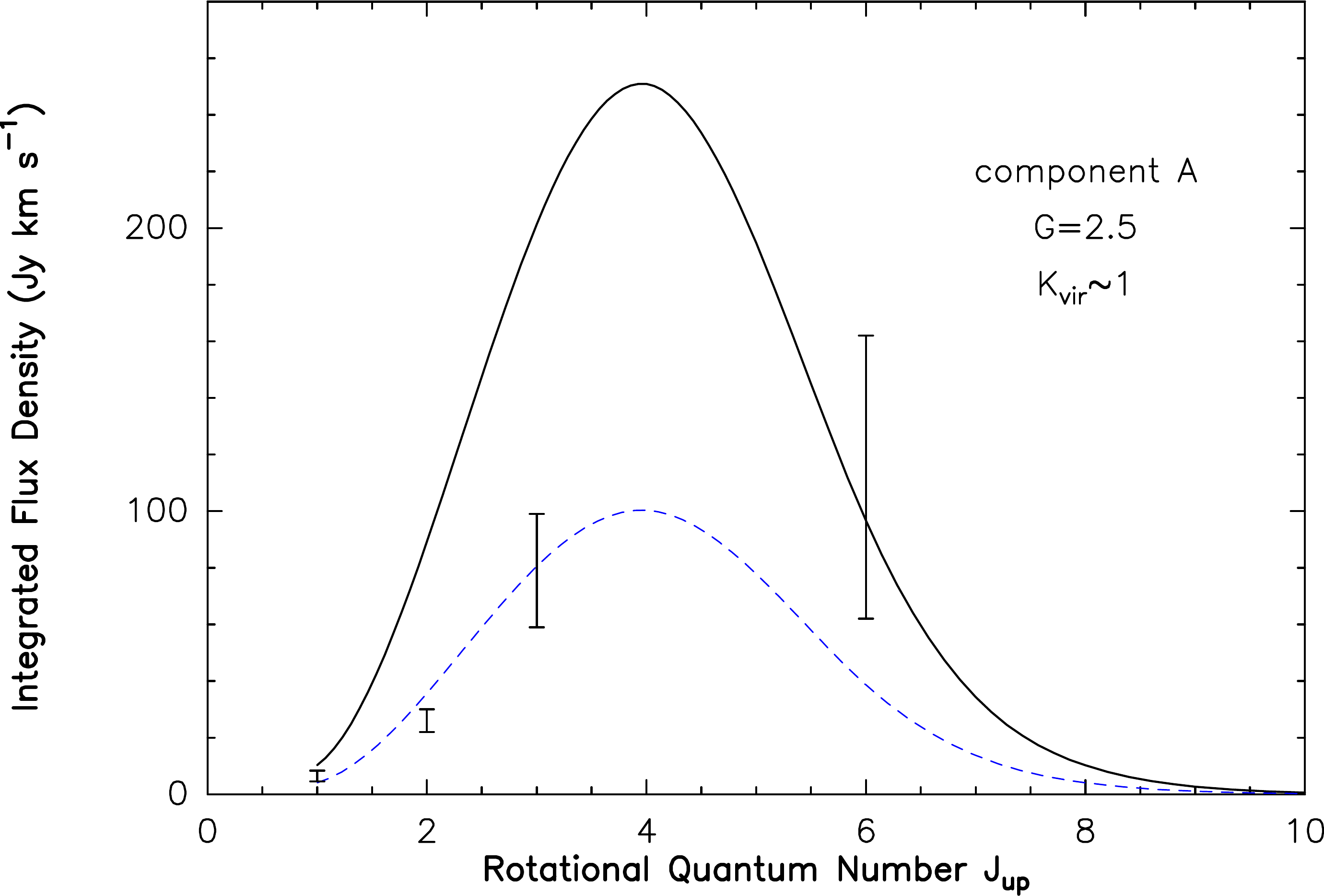}{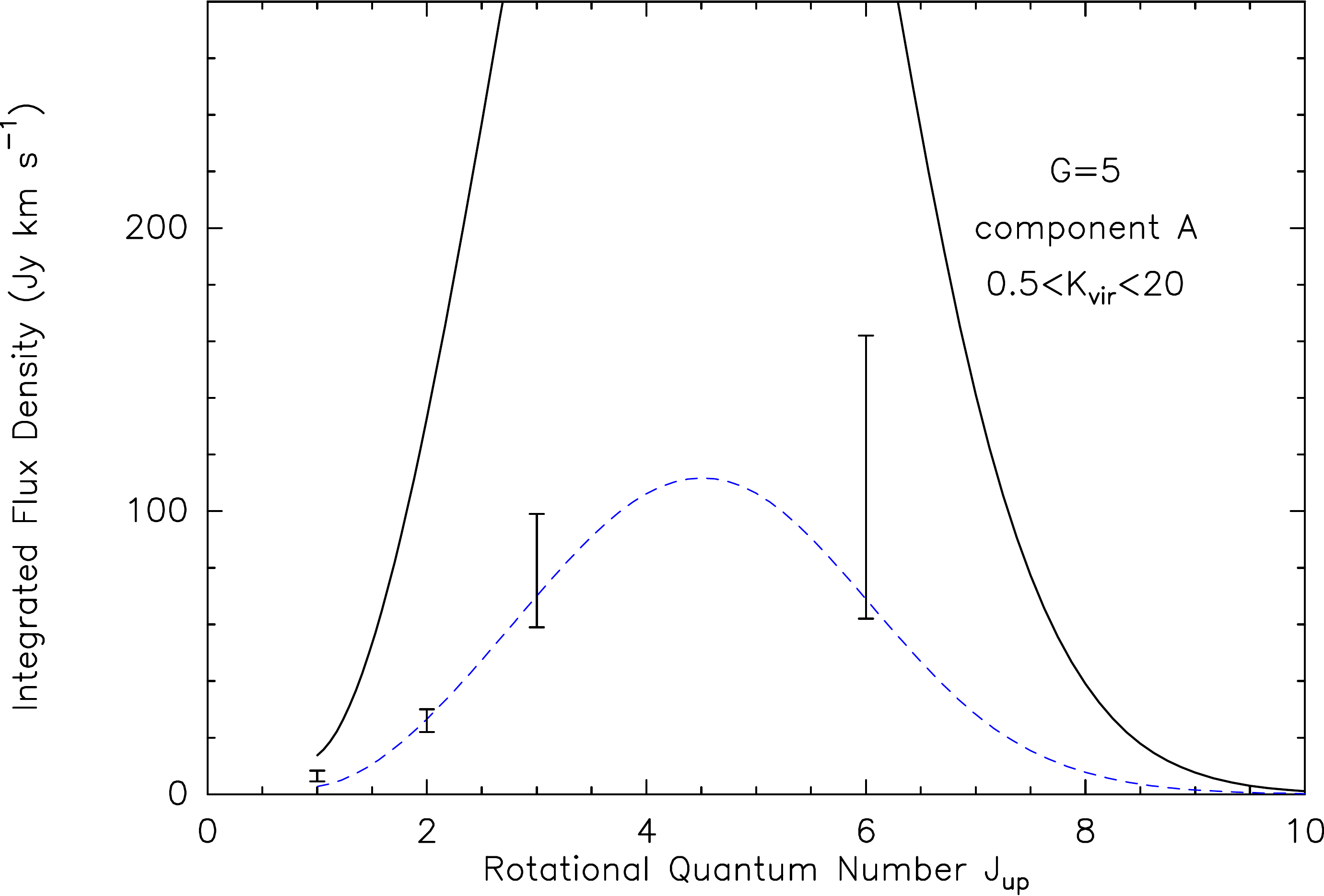}
\caption{Solid line:  the $^{13}$CO  SLED of  component (A),  the most
  massive   of  the  HCO$^{+}$/HCN/CO SLED  decomposition  of  the
  NGC\,6240  (see Table  3), for  [CO/$^{13}$CO]=80. Dashed  line: its
  scaled-down version by a factor of G so that it comes roughly to the
  level  of  the  observed   $^{13}$CO  line  strengths.   Note  that,
  counterintuitively,  it  is   the  super-virial  decomposition  that
  presents  the largest  discrepancy  (i.e. the  higher densities  help
  maintain high line  optical depths despite the  higher $\rm K_{vir}$
  values that act to lower them).  }
\end{figure}


\begin{figure}
\centering
\epsscale{1.10}
\plottwo{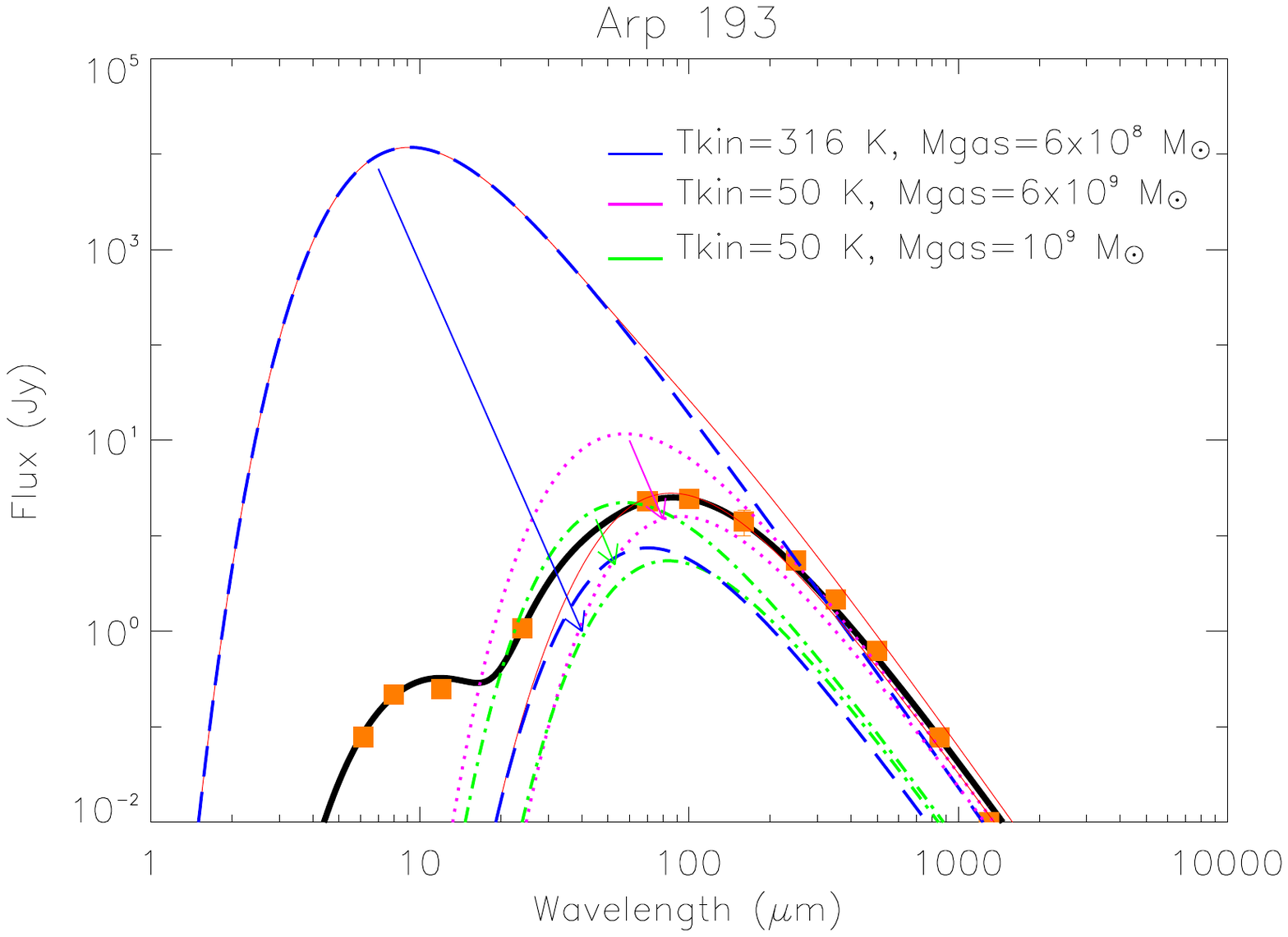}{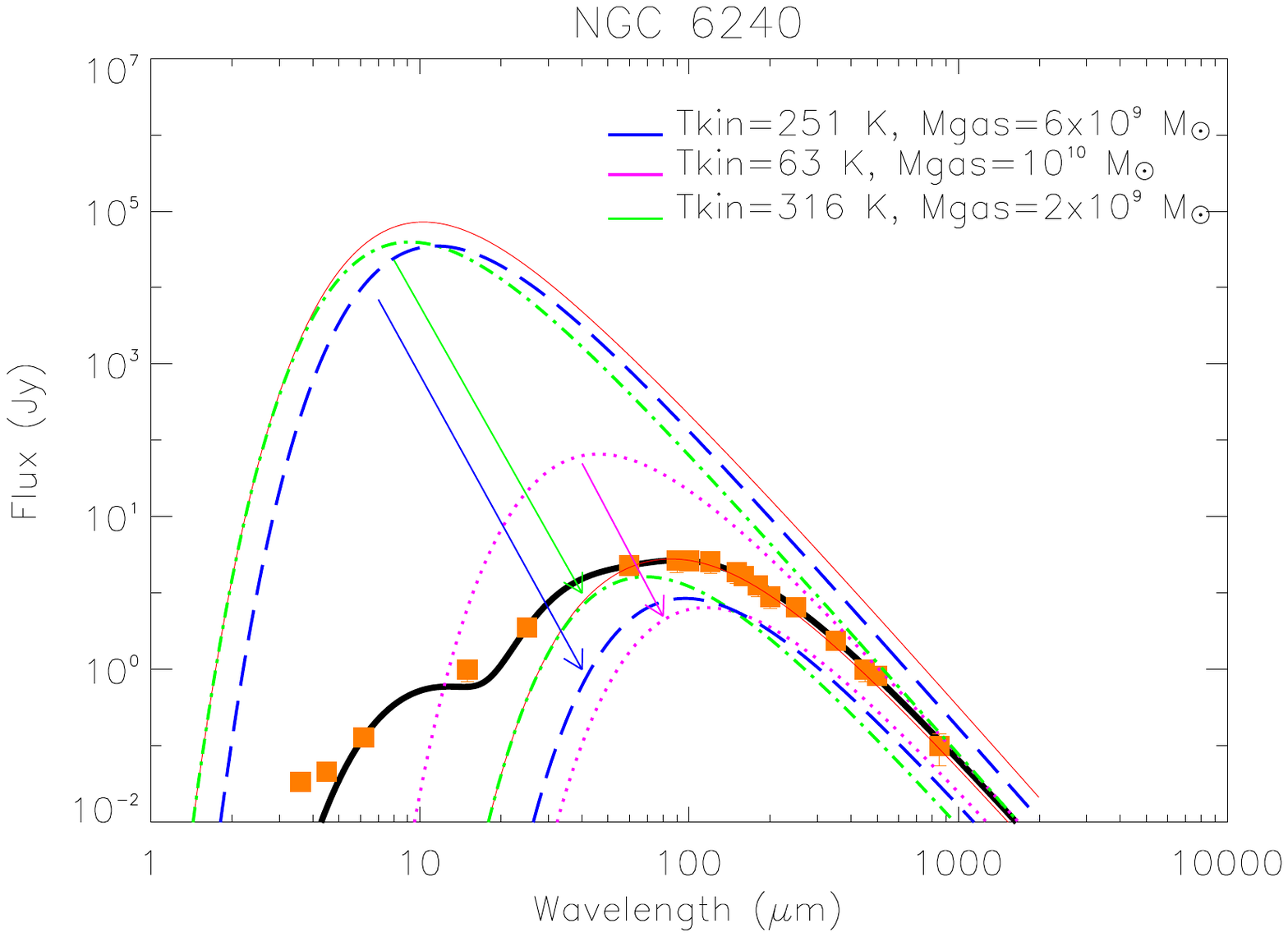}
\caption{Dust SEDs corresponding to the molecular SLEDs decompositions
  of the two galaxies  (see 4.1).  The blue-dashed, magenta-dotted and
  green-dash/dotted  curves show  the  SLED-deduced dust  SEDs of  the
  three gas  components (their $\rm  T_{kin}$ and $\rm  M_{gas}$ shown
  inside the  plot) for the $\rm K_{vir}$$\sim  $1 decompositions (see
  Table 3).  For T$_{\rm  dust}$=T$_{\rm kin}$ the corresponding total
  dust SED  greatly exceeds the observed one  (black line), especially
  for NGC\,6240.   For Arp\,193 this also  happens but is  due only to
  the warmest  (and least massive)  of its three components.   In such
  cases only  $\rm T_{dust}$$<$$\rm T_{kin}$ allows  the computed dust
  SED (solid red line) to  match the far-IR/submm part of the observed
  one  (black line,  with orange  rectangles  denoting observations).
The arrows mark the lowering  of $\rm T_{dust}$ from the $\rm T_{kin}$
for each of the three gas~components.}
\end{figure}


\end{document}